\documentclass[twocolumn]{aastex61}
\usepackage{hyperref}
\usepackage{color}
\usepackage{comment}
\usepackage{ifthen}
\usepackage{grffile}
\usepackage[utf8]{inputenc}
\usepackage{bm}
\usepackage{amsmath,mathtools}

\specialcomment{notes}{\begingroup\sffamily\color{red}}{\endgroup}
\includecomment{comment}

\begin{document}

\title{The Unusual Initial Mass Function of the Arches Cluster}

\affiliation{Institute for Astronomy, University of Hawaii, 2680 Woodlawn Drive, Honolulu, HI 96822, USA}
\author{Matthew W. Hosek Jr.}
\correspondingauthor{Matthew W. Hosek Jr.}
\email{mwhosek@hawaii.edu}

\author{Jessica R. Lu}
\affiliation{Department of Astronomy, 501 Campbell Hall, University of California, Berkeley, CA, 94720}
\affiliation{Institute for Astronomy, University of Hawaii, 2680 Woodlawn Drive, Honolulu, HI 96822, USA}

\author{Jay Anderson}
\affiliation{Space Telescope Science Institute, 3700 San Martin Drive, Baltimore, MD 21218, USA}

\author{Francisco Najarro}
\affiliation{Departamento de Astrof\'isica, Centro de Astrobiolog\'ia, (CSIC-INTA), Ctra. Torrej\'on a Ajalvir, km 4, 28850 Torrej\'on de Ardoz, Madrid, Spain}

\author{Andrea M. Ghez}
\affiliation{UCLA Department of Physics and Astronomy, Los Angeles, CA 90095}

\author{Mark R. Morris}
\affiliation{UCLA Department of Physics and Astronomy, Los Angeles, CA 90095}

\author{William I. Clarkson}
\affiliation{Department of Natural Sciences, University of Michigan-Dearborn, 4901 Evergreen Road, Dearborn, MI 48128}

\author{Saundra M. Albers}
\affiliation{Department of Astronomy, 501 Campbell Hall, University of California, Berkeley, CA, 94720}

\begin{abstract}
As a young massive cluster in the Central Molecular Zone, the Arches cluster is a valuable probe of the stellar Initial Mass Function (IMF) in the extreme Galactic Center environment. We use multi-epoch Hubble Space Telescope observations to obtain high-precision proper motion and photometric measurements of the cluster, calculating cluster membership probabilities for stars down to $\sim$1.8 M$_{\odot}$ between cluster radii of 0.25 pc -- 3.0 pc. We achieve a cluster sample with just $\sim$6\% field contamination, a significant improvement over photometrically-selected samples which are severely compromised by the differential extinction across the field. Combining this sample with K-band spectroscopy of 5 cluster members, we forward model the Arches cluster to simultaneously constrain its IMF and other properties (such as age and total mass) while accounting for observational uncertainties, completeness, mass segregation, and stellar multiplicity. We find that the Arches IMF is best described by a 1-segment power law that is significantly top-heavy: $\alpha$ = 1.80 $\pm$ 0.05 (stat) $\pm$ 0.06 (sys), where dN/dm $\propto$ m$^{-\alpha}$, though we cannot discount a 2-segment power law model with a high-mass slope only slightly shallower than local star forming regions ($\alpha$ = 2.04$^{+0.14}_{-0.19}$ $\pm$ 0.04) but with a break at 5.8$^{+3.2}_{-1.2}$ $\pm$ 0.02 M$_{\odot}$. In either case, the Arches IMF is significantly different than the standard IMF. Comparing the Arches to other young massive clusters in the Milky Way, we find tentative evidence for a systematically top-heavy IMF at the Galactic Center.
\end{abstract}

\section{Introduction}
A fundamental quantity in star formation is the Initial Mass Function (IMF), which describes the distribution of stellar masses created during star formation. Though its functional form is debated \citep[e.g.][]{Chabrier:2005so}, the IMF is often represented as a multi-part power-law given by dN/dm $\propto$ m$^{-\alpha}$, where:
\begin{equation}
\alpha =
\begin{cases}
0.3 \pm 0.4, & \text{for } 0.01 < m / M_{\odot} \lesssim 0.08 \\
1.3 \pm 0.3, & \text{for } 0.08 < m / M_{\odot} \leq 0.5 \\
2.3 \pm 0.36, & \text{for } 0.5 < m / M_{\odot} \leq 150 \\
\end{cases}
\end{equation}
as discussed in \citet{Kroupa:2002lq}. Stellar populations in the Milky Way and nearby galaxies have been found to be consistent with this ``local IMF'', leading to the suggestion that it may be a universal property of star formation \citep[see reviews by][and references therein]{Bastian:2010dp, Offner:2014vn}. Thus, the local IMF is often used to describe stellar populations throughout the universe.

However, it is unknown whether the local IMF is applicable to environments other than those found in local star formation regions. Of particular interest are starburst environments, which exhibit extremely high gas densities and temperatures, radiation fields, and turbulence \citep[e.g.][]{Swinbank:2011rg}. Some studies predict that the increased thermal Jeans mass results in an overabundance of high-mass stars and a ``top-heavy'' IMF \citep[e.g.][]{Larson:2005sp, Bonnell:2006os, Klessen:2007nx, Bonnell:2008jl, Papadopoulos:2011qu, Narayanan:2013ec}. Alternatively, others claim that the IMF is set by the mass distribution of pre-stellar cores within a molecular cloud (the core mass function, or CMF), which itself is set by turbulence \citep[e.g.][]{Padoan:2002tw, Hopkins:2012eu}. These theories predict that the increased turbulence in starburst environments would favor the formation of low-mass stars and a ``bottom-heavy'' IMF \citep[][]{Hopkins:2013oz, Chabrier:2014oq}. However, recent simulations suggest that CMF cannot be directly mapped to the IMF \citep[e.g.][]{2016MNRAS.462.4171B, Liptai:2017xc}. A third set of studies contend that the IMF is driven by local processes such as radiative feedback \citep[e.g.][]{Bate:2009jw, Offner:2009ng, Krumholz:2011ye, Krumholz:2012qf}, and is largely independent of environment \citep[e.g.][]{Guszejnov:2016gs}. Thus, understanding how the IMF behaves in starburst environments yields critical insight into the underlying physics driving star formation \citep[e.g.][]{Krumholz:2014ne}.

There is some observational evidence that the IMF changes in starburst environments, though these results are debated. Studies of massive elliptical galaxies have found that the IMF becomes increasingly bottom-heavy with increasing velocity dispersion and/or $\alpha$-element enhancement, conditions that reflect starburst-like conditions \citep[e.g.][]{Conroy:2012fk, Conroy:2013ju, Cappellari:2012zh, 2013MNRAS.432.1862C, La-Barbera:2013wy, Spiniello:2014oq, Li:2017oq}. Further studies suggest that the cores of massive galaxies, which are thought to have formed rapidly in starburst-like environments at high redshift \citep[e.g.][]{Oser:2010tp}, are systematically bottom-heavy relative to the rest of the galaxy \citep[e.g.][]{Martin-Navarro:2015kq, van-Dokkum:2017bx, Conroy:2017qc, Parikh:2018bq}. However, these results rely on modeling stellar populations from unresolved stellar spectra, which is prone to systematic effects such as elemental abundance gradients \citep[e.g.][]{McConnell:2016am, Zieleniewski:2015kc, Zieleniewski:2017ye, Vaughan:2018qp}. Overall, the consistency of IMF determinations for a single galaxy using spectroscopic, kinematic, and lensing methods has not yet been established, with some galaxies showing agreement and others showing significant discrepancies \citep{Lyubenova:2016xa, Newman:2017ay}. This highlights the difficulty of measuring the IMF from these complex and unresolved stellar populations.

Massive star clusters in starburst galaxies (also known as super star clusters, or starburst clusters) also offer a probe into starburst environments. Still unresolved with current observing facilities, their mass functions are inferred from the light-to-mass ratios \citep[e.g.][]{Ho:1996uv}. This analysis also faces many challenges, including the need for virial equilibrium, uncertainties in stellar models and extinction corrections, the impact of mass segregation and multiplicity, and anisotropy in the velocity dispersion \citep[e.g.][]{Bastian:2007pu}. A range of both bottom-heavy and top-heavy IMFs have been reported for these clusters, perhaps as a result of these difficulties \citep[][]{Larsen:2004xw, McCrady:2005tt, Bastian:2006kz}.

Ideally, one would directly measure the IMF of starburst environments using resolved stellar populations. Such investigations are possible at the Milky Way Galactic Center (GC), which has been shown exhibit similar densities, temperatures, and kinematics to those in starburst galaxies \citep{Kruijssen:2013lh, Ginsburg:2016nr}. The GC contains several young massive clusters whose youth and high mass make them ideal tools for measuring the IMF \citep{Morris:1996yq}. The Young Nuclear Cluster (YNC; $\sim$2.5 -- 5.8 Myr, M $\gtrsim$ 2x10$^4$ M$_{\odot}$), which lies within the central parsec of the galaxy, has been found to have a top-heavy IMF with $\alpha$ = 1.7 $\pm$ 0.2 \citep[][]{Lu:2013wo}. The Arches cluster \citep[2 -- 4 Myr, M $\sim$ 4--6 x 10$^4$ M$_{\odot}$;][]{Martins:2008hl, Clarkson:2012ty}, located within the Central Molecular Zone (CMZ) and at a projected distance of $\sim$26 pc from the central supermassive black hole, offers an additional opportunity to probe the IMF in this extreme environment.

Despite many efforts, the IMF of the Arches cluster has not yet been established. This is due to two significant challenges: mass segregation and differential extinction. As a result of mass segregation, the \emph{present-day mass function} (PDMF) of the inner region (r $\lesssim$ 0.5 pc) has been measured to be top-heavy \citep{Figer:1999lo, Stolte:2002zr, Stolte:2005mz, Kim:2006fy}, while the outer regions (r $\gtrsim$ 0.5 pc) have been found to be either consistent with the local IMF or bottom-heavy \citep{Espinoza:2009bs, Habibi:2013th}. Dynamical modeling is required to determine whether the observed PDMF is consistent with the local IMF \citep[e.g.][]{Kim:2000wd, Harfst:2010ys, Park:2018tg}, though the uncertainty in cluster orbit \citep[][]{Stolte:2008qy} and initial conditions requires that a large parameter space must be considered.

In addition, inferring the IMF from the PDMF depends heavily on the PDMF at large cluster radii, where the differences between dynamical models are the largest \citep[e.g., Figure 13 of][]{Habibi:2013th}. However, significant differential extinction \citep[$\Delta$A$_V$ $\sim$ 15 mag;][]{Habibi:2013th} makes it challenging to separate the cluster from field populations via photometry, especially at large radii where field-star contamination can be high \citep[e.g.][]{Stolte:2005mz}. Measurements of the internal velocity dispersion of the cluster indicate that its mass function is top-heavy and/or truncated at low masses \citep{Clarkson:2012ty}, but this has yet to be confirmed by direct star counts.

In this paper, we combine multi-epoch \emph{Hubble Space Telescope} (\emph{HST}) WFC3-IR observations with Keck OSIRIS K-band spectroscopy to measure the IMF of the Arches cluster for M $>$ 1.8 M$_{\odot}$. We describe our observations in $\mathsection$\ref{sec:obsAll} and our methods for calculating cluster membership probabilities, correcting for extinction, and measuring observational completeness in $\mathsection$\ref{sec:methodsAll}. In $\mathsection$\ref{sec:model} we detail our forward modeling technique for constraining the IMF, and in $\mathsection$\ref{sec:resultsAll} we present our result that the Arches cluster IMF is inconsistent with the local IMF. We compare our result to to past Arches IMF measurements and place it in context with other young massive clusters in the Milky Way in $\mathsection$\ref{sec:discussion}. We present our conclusions in $\mathsection$\ref{sec:conclusions}.

\section{Observations and Measurements}
\label{sec:obsAll}

\subsection{HST Photometry and Astrometry}
\label{sec:photometry}
\begin{figure}
\begin{center}
\includegraphics[scale=0.40]{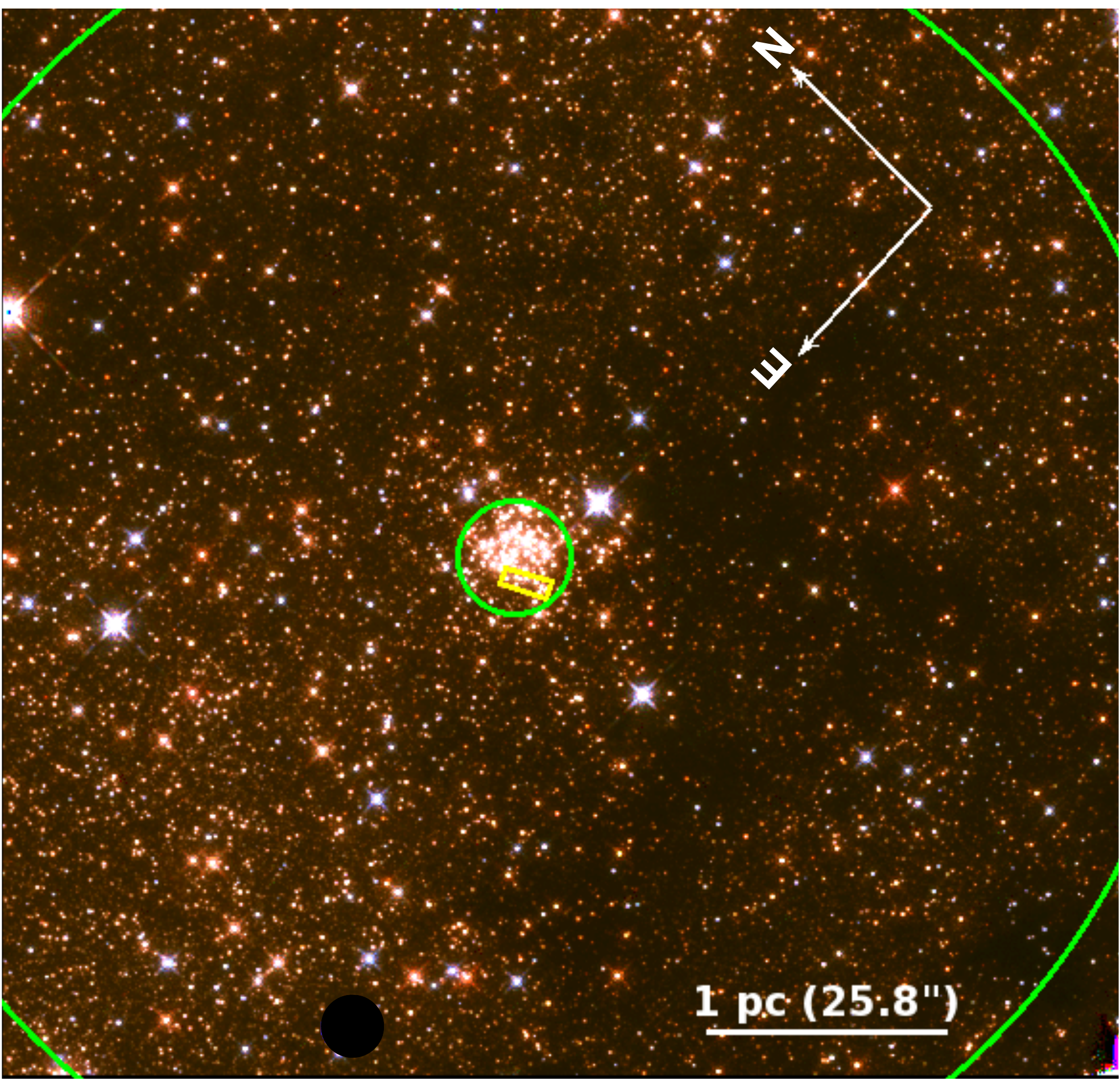}
\caption{Three color \emph{HST} image of the Arches Cluster, with F127M~=~blue, F139M = green, and F153M = red. The inner and outer green circles represent cluster radii of 0.25 pc and 3.0 pc, which define the inner and outer boundaries of our \emph{HST} sample, respectively. The yellow box near the center of the cluster corresponds to the Keck OSIRIS field, where K-band spectroscopy of 5 cluster members were obtained. The hole in the lower left side of the image is due to a known defect in the WFC3IR chip.}
\label{fig:Arches}
\end{center}
\end{figure}

Astrometry and photometry of the Arches cluster were obtained from observations with the infrared channel of the Wide Field Camera 3 (WFC3-IR) on the Hubble Space Telescope (\emph{HST}) for 4 epochs between 2010 and 2016. The 2010 epoch contains images in the F127M, F139M, and F153M filters (GO-11671, PI: Ghez, A.M.), while the 2011, 2012, and 2016 epochs have images only in the F153M filter (GO-12318, GO-12667, PI: Ghez, A.M.; GO-14613, PI: Lu., J.R.). A detailed description of the 2010 -- 2012 observations is provided in \citep[][hereafter H15]{Hosek:2015cs}. The 2016 observations were designed to mimic the earlier F153M epochs in order to maximize the astrometric precision between the data sets. These observations have a field of view (FOV) of 120" x 120", providing coverage of at least 30\% of the cluster area within successive circular annuli of width 0.25 pc out to 3 pc (Figure \ref{fig:Arches}).

We extract high-precision astrometry and photometry using the \texttt{FORTRAN} codes \texttt{img2xym\_wfc3ir}, a version of the \texttt{img2xym\_WFC} package for WFC3-IR \citep{Anderson:2006il}, and \texttt{KS2}, a generalization of the software developed for the Globular Cluster Treasury Program \citep[][ see also \citeauthor{Bellini:2018ow} 2018]{Anderson:2008qy}. A detailed description of this procedure and the analysis of the subsequent astrometric and photometric errors is provided in Appendix A of H15. In short, point-spread function (PSF) fit astrometry and photometry is extracted using a grid of spatially-varying PSF models across the field. No significant differences in measurement precision were found for the 2016 epoch compared to the previous epochs, with average astrometric and photometric errors of 0.16 mas (1.3x10$^{-3}$ pixel) and 0.008 mag, respectively, for the brightest non-saturated stars (error on the mean calculated from 21 individual measurements). The photometry is calibrated to the Vega magnitude system using the improved \texttt{KS2} zero-points derived in \citet[][hereafter H18]{Hosek:2018lr}, which uses significantly more stars than the original zero-point derivation in H15.

\begin{figure}
\begin{center}
\includegraphics[scale=0.35]{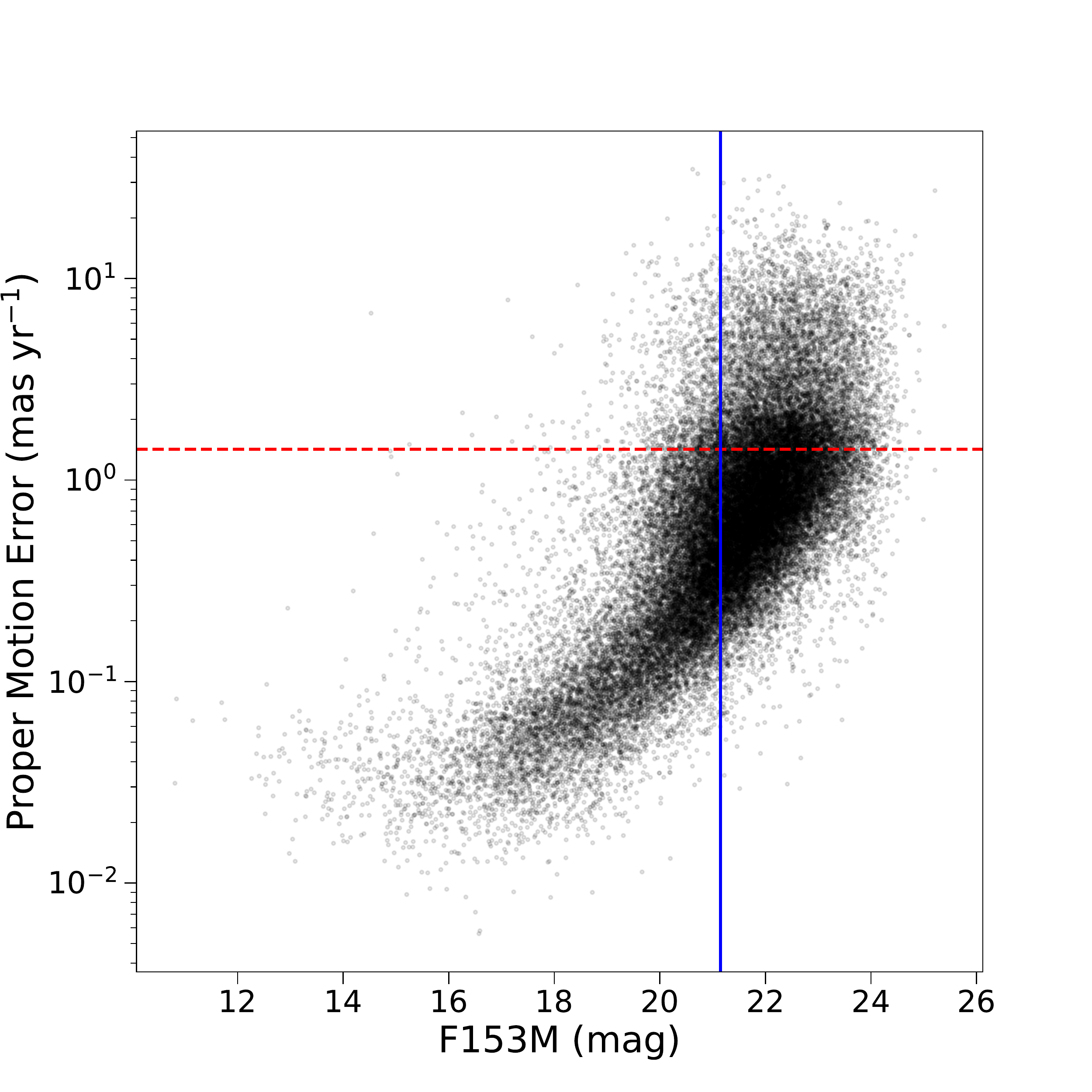}
\caption{Proper motion error as a function of F153M magnitude in the final star catalog. For each star, the error shown is the average between the X and Y directions. The red dotted line denotes the proper motion error limit of 1.42 mas yr$^{-1}$ required for membership analysis ($\mathsection$\ref{sec:Membership}). The solid blue line shows the completeness limit of F153M = 21.18 mag, which corresponds to $\sim$1.8 M$_{\odot}$ ($\mathsection$\ref{sec:comp}). These errors are $\sim 3 \times$ lower than those reported in H15 due to the increased time baseline.}
\label{fig:pm_err}
\end{center}
\end{figure}

The stellar positions in each epoch are transformed into a master astrometric reference frame using a 2nd-order polynomial transformation in both X and Y (12 free parameters). The master frame is constructed such that there is no net motion of the cluster, as only high-probability cluster members ($\geq$0.7) in the H15 catalog are used as reference stars. An iterative process is used to match stars, calculate initial proper motions, and then rematch stars using those proper motions to identify stars across the epochs. The star matching is done by position, using a search radius of 0.5 pix (0.06"). Proper motions are calculated for stars detected in at least 3 F153M epochs using a linear fit to the X and Y positions as a function of time, weighted by their astrometric errors. The final star catalog contains $\sim$45,000 stars with proper motion errors 3 times smaller than H15 on account of the increased time baseline, reaching a precision of $\sim$0.03 mas yr$^{-1}$ at the bright end (Figure \ref{fig:pm_err}).

\subsection{Keck OSIRIS Spectroscopy}
\label{sec:spectroscopy}
K-band spectroscopy of a sample of Arches cluster members was obtained using the OH-Suppressing Infrared Integral Field Spectrograph \citep[OSIRIS;][]{Larkin:2006hh} with Laser Guide Star Adaptive Optics \citep{Wizinowich:2006dk} on the Keck I telescope on 2014 May 16. The Kbb filter was used with the 0.10" pixel scale, which provides a spectral coverage of 1.965~$\mu$m~--~2.381~$\mu$m at R $\sim$ 3800 over a 1.6" x 6.4" FOV. A single field was observed near the core of the cluster  (J2000: $\alpha$~=~17:45:50.7, $\delta$~=~-28:49:23.4; Figure \ref{fig:Arches}) at a position angle of 28$^{\circ}$, using 10 dithered exposures of 900 s for a total integration time of 9000 s. This field was chosen to maximize the number of non-WR stars (F153M $\geq$ 14.5 mag, see $\mathsection$\ref{sec:sample}) while avoiding the densest inner region of the cluster. The spectroscopic sample contains five stars, as described in Table \ref{tab:spectraStars}.

\begin{deluxetable*}{l c c c c c c c c}
\tablewidth{0pt}
\tabletypesize{\footnotesize}
\tablecaption{OSIRIS Spectroscopic Sample}
\tablehead{
\colhead{Name\tablenotemark{a}} & \colhead{RA\tablenotemark{b}} & \colhead{DEC\tablenotemark{b}} & \colhead{Spectral Type\tablenotemark{c}} & \colhead{F127M}  & \colhead{F153M} & \colhead{A$_{Ks}$\tablenotemark{d}} & \colhead{T$_{eff}$}  & \colhead{log $g$} \\
& (J2000) & (J2000) & (literature) & \colhead{(mag)} & \colhead{(mag)} & \colhead{(mag)} & \colhead{(K)}   &  \colhead{(cgs)}
}
\startdata
47 & 17:45:50.68 & -28:49:24.39 & O4-5 Ia & 17.01 $\pm$ 0.01 & 14.85 $\pm$ 0.01  & 2.43 & 34750$^{+3000}_{-1500}$ & 3.50$^{+0.30}_{-0.15}$ \\
44 & 17:45:50.62 & -28:49:24.77 & -- & 17.18 $\pm$ 0.01 & 14.91 $\pm$ 0.01  & 2.43 & 34500$^{+3000}_{-1500}$ & 3.75$^{+0.15}_{-0.25}$  \\
53 & 17:45:50.64 & -28:49:24.14 & O4-5 Ia  & 17.10 $\pm$ 0.01 & 14.91 $\pm$ 0.01  & 2.43 & 37000$^{+2000}_{-2000}$  & 3.50$^{+0.30}_{-0.10}$  \\
55 & 17:45:50.73 & -28:49:24.54 & O5.5-6 I-III & 17.09 $\pm$ 0.01 & 14.97 $\pm$ 0.01 & 2.40 & 34500$^{+3000}_{-1500}$ & 3.85$^{+0.25}_{-0.15}$ \\
60 & 17:45:50.74 & -28:49:21.08 & O4-5 Ia  & 17.20 $\pm$ 0.01 & 15.03 $\pm$ 0.01  & 2.39 & 36000$^{+2500}_{-1500}$  & 3.60$^{+0.20}_{-0.15}$  \\
\enddata
\tablenotetext{a}{As defined in the catalog from \citet{Figer:2002nr}}
\tablenotetext{b}{Measured in 2010 F153M epoch}
\tablenotetext{c}{From \citet{Clark:2018ij}}
\tablenotetext{d}{Derived using the extinction map in $\mathsection$\ref{sec:extinction}}
\label{tab:spectraStars}
\end{deluxetable*}

The OSIRIS data cubes were reduced using version 4.1.0 of the OSIRIS data reduction pipeline\footnote{https://github.com/Keck-DataReductionPipelines/OsirisDRP/releases} \citep[ODRP;][]{2004SPIE.5492.1403K}. The ODRP corrects for dark current, electronic biases and crosstalk, and cosmic rays, and properly extracts the wavelength-calibrated spectrum at each spaxel (spatial pixel). The science data cubes were averaged together using the ``Mosaic Frames" module to create the master science data cube. One-dimensional science spectra were extracted using a 3x3 aperture box centered on the spaxel with the highest integrated flux for the star. This aperture size was chosen to maximize the signal-to-noise while minimizing contamination from nearby stars.

After extraction, the raw science spectra need to be corrected for contamination from sky features such as continuum, OH emission lines, and telluric absorption lines. The standard set of calibration observations (sky frames and telluric standards) were obtained at the telescope, but we found that the sky features were better corrected using the \texttt{Skycorr}\footnote{http://www.eso.org/sci/software/pipelines/skytools/skycorr} \citep{Noll:2014gr} and
\texttt{molecfit}\footnote{https://www.eso.org/sci/software/pipelines/skytools/molecfit} \citep{Smette:2015ge, Kausch:2015fe} software packages. \texttt{Skycorr} removes sky emission lines by fitting physically-related OH line groups in a reference sky spectrum and scaling them to match the science spectrum \citep[e.g.][]{Davies:2007dp}. The sky continuum is measured by a linear interpolation of the wavelength channels without line emission, and then combined with the OH line model to produce the final sky spectrum that is subtracted from the science spectrum. In this case, a reference sky spectrum for each star is extracted using a box annulus formed by a 5x5 and 7x7 spaxel box centered on the star itself, and then rescaled to science spectrum aperture size. Once \texttt{Skycorr} has removed the sky emission and continuum, the telluric absorption lines are modeled using \texttt{molecfit}, which uses a radiative transfer code and an atmospheric profile based on the date and location of the observations to predict atmospheric lines caused by molecules such as H$_2$O, CO$_2$, and CH$_4$. The telluric model is then divided out of the science spectrum to produce the final reduced science spectrum.

\begin{figure*}
\begin{center}
\includegraphics[scale=0.35]{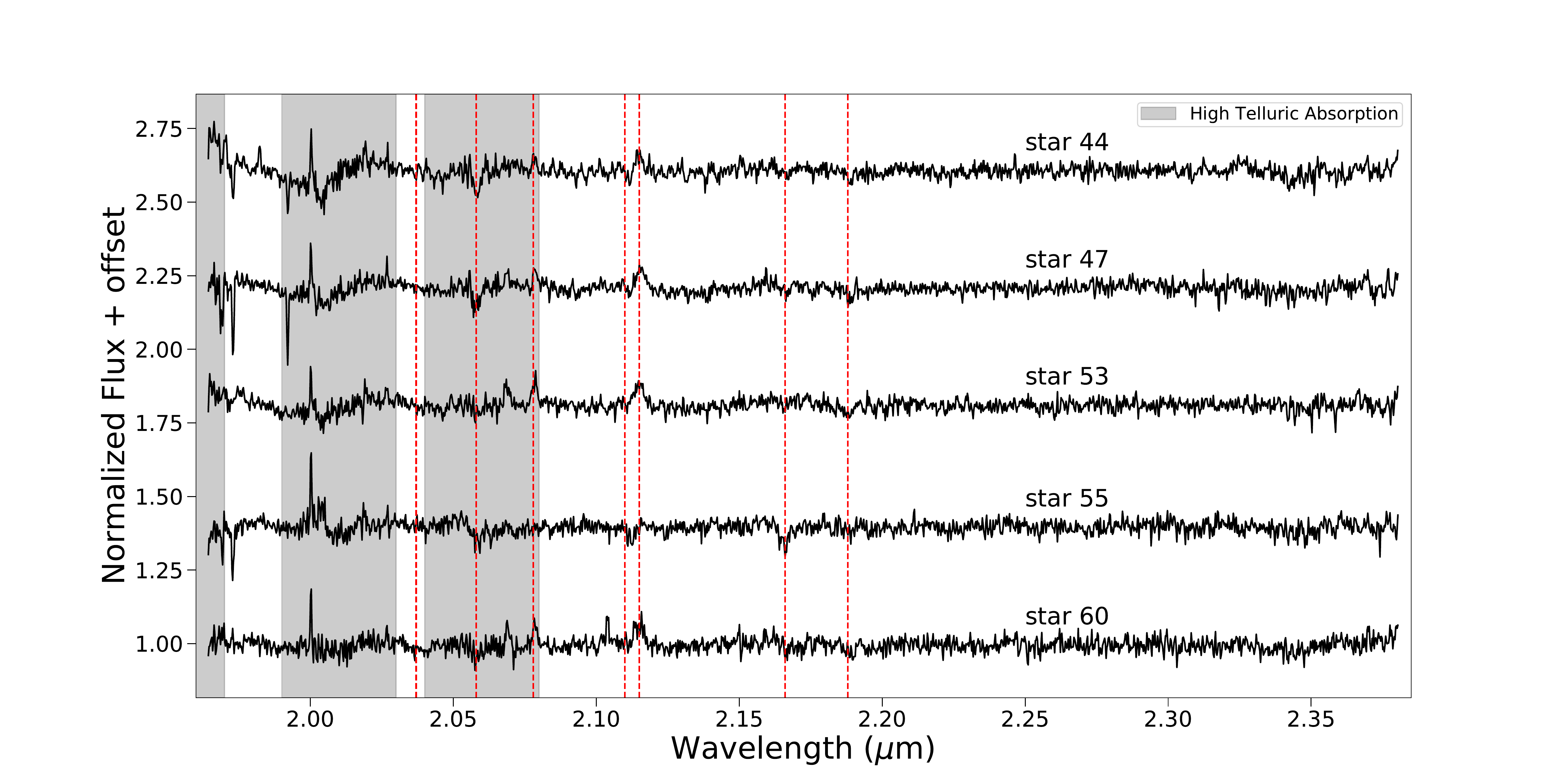}
\caption{Reduced OSIRIS spectra of Arches cluster members. The gray regions mark wavelengths with high telluric absorption, while the red dotted lines denote several useful spectral features.}
\label{fig:spectra}
\end{center}
\end{figure*}

However, as discussed by \citet{Lockhart:2017hp}, OSIRIS introduces a shape to the stellar continuum due to its varying sensitivity as a function of wavelength that cannot be modeled by \texttt{molecfit}. This requires an extra step of creating an OSIRIS ``flat'' free of sky, telluric, and stellar-flux contributions. We construct this flat using the observed telluric standards, empirically subtracting the sky and using \texttt{molecfit} to remove the telluric lines. In the A0 V spectrum, the only remaining feature is the Br-$\gamma$ line. To remove this line, we combine the A0 V and G2 V spectra using the technique described in \citet[][]{Do:2009jk}, replacing the A0 V spectrum between 2.155 $\mu$m and 2.175 $\mu$m with the spectrum of the G2 V star after it has been divided by the solar spectrum. Finally, we smooth the resulting spectrum using a median filter (kernel size = 51 pix) to create the OSIRIS flat. The science spectra are divided by this flat and normalized to produce the final science spectra (Figure \ref{fig:spectra}).

\section{Methods}
\label{sec:methodsAll}

\subsection{Proper-Motion Based Cluster Membership}
\label{sec:Membership}
\begin{figure}
\begin{center}
\includegraphics[scale=0.3]{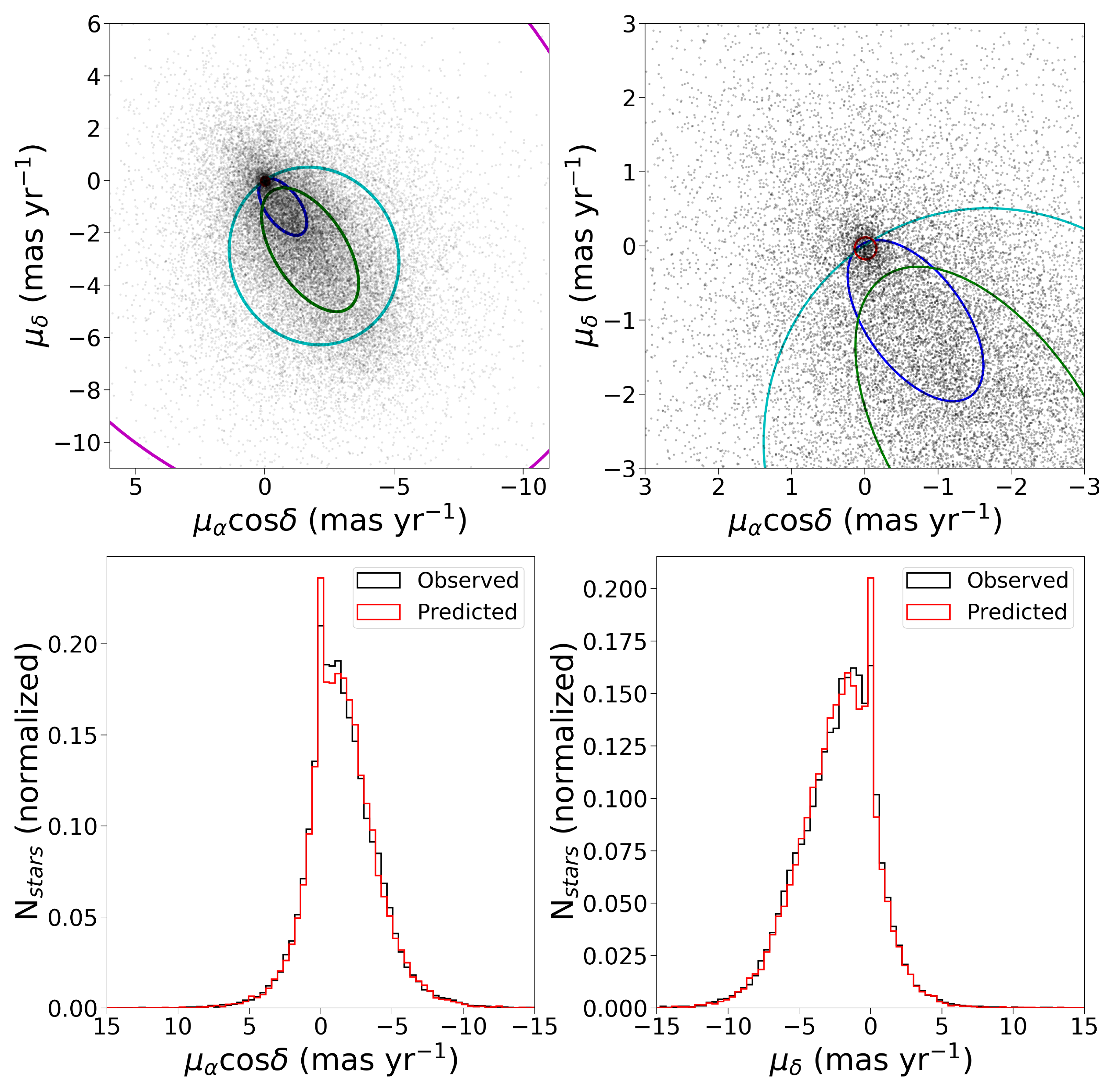}
\caption{The Gaussian Mixture Model fit to the observed cluster and field proper motion distributions. Top: The vector-point diagram of the proper motions with the 1$\sigma$ gaussian contours overlaid. The red gaussian corresponds to the cluster, while the blue, green, cyan, and magenta Gaussians describe the field population. The right panel is a zoomed-in version of the left panel, focusing on the cluster distribution. Bottom: The observed (black) vs. predicted (red) proper motion distributions in the RA and DEC directions (left and right panels, respectively). Good agreement is found between the observations and model.}
\label{fig:mem_dist}
\end{center}
\end{figure}

Cluster membership probabilities are calculated using the proper motions derived in $\mathsection$\ref{sec:photometry} and the Gaussian Mixture Model technique described in H15. This approach provides the flexibility needed to fit the complex kinematics of the cluster and field populations while taking the proper motion errors into account. To reduce outliers, an error cut of 1.42 mas yr$^{-1}$ (1/3 of the difference between the average cluster and field population proper motions in H15) is adopted, resulting in a membership catalog of 29,895 stars. This is significantly larger than the sample analyzed in H15 ($\sim$6000 stars) because we adopt a proper motion error cut that is 2.2x larger, do not impose a magnitude error cut, and generally have improved proper motion errors due to the extra epoch of data. As a result, a 5-Gaussian mixture model is required to fit the cluster and field populations (Figure \ref{fig:mem_dist}), as opposed to the 4-Gaussian model used in H15. This is confirmed by the Bayesian Information Criterion (see Equation $\ref{eq:BIC}$ and $\mathsection$\ref{sec:IMFcomp} for description), which significantly favors the 5-Gaussian model.

Individual cluster membership probabilities are calculated as

\begin{equation}
P_{pm}^i = \frac{\pi_c P_{c}^i}{\pi_c P_{c}^i + \sum_k^K{\pi_k P_{k}^i}}
\end{equation}

\noindent where $\pi_c$ and $\pi_k$ are the fraction of total stars in the cluster and $k$th field Gaussian, respectively, and P$_{c}^i$ and P$_{k}^i$ are the probability of $i$th star being part of the cluster and $k$th field Gaussian, respectively. A table describing the parameters of the Gaussian Mixture Model fit is provided in Appendix \ref{app:GMM}.

\subsection{Extinction Correction}
\label{sec:extinction}
Red Clump (RC) stars are used to correct for differential extinction across the field. The intrinsic magnitude and colors of these stars do not vary significantly with age or metallicity, making them useful ``standard crayons'' with which to measure extinction \citep{Girardi:2016ho}. While not associated with the Arches cluster itself, RC stars are numerous in the Galactic bulge and have a density distribution that is sharply peaked at the GC \citep{Wegg:2013pd}. Thus, we assume that the extinction of the RC stars is similar to that of the cluster, and so an extinction map derived using RC stars can be used for cluster stars. This approach was validated in H15, who showed that an RC extinction map significantly reduced the differential extinction in proper-motion selected Arches members.

We improve the extinction map presented in H15 by using a refined sample of RC stars identified using an unsharp-masking technique \citep[e.g.][]{De-Marchi:2016qw} and adopting a revised version of the optical/near-infrared extinction law presented in H18 (Appendix \ref{app:EL}). The advantage of the unsharp masking technique is that it increases the contrast of high-density features, such as the RC population, while reducing low-frequency noise. We select RC stars using the criteria described in H18: we calculate a best-fit line to the high-density RC feature in the CMD after unsharp masking and identify stars within $\Delta$F153M = 0.3 mags of the best fit line as the RC population (see Figure 7 from H18). This width is selected to encompass the RC feature, and is likely caused by the distribution of stellar distances, metallicities, and ages within the population, all of which alter their location in the CMD. In addition, we consider only stars with P$_{clust} \leq$ 0.02 in order to eliminate cluster members from the sample (which is necessary since the populations overlap in CMD space), and require a photometric error better than 0.05 mags in both the F127M and F153M filters in order to remove field interlopers that scatter into the selection space. Ultimately, 875 RC stars are used in the final extinction map.

\begin{figure}
\begin{center}
\includegraphics[scale=0.5]{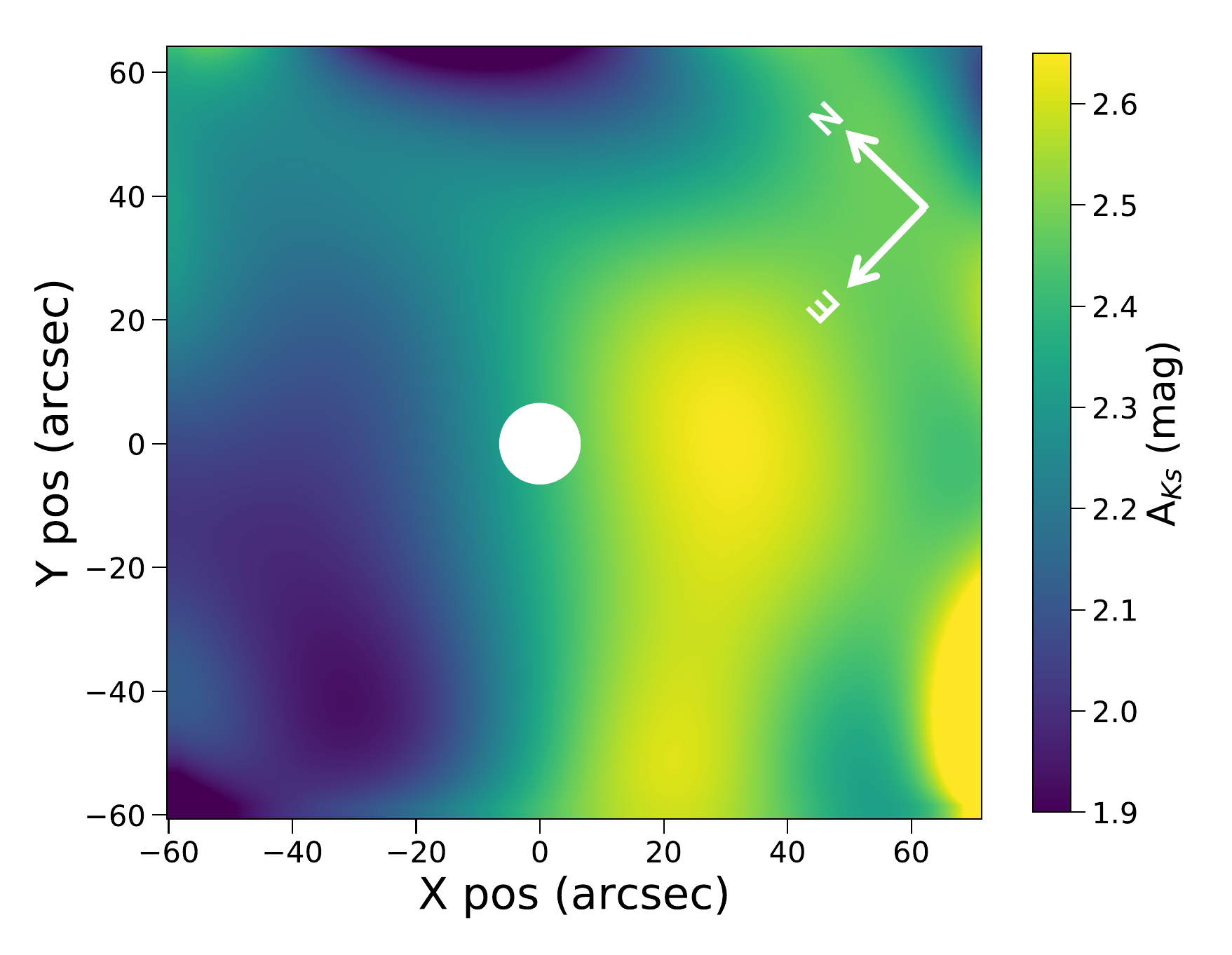}
\caption{The RC-interpolated extinction map for the Arches cluster field, with the positions shown in arcseconds relative to the cluster center. No measurement is made for $r_{cl} <$ 0.25 pc due to the low HST completeness in the area.}
\label{fig:redmap}
\end{center}
\end{figure}

The Arches extinction map is created using a spatial interpolation of the RC star sample with a fifth-order bivariate spline\footnote{The interpolation is calculated using the \emph{scipy.interpolate.bisplrep} routine in \texttt{python}.} (Figure \ref{fig:redmap}). All pixels with $r_{cl} <$ 0.25 pc are removed from the map, since high stellar crowding prevents an adequate number of RC stars from being detected at these radii. Ignoring the extreme values at the edge of the field where the interpolation becomes invalid, the extinction map values range 
from 1.9 mag $<$ A$_{Ks}$ $<$ 2.65 mag, with a median extinction of A$_{Ks}$ = 2.38 mag for stars with P$_{pm} \geq$ 0.5. We will adopt this as an initial estimate for the average extinction of the cluster and include a term in the IMF analysis to capture residual differential extinction in the cluster due to errors in the extinction map ($\mathsection$\ref{sec:model}).

\subsection{Completeness}
\label{sec:comp}
Observational completeness is determined using artificial star planting and recovery tests. We plant a total of 675,000 artificial stars and run them through the same detection pipeline as the real stars. These stars are generated in three sets. The first set contains 400,000 artificial stars with magnitudes drawn from the observed CMD, perturbed by a random amount drawn from a Gaussian distribution with a width equal to the photometric uncertainty. These stars are planted uniformly across the field. The second set contains 175,000 artificial stars that are assigned to a grid of magnitudes and colors in order to cover sparsely populated regions of the CMD (e.g., the brightest and faintest observed magnitudes), in order to improve the confidence of the completeness corrections in these regions. These stars are also given a uniform spatial distribution. The final set of 100,000 artificial stars are generated based on the brighter stars in the observed CMD (F153M $\leq$ 18 mag) and planted according to the radial profile of the Arches cluster from H15. This increases the confidence of the completeness correction near the cluster center, where the effects of stellar crowding are strongest.

\begin{figure*}
\begin{center}
\includegraphics[scale=0.37]{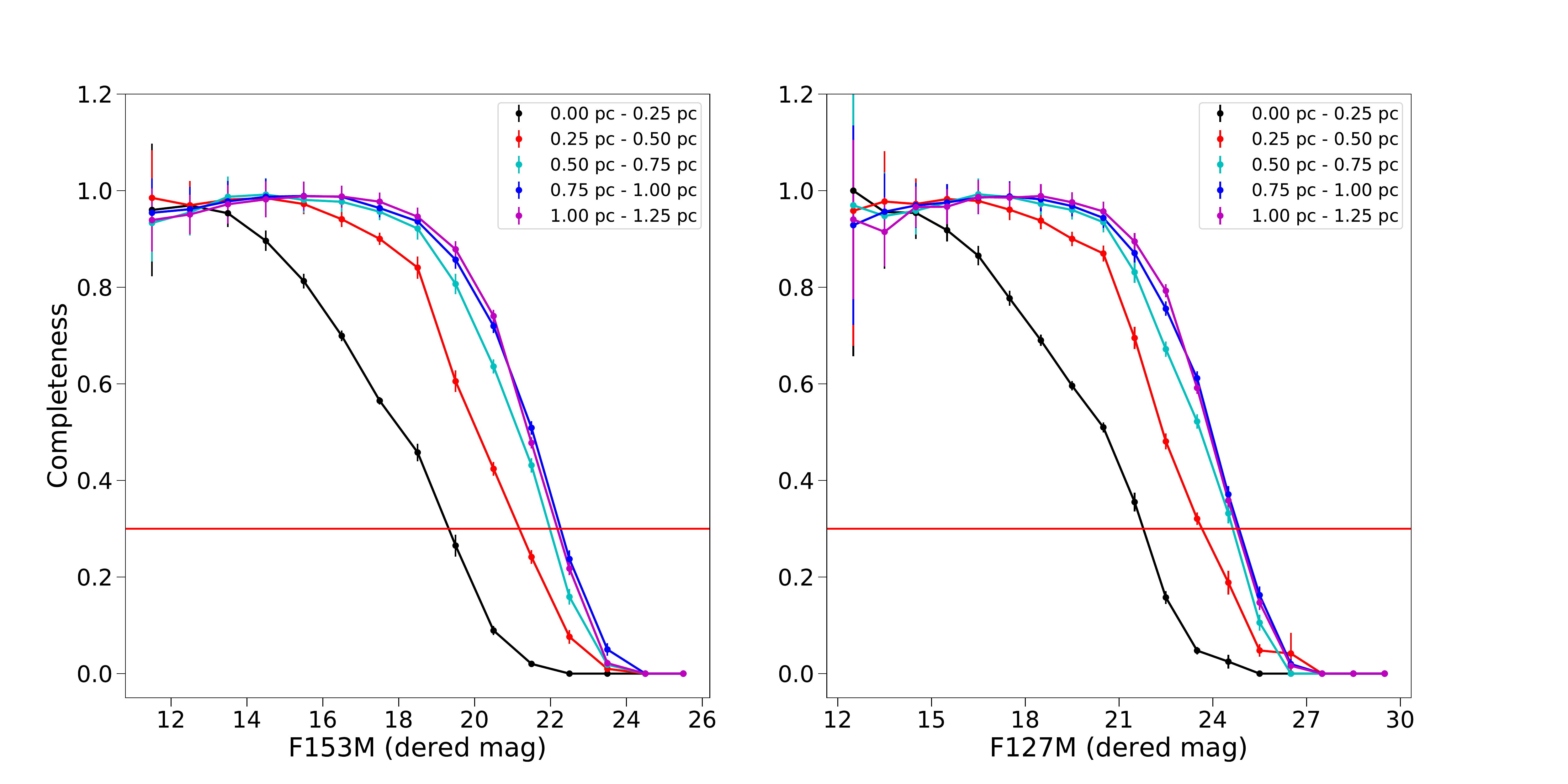}
\caption{Observational completeness as a function of cluster radius and differentially de-reddened F153M (left panel) and F127M (right panel) magnitude. At the average color of the cluster in the CMD, the F153M curve sets the completeness. Due to the low completeness in the innermost radius bin (0 pc - 0.25 pc), we exclude stars at these radii from the IMF analysis. We require a minimum requirement of 30\% completeness across the sample (red horizontal line), and thus adopt an F153M magnitude cut at F153M = 21.18 mag ($\mathsection$\ref{sec:sample}).}
\label{fig:completeness}
\end{center}
\end{figure*}

After the artificial stars are extracted by the detection pipeline, their photometric and astrometric errors are lower than the real data errors because they don't account for PSF uncertainty. Following H15, a magnitude-dependent error term is added in quadrature to the artificial star errors so their distribution matches those of the real star errors. Proper motions are then calculated and photometry differentially de-reddened for the artificial stars in the same manner as the real stars. To be successfully recovered, an artificial star must detected within 0.5 mags of its planted magnitude and 0.5 pixels of its planted position in at least three of the four F153M epochs and the F127M epoch, and have a proper motion error $\leq$ 1.42 mas yr$^{-1}$. The resulting F127M and F153M completeness curves as a function of differentially de-reddened magnitude in different cluster radius bins (0 pc $\leq$ R $\leq$ 3 pc, in steps of 0.25 pc) are shown in Figure \ref{fig:completeness}.

For the IMF analysis, we calculate the completeness for each star based on its cluster radius and position in the CMD. Within a given radius bin, the CMD is binned in steps of 0.15 mags in F153M (range: 24.5 mag -- 12.3 mag) and 0.2 mags in F127M - F153M (range: 0 mag -- 5 mags). The completeness in each bin is assigned to the lowest value from the F127M and F153M completeness curves at the respective F153M and F127M magnitudes at the center of the bin. At the average color of the cluster, the F153M curve sets the completeness limit.

\subsection{Final Sample}
\label{sec:sample}
Starting with the cluster membership catalog described in $\mathsection$\ref{sec:Membership} (29,895 stars), we apply a series of cuts in order to produce a high-quality sample for the IMF analysis. We require:

\begin{itemize}
\item P$_{pm} \geq$ 0.3, in an effort to reduce the number of field stars in our sample.
\item A minimum of 30\% completeness as determined in $\mathsection$\ref{sec:comp}. Due to the limited HST completeness at small cluster radii, we only consider stars with $r_{cl} >$ 0.25 pc. We thus achieve a depth of F153M $\leq$ 21.18 mag, corresponding to M $\geq \sim$1.8 $M_{\odot}$. We note that the analysis is not sensitive to this choice of the completeness limit; adopting a minimum completeness of 50\% does not significantly impact the results, other than changing the lower mag limit to F153M $\leq$ 20 mag (M $\geq \sim$2.5 $M_{\odot}$).
\item A minimum of 30\% area coverage within successive circular annuli of width 0.25 pc. As discussed in H15, this is achieved for $r_{cl} \leq$ 3.0 pc.
\item All F153M measurements for a given star to agree with its median F153M magnitude within 0.5 mags. This was found to remove situations where a faint star is misidentified as a nearby bright star.
\item WR stars will be removed from our sample, given the uncertainty in their stellar models and thus stellar masses. We use the population of spectroscopically-identified WR stars in the Arches cluster \citep{Figer:2002nr, Martins:2008hl,Clark:2018ij} determine their F153M magnitudes at the average cluster extinction of A$_{Ks}$ = 2.38 mag. The faintest of these stars, star B1 in \citet{{Clark:2018ij}}, is found to have a differentially de-reddened magnitude of F153M = 14.1 mags (observed F153M = 14.01 mag; the star is less extinguished than the cluster average), and so we adopt a conservative magnitude cut of F153M $\geq$ 14.5 mag.
\end{itemize}

Finally, a photometric color-cut is used to remove obvious field contaminants from the sample. High-probability cluster members (P$_{pm} \geq$ 0.6) are corrected for differential extinction as described in $\mathsection$\ref{sec:extinction}, and a 3$\sigma$ clipping algorithm is used to calculate the average F127M - F153M color and standard deviation as a function of F153M magnitude. For the entire sample, stars with differentially de-reddened colors larger than 2$\sigma$ to the blue or 3$\sigma$ to the red of the cluster sequence are automatically assigned P$_{pm}$ = 0, while all others are unchanged. This color-cut is more conservative to the red in order to account for the fact that some stars may have intrinsic reddening due to circumstellar disk material due to the cluster's young age \citep[e.g.][]{Stolte:2015rr}.

\begin{figure*}
\begin{center}
\includegraphics[scale=0.35]{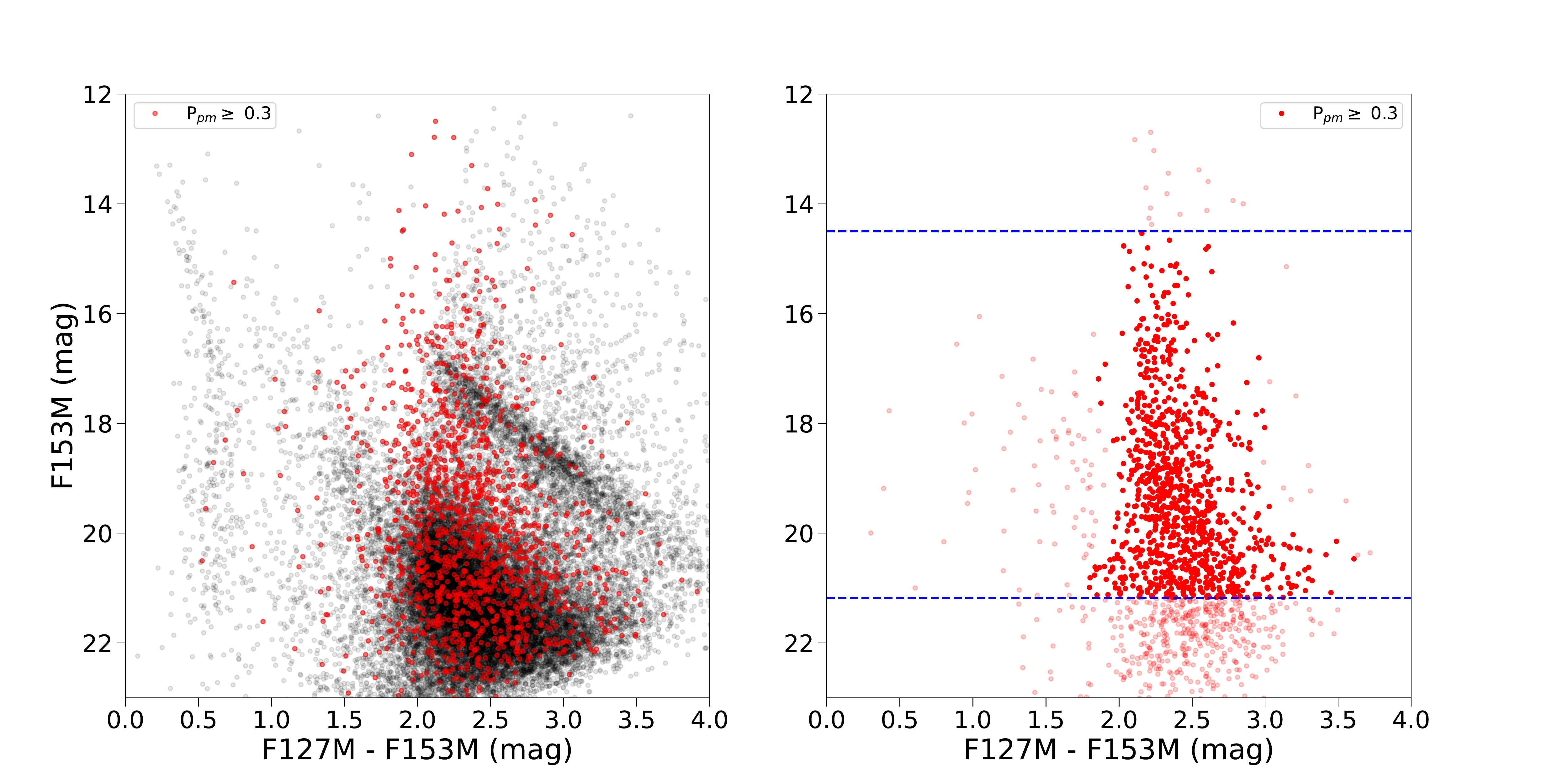}
\caption{Left: The observed CMD of the proper-motion selected sample (P$_{pm} \geq$ 0.3; in red) versus the field stars (black). Due to the significant overlap between the populations, proper-motion analysis is required to obtain an accurate cluster sample. Right: The differentially de-reddened CMD of the stars used in the IMF analysis. The bright red points are stars with P$_{pm} \geq$ 0.3 and F153M magnitudes within the adopted magnitude limits (blue dashed line). Stars eliminated by the color or magnitude cuts are shown as the faded red points. The cluster sequence significantly tightens after the differential extinction correction, though a term for residual differential extinction is still required in the IMF analysis.}
\label{fig:sample}
\end{center}
\end{figure*}

After these cuts, we are left with a sample of 981 stars with $\sum$P$_{pm}$ = 638.0. The CMD of this sample before and after the differential extinction correction is shown in Figure \ref{fig:sample}, and a summary of the cuts and their impact on the sample size is given in Table 2.

\begin{deluxetable}{l l c c}
\tablecaption{Sample Selection}
\label{tab:cuts}
\tabletypesize{\scriptsize}
\tablehead{
\colhead{Selection Description} & \colhead{Criterion} & \colhead{N$_{stars}$} & \colhead{$\sum$P$_{pm}$}
}
\startdata
{\bf Original Sample} &    & {\bf 29895} & {\bf 1290.7} \\
\hline
&  & \multicolumn{1}{c}{Cut from Sample} &\\
\hline
Membership & P$_{pm} \geq 0.3$ & 28237 & \\
Completeness & $\geq$ 0.3 & 539 & \\
F153M Mag Diff &  $\leq$ 0.5 mags & 45 & \\
WR stars & F153M $\geq$ 14.5 mag & 16 & \\
Color cut & see $\mathsection$\ref{sec:sample} & 78 & \\
\hline
\hline
{\bf Final Sample} &    & {\bf 980}  & {\bf 636.7}  \\
\enddata
\end{deluxetable}

Despite these efforts, some field contamination inevitably remains in our sample. This is due to stars with similar proper motions and colors as the cluster, and so their membership probabilities are artificially inflated. In $\mathsection$\ref{sec:IMF}, we derive revised cluster membership probabilities  \emph{after} the IMF analysis using the best-fit cluster and field model in order to take full advantage of the photometric information. We find that the number of cluster stars based on P$_{pm}$ is $\sim$6\% larger than the number of cluster stars based on the revised membership probabilities, and thus conclude that the sample contains approximately this amount of field contamination.

\subsection{Spectroscopic Analysis}
\label{sec:specResults}
Effective temperatures and surface gravities are derived for the spectroscopic stars by comparing the spectra to non-LTE CMFGEN model atmospheres \citep{Hillier:1998df, Hillier:2001df}. Non-LTE treatment is required due to the high temperatures of the stars and the presence of significant stellar winds, as evidenced by the Br-$\gamma$ emission inferred from the weak Br-$\gamma$ photospheric absorption line. Uncertainties in the stellar parameters are estimated by adjusting the models until they no longer provide good fits to the main diagnostic lines. Throughout the analysis we assume a terminal velocity (V$_{inf}$) of 2000 km s$^{-1}$, since this cannot be constrained from the spectra.

\begin{figure}
\begin{center}
\includegraphics[scale=0.5]{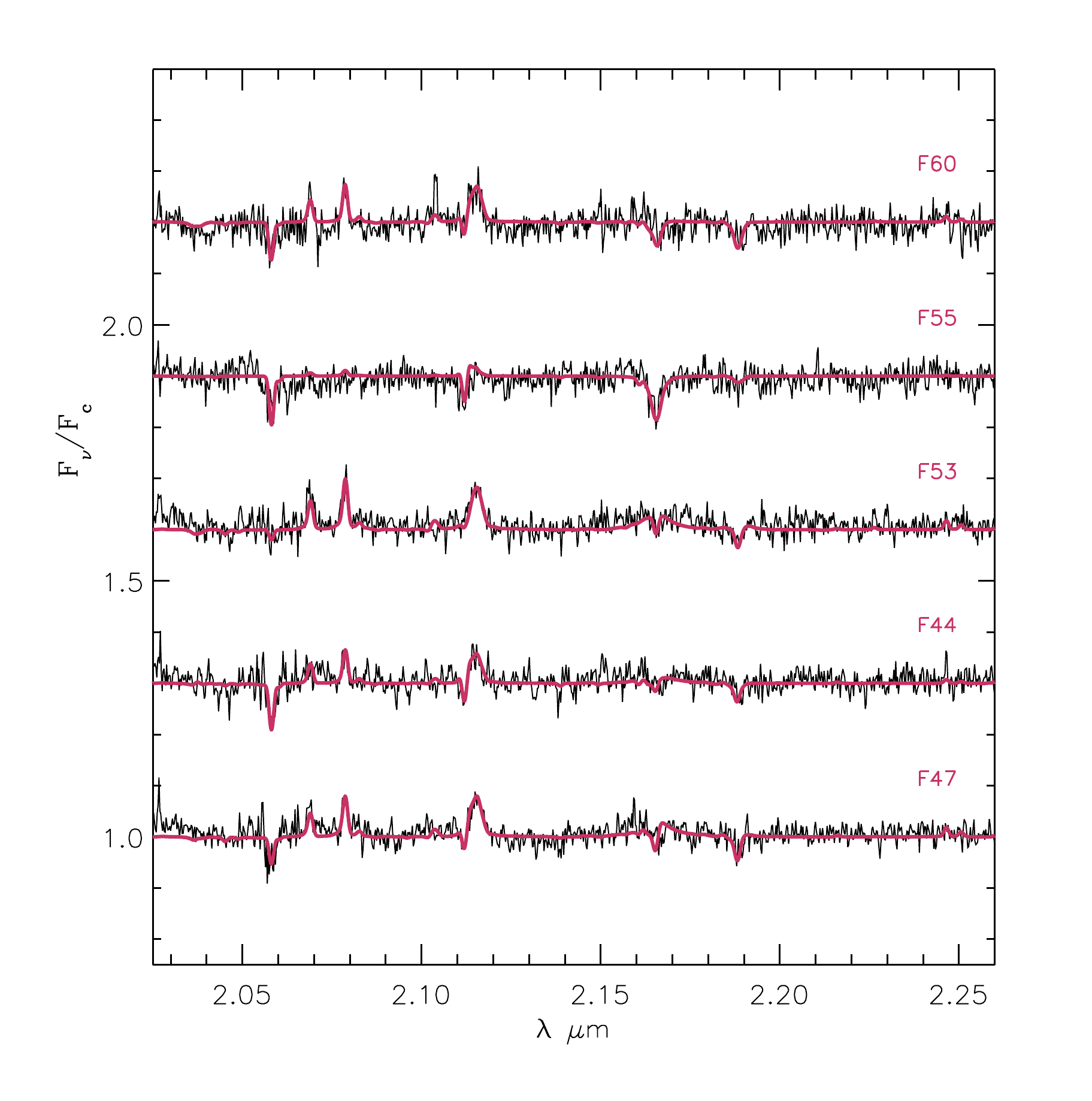}
\caption{Best-fit CMFGEN models (red) compared to the observed spectra (black).}
\label{fig:spec_fits}
\end{center}
\end{figure}

The best-fit model spectra are shown in Figure \ref{fig:spec_fits} and the corresponding T$_{eff}$ and log $g$ values are reported in Table \ref{tab:spectraStars}. T$_{eff}$ is constrained to within $\pm$3000 K or better, and is determined primarily from the HeII/HeI line ratios as well as the absorption component of the HeI 2.113 $\mu$m line. Stars 47, 55, and 60 were recently classified as O4-5~Ia stars and star 53 as an O5.5-6~I-III star by \citet{Clark:2018ij}. Our derived temperatures are consistent with the observed T$_{eff}$ vs. spectral type relation for galactic O-type stars within uncertainties \citep{Martins:2005ac}. The log $g$ values are less well constrained since they rely on the weak Br$\gamma$ lines, and thus are not used in the IMF analysis.

\section{Modeling the Cluster}
\label{sec:model}
We use a forward modeling approach to derive the IMF of the Arches cluster, comparing the observations to a cluster and field model within a Bayesian framework. The methodology described in \citet{Lu:2013wo} is expanded to simultaneously fit the IMF and other cluster parameters while taking into account degeneracies between cluster parameters, observational uncertainties, stellar multiplicity, and the empirical field population. Two IMF models are used: a 1-segment power law and a 2-segment power law. In the 1-segment IMF model, the free parameters are the the high-mass IMF slope $\alpha_1$, the cluster mass (M$_{cl}$), age (log $t$), distance ($d$), average extinction (A$_{Ks}$), and residual differential extinction after the extinction map correction ($\Delta$A$_{Ks}$). The 2-segment IMF model has additional free parameters m$_{break}$ and X$_\alpha$, where m$_{break}$ is the mass at which the IMF slope is $\alpha_2$ = X$_{\alpha}$ * $\alpha_1$ for m $\leq$ m$_{break}$ and $\alpha_1$ for m $>$ m$_{break}$. We require that 0 $\leq$ X$_{\alpha}$ $\leq$ 1 to enforce that $\alpha_2 <= \alpha_1$ (i.e., the low-mass IMF slope is more shallow than the high-mass IMF slope). The model parameters and their adopted priors are presented in Table \ref{tab:Params}.

\begin{deluxetable}{l c c c}
\tablewidth{0pt}
\tabletypesize{\footnotesize}
\tablecaption{IMF Model Parameters}
\tablehead{
\colhead{Parameter} & \colhead{Description} & \colhead{Prior\tablenotemark{a}} & \colhead{Units}
}
\startdata
$\alpha_1$ & High-mass IMF slope & U(1.0, 3.0) & --- \\
X$_{\alpha}$ & $\alpha_2$ / $\alpha_1$\tablenotemark{b} & U(0, 1) & --- \\
$m_{break}$ & Break mass\tablenotemark{b} & U(2, 14) & M$_{\odot}$ \\
M$_{cl}$ & Mass\tablenotemark{c} & U(3000, 50000) & M$_{\odot}$ \\
log $t$ & Age & U(6.2, 7.0) & log(years) \\
$d$ & Distance & G(8000, 250) & parsecs\\
A$_{Ks}$ & Average extinction & U(1.5, 2.7) & A$_{Ks}$ (mags) \\
$\Delta$A$_{Ks}$ & Differential extinction & U(0, 0.5) & A$_{Ks}$ (mags)  \\
\enddata
\tablenotetext{a}{Uniform distributions: U(min, max), where min and max are bounds of the distribution; Gaussian distributions: G($\mu$, $\sigma$), where $\mu$ is the mean and $\sigma$ is the standard deviation}
\tablenotetext{b}{Only used in 2-segment IMF model}
\tablenotetext{c}{Formally, M$_{cl}$ is the cluster mass between m$_{min}$ and m$_{max}$ (0.8 M$_{\odot}$ and 150 M$_{\odot}$, respectively) since this is the mass range over which the IMF is sampled when constructing the cluster}
\label{tab:Params}
\end{deluxetable}

To create a synthetic cluster, a population of stellar masses is stochastically generated based on the input IMF and the total cluster mass. We use the numerical formulation described by \citet{Pflamm-Altenburg:2006ai} to efficiently generate masses from the IMF between 0.8 M$_{\odot}$ and 150 M$_{\odot}$. Note that this is the mass range over which M$_{cl}$ is valid, since masses above and below these values are not generated in the synthetic cluster. The multiplicity of each star is determined using the mass-dependent multiplicity fraction, companion star fraction, and mass ratio empirically derived by \citet{Lu:2013wo} from studies of nearby young clusters in the literature. Stars and their companions are generated in batches until the cumulative stellar mass is larger than the designated mass of the cluster. Then, the population is trimmed to the star at which the cumulative mass is closest to the overall cluster mass, and then 1 additional star is drawn from the IMF and added to the sample.

Stellar evolution models are used to determine the physical properties of each star in the population. For a given age, a stellar evolution model provides the effective temperature (T$_{eff}$) and surface gravity (log $g$) at each stellar mass. We use two sets of stellar evolution models: the Pisa evolution models \citep{Tognelli:2011fr} for the pre-main sequence stars and the most recent Geneva models with rotation \citep{Ekstrom:2012qm} for the main sequence and evolved stars. The Pisa models have been shown to be consistent with observations of eclipsing binaries \citep{Stassun:2014sw} and nearby moving groups \citep{Herczeg:2015ph} for stars above 1 M$_{\odot}$, and are advantageous in that they model pre-main sequence stars to high masses ($\sim$7 M$_{\odot}$). High mass pre-main sequence stars are necessary due to the young age of the Arches cluster. The Geneva models have been shown to match observations for all but the most massive stars \cite[M $>$ 60 M$_{\odot}$;][]{Martins:2013fh}, where stellar evolution models become uncertain.

The physical properties are fed into a stellar atmosphere model, which returns a spectral energy distribution (SED) for each star. We assume solar metallicity, consistent with spectroscopic studies of the bright WR stars which find the Arches metallicity to be solar \citep{Najarro:2004ij} or slightly super-solar \citep[Z = 1.3 - 1.4 Z$_{\odot}$;][]{Martins:2008hl}. Two sets of atmosphere models are used: an ATLAS9 grid \citep{Castelli:2004yq} for T$_{eff} >$ 5500 K and a PHOENIX grid \cite[version 16;][]{Husser:2013ts} for T$_{eff} <$ 5000 K. An average between the two model grids is used in the transition region between 5000 K -- 5500 K. Both model grids assume local thermodynamic equilibrium (LTE), an assumption that begins to fail for massive stars. However, synthetic photometry calculated with ATLAS9 models compared to non-LTE CMFGEN models \citep{Fierro:2015sh} show differences of $\leq$$\sim$0.017 mags in F153M up to temperatures of 31,000 K.

The choice of stellar evolution and atmosphere models is an unavoidable source of systematic uncertainty in our analysis. To assess the impact of our model selections, we also run our IMF analysis using the recent MIST v1.0 evolution models \citep{Choi:2016en, Dotter:2016ve}, which are computed using the Modules for Experiments in Stellar Astrophysics (MESA) code \citep{Paxton:2011ly, Paxton:2013dw, Paxton:2015gf}. These analyses are discussed in $\mathsection$\ref{sec:caveats}.

We use \emph{Pysynphot} \citep{STScI-Development-Team:2013fd} to calculate synthetic photometry for the individual stars in the cluster population. The SEDs are reddened to the model A$_{Ks}$ according to the extinction law and then convolved with the WFC3IR F127M and F153M filter transmission functions. Multiple systems are treated as unresolved, with the total flux in each filter calculated as the sum of the system components. To simulate differential extinction, the photometry of each star system is perturbed by a random amount drawn from a Gaussian distribution centered at 0 with a width corresponding to the given $\Delta$A$_{Ks}$ in that particular filter.

Finally, the synthetic stars are assigned cluster radii based on the observed radial density profile of the Arches. We combine the radial profile for R $<$ 0.25 pc from \citep{Espinoza:2009bs} with the magnitude-dependent profiles between 0.25 pc $\leq$ R $\leq$ 3.0 pc from H15 (one profile for F153M $>$ 17 mag, the other for F153M $\leq$ 17 mag) for complete radial coverage over our data range. Each star's cluster radius is drawn from the following probability density distribution:

\begin{equation}
P(r) = \begin{cases}
\frac{1}{c_b} \Sigma_b (r) 2\pi r \; a(r) \; dr, & \text{$F153M \leq$ 17 mag} \\
\frac{1}{c_f} \Sigma_f (r) 2\pi r \; a(r) \; dr, & \text{$F153M >$ 17 mag} \\
\end{cases}
\end{equation}

where $\Sigma_b (r)$ and $\Sigma_f (r)$ are the bright-star (F153M $\leq$ 17 mag) and faint-star (F153M $>$ 17 mag) radial profiles, respectively, $c_b$ and $c_f$ are constants such that $\int_{r = 0 pc}^{r = 3 pc} P(r) = 1$, and a(r) is the fraction of the observed area at radius $r$ ($a(r) = 1.0$ for 0 $<$ r $\leq$ 2.3 pc, $a(r)$ $<$ 1.0 for r $>$ 2.3 pc). Thus we are able to simulate mass segregation in the synthetic cluster, and can properly account for the fact that all stars with $r <$ 0.25 pc are removed from the observed sample due to low completeness. The synthetic cluster stars are then binned using the same radius, color, and magnitude bins as the completeness calculations ($\mathsection$\ref{sec:comp}) in preparation for the IMF analysis.

\subsection{Bayesian Analysis}
\label{sec:bayes}
For a cluster model $\Theta$, we adopt a likelihood function with four components:

\begin{equation}
\begin{multlined}
\mathcal{L}(\bm{k_{obs}}, N_{cl}, N_{WR}, T_{eff} | \Theta) = \\
p(\bm{k_{obs}} | \Theta) \cdot p(N_{cl} | \Theta) \cdot p(N_{W} | \Theta) \cdot p(\{T_{eff}, m_{obs}\} | \Theta)
\end{multlined}
\label{eq:like}
\end{equation}

\noindent where $p(\bm{k_{obs}} | \Theta)$ is the probability of obtaining the observed distribution of stars in CMD space, with $\bm{k_{obs}}$ representing the set of observed F153M magnitudes and F127M~-~F153M colors; $p(N_{cl} | \Theta)$ is the probability of detecting the number of observed cluster stars $N_{cl}$; $p(N_{W} | \Theta)$ is the probability of the detecting the observed number of WR stars; and $p(\{T_{eff}, m_{obs}\} | \Theta)$ is the probability of measuring the observed $T_{eff}$ values for the spectroscopic stars given their F153M magnitudes m$_{obs}$.

To calculate $p(\bm{k_{obs}} | \Theta)$ we must first calculate the CMD probability distribution for the cluster model and the field. The intrinsic CMD probability distribution for cluster stars generated by the model $\Theta$, $p(\bm{k_{int}} | \Theta)_{cl}$, is calculated according to the procedure described in $\mathsection$\ref{sec:model}. Here, $\bm{k_{int}}$ is the distribution of synthetic star mags and colors in the model cluster. To reduce the impact of stochastic effects in the synthetic CMD, the model cluster is generated with a total mass of 5x10$^6$ M$_{\odot}$ ($\sim$500 times more massive than the expected mass of the Arches), regardless of the M$_{cl}$ designated by the model. To calculate the observed CMD probability distribution for the model cluster, we apply the observational completeness and make the same magnitude cuts as the observed sample ($\mathsection$\ref{sec:sample}):

\begin{equation}
\label{eq:clust}
p(\bm{k_{int}} | \Theta)_{cl, obs} = \frac{\sum_{r=0}^{N_r} p(\bm{k_{int, r}} | \Theta)_{cl} * C(r)}{\sum_{k=0}^{N_k}\sum_{r=0}^{N_r} p(\bm{k_{int, r}} | \Theta)_{cl} * C(r)}
\end{equation}

\noindent where $p(\bm{k_{int, r}} | \Theta)_{cl}$ and $C(r)$ are the intrinsic model cluster CMD probability distribution and observational completeness at a cluster radius $r$, $N_r$ is the number of radius bins, and $N_k$ is the total number of magnitude-color bins in the CMD itself.

In addition to the synthetic cluster, we construct a CMD probability distribution for the field stars. We select all stars with P$_{pm} \leq$ 0.03 and then apply the same differential extinction correction, magnitude, and color cuts as the IMF analysis sample and then normalize to calculate the field CMD probability distribution $p(\bm{k_{obs, f}})$:

\begin{equation}
\label{eq:field}
p(\bm{k_{obs, f}}) = \frac{\bm{k_{obs, f}}}{\sum_{k=0}^{N_{k}} \bm{k_{obs, f}}}
\end{equation}

\noindent where $\bm{k_{obs, f}}$ is the observed field CMD. Note that we do not apply a completeness correction since the CMD is already ``observed'' and thus it is already inherently included, and that $p(\bm{k_{obs, f}})$ is not dependent on the cluster model.

With the cluster and field CMD probability distributions in place, we can calculate the probability of observing the $i$th star given its color and magnitude ($p(k_{obs,i} | \Theta)$). We infer that the field membership probability for a given star is P$_{f}$ = 1 - P$_{pm}$. To incorporate observational error, we assume that $k_i$ = $k'_i + \epsilon_i$, where $\epsilon_i$ is drawn from a normal distribution centered at zero and with standard deviation drawn from the set of observational errors $\sigma_{k,i}$. Thus:

\begin{equation}
\label{eq:cmdlike}
\begin{multlined}
p(k_{obs,i} | \Theta) = \int^{\infty}_{-\infty} (P_{pm} * p(\bm{k_{int}} | \Theta)_{cl, obs} + P_f * p(\bm{k_{obs,f}})) * \\
\shoveleft[1cm]\frac{1}{\sqrt{2\pi\sigma_{k,i}}}e^{\frac{-(k'_i - k_i)^2}{2\sigma_{k,i}^2}} dk'_i
\end{multlined}
\end{equation}

The final CMD likelihood is calculated by multiplying the individual likelihoods for the observed stars together:
\begin{equation}
p(\bm{k_{obs}} | \Theta) = \prod_{i=1}^{N_{obs}} p(k_{obs,i} | \Theta)
\end{equation}

\noindent where $N_{obs}$ is the number of stars in the sample.

The second component of the likelihood, $p(N_{cl} | \Theta)$, is calculated from the number of cluster stars we would predict to observe given the cluster model. Returning to the intrinsic synthetic cluster CMD $\bm{k_{int}}$, we perturb the photometry of each star by a random amount drawn from the photometric error of the observations at its magnitude and then apply the magnitude cuts and observational completeness. Following \citet{Lu:2013wo}, we linearly scale the number of stars in the simulated cluster after it is convolved with the observational completeness ($N_{sim}$) to the cluster model mass in order to obtain the expected number of observed stars $N_e$:

\begin{equation}
N_{e} = N_{sim} * \left(\frac{M_{cl}}{5*10^6}\right)
\end{equation}

\noindent where M$_{cl}$ is the cluster model mass. The probability of obtaining the observed number of cluster stars $N_{cl} = \sum P_{pm}$ is calculated from a Poisson distribution:

\begin{equation}
p(N_{cl} | \Theta) = \frac{N_{e}^{N_{cl}} e^{-N_{e}}} {N_{cl}!}
\end{equation}

The purpose of applying the observational errors to $\bm{k_{int}}$ for this calculation is to account for any potential Malmquist bias that is introduced by our magnitude cuts. Note that this is not done in Equation \ref{eq:clust} for the CMD component of the likelihood since the observational errors are already accounted for in Equation \ref{eq:cmdlike}.

The third component of the likelihood is based on the predicted number of WR stars in the cluster model, which serves as a constraint on the cluster age \citep[e.g.][]{Lu:2013wo}. The brightest stars in the inner region of the cluster ($r_{cl} <$ 0.75 pc) were cataloged by \citet{Figer:2002nr}, and later spectroscopic studies identified 13 WR stars among this sample \citep{Martins:2008hl, Clark:2018ij}. In the cluster model, we calculate the number of predicted WR stars within this radius range and, similarly scaling that number to cluster model mass, calculate the probability of obtaining the observed number of WR stars:

\begin{equation}
p(N_{W} | \Theta) = \frac{N_{W_0}^{N_{W}} e^{-N_{W_0}}} {N_{W}!}
\end{equation}

\noindent where $N_{W}$ = 13 and is the number of WR stars in the observations within $r_{cl} <$ 0.75 pc, and $N_{W_0}$ is the number of WR stars predicted by the scaled cluster model in that same radius range.

The final component of the likelihood comes from from the T$_{eff}$ measurements from the spectroscopic sample. For each star, we calculate $\overline{T_{eff_0}}$ and $\sigma_{Teff_0}$, which represent the median $T_{eff}$ and its standard deviation for all stars in the cluster model with ($m_{obs}$ - $\sigma_{m_{obs}}$) $\leq$ $m$ $\leq$  ($m_{obs}$ + $\sigma_{m_{obs}}$) and ($col_{obs}$ - $\sigma_{col_{obs}}$) $\leq$ $col$ $\leq$  ($col_{obs}$ + $\sigma_{col_{obs}}$) where $m_{obs}$, $\sigma_{m_{obs}}$, $col_{obs}$, $\sigma_{col_{obs}}$ are the F153M magnitude and F127M - F153M color of the observed star and its respective errors. The likelihood of measuring $T_{eff}$ for the star is then:

\begin{equation}
p(T_{eff}, m_{obs} | \Theta) = \frac{1}{\sigma_{tot} \sqrt{2\pi}} * e^{-(T_{eff} - T_{eff_0})^2 / (2\sigma_{tot}^2)}
\end{equation}

\noindent where $T_{eff}$ and $\sigma_{T_{eff}}$ is the measured effective temperature and associated error of the star and $\sigma_{tot} = \sqrt{\sigma_{T_{eff}}^2 + \sigma_{Teff_0}^2}$. The likelihood of the spectroscopic sample is calculated by multiplying the individual likelihoods together:

\begin{equation}
p(\{T_{eff}, m_{obs}\} | \Theta)  = \prod_{i=1}^{N_{spec}} p(T_{eff_i}, m_{obs_i} | \Theta)
\end{equation}

\noindent where $N_{spec}$ is the number of stars in the spectroscopic sample.

We derive the best-fit cluster model using Bayes theorem:

\begin{equation}
P(\Theta | \bm{k_{obs}}, N_{cl}, N_{WR}, T_{eff}) = \frac{\mathcal{L}(\bm{k_{obs}}, N_{cl}, N_{WR}, T_{eff} | \Theta) P(\Theta)}{P(\bm{k_{obs}}, N_{cl}, N_{WR}, T_{eff})}
\end{equation}

\noindent where $P(\Theta | \bm{k_{obs}}, N_{cl}, N_{WR}, T_{eff})$ is the posterior probability for the given model $\Theta$, $\mathcal{L}(\bm{k_{obs}}, N_{cl}, N_{WR}, T_{eff} | \Theta)$ is the likelihood equation, $P(\Theta)$ is the priors on the model free parameters, and $P(\bm{k_{obs}}, N_{cl}, N_{WR}, T_{eff})$ is the sample evidence. To sample the parameter space to find the best-fit model we use \emph{Multinest}, a publicly available multimodal sampling algorithm shown to be more efficient that Markov Chain Monte Carlo algorithms when exploring complex parameter spaces \citep{Feroz:2009lq}. We adopt an evidence tolerance of 0.5, a sampling efficiency of 0.8, and 1000 live points to run the analysis. The algorithm is run using the python wrapper module \emph{PyMultinest} \citep{Buchner:2014wa}.

We test the accuracy of this procedure by running the analysis on simulated clusters of known properties. A discussion of how the simulated clusters are created and the results of the tests are provided in Appendix \ref{app:artificialTest}. We find that the analysis is able to recover the input values to within 1$\sigma$ for all parameters for both the 1-segment and 2-segment IMF models.

\subsection{Model-Dependent Membership Probabilities and Stellar Properties}
\label{sec:modelDependent}
After the best-fit cluster model is determined, we calculate revised cluster membership probabilities that take full advantage of the available kinematic and photometric information. The cluster model provides the distribution of cluster stars in CMD space, from which stars with proper motions similar to the cluster but with photometry similar to the field can be de-weighted. First, we calculate the expected cluster CMD $\bm{k_{\Theta, cl}}$ and field star CMD $\bm{k_{f}}$:

\begin{equation}
\begin{aligned}
\bm{k_{\Theta, cl}} &= \sum_{i=0}^{N_{obs}} P_{pm, i} * p(\bm{k_{int}} | \Theta)_{cl, obs}  \\
\bm{k_{f}} &= \sum_{i=0}^{N_{obs}} P_{f, i} * p(\bm{k_{obs, f}})
\end{aligned}
\end{equation}

\noindent where $p(\bm{k_{int}} | \Theta)_{cl, obs} $ and  $p(\bm{k_{obs, f}})$ are as defined in Equations \ref{eq:clust} and \ref{eq:field}. The revised membership probability for a given star then becomes:

\begin{equation}
\begin{multlined}
P_{clust, i} = \int^{\infty}_{-\infty} \left(\frac{\bm{k_{\Theta,cl}}}{(\bm{k_{\Theta,cl}} + \bm{k_{f}})}\right) * \frac{1}{\sqrt{2\pi\sigma_{k,i}}}e^{\frac{-(k'_i - k_i)^2}{2\sigma_{k,i}^2}} dk'_i
\end{multlined}
\end{equation}

P$_{clust}$ is thus a combination of the proper motion membership (which sets the scale of cluster and field CMD components) as well as the cluster and field CMDs themselves.

We also use the best-fit cluster model to infer the intrinsic properties (e.g. mass) for each star in the observed sample. These values are often estimated by tracing the star to a theoretical cluster isochrone along the reddening vector, but this approach is challenging near the pre-main sequence turn-on where multiple intersections between the reddening vector and isochrone can occur. Instead, we calculate a probability distribution for the desired stellar property from $\bm{k_{int}}$, based on the stars located at the observed star's location in the CMD. For example, the mass probability distribution within a given CMD bin $k$ is:

\begin{equation}
p(m | \Theta)_k = \frac{\sum_i^{N_i} m_{i, b, k}}{\sum_b^{N_b} \sum_i^{N_i} m_{i, b, k}}
\end{equation}

\noindent where m$_{i, b, k}$ is the mass of the $i$th star in mass bin $b$ in the CMD bin $k$. $N_i$ is the number of stars in mass bin $b$, and $N_b$ is the total number of mass bins. The mass bins are chosen to be 20 equal log-spaced bins between 0.8 M$_{\odot}$ and 70 M$_{\odot}$, which are the minimum and maximum masses in the cluster model\footnote{Though the IMF is sampled from 0.8 M$_{\odot}$ --150 M$_{\odot}$ to create the cluster, only synthetic stars within the F153M magnitude limits are considered in this analysis. This corresponds to a mass range between 1.8 M$_{\odot}$ -- 51 M$_{\odot}$ for the best-fit isochrone, but differential extinction scatters lower- and higher-mass stars into the sample.}.

For a given star, we calculate its mass probability distribution by multiplying $p(m | \Theta)_k $ by the position of the star in the CMD convolved with its photometric error:

\begin{equation}
\phi(m)_i = \int^{\infty}_{-\infty} p(m | \Theta)_k * \frac{1}{\sqrt{2\pi\sigma_{k,i}}}e^{\frac{-(k'_i - k_i)^2}{2\sigma_{k,i}^2}} dk'_i
\end{equation}

We construct the observed initial mass function $\Phi_{obs}$ by summing the mass probability distributions over the sample, taking into account each star's revised cluster membership probability, observational completeness, and area completeness:

\begin{equation}
\Phi_{obs} = \sum_i^{N_i} \phi(m)_i * \frac{P_{clust, i}}{C(r) * a(r)}
\end{equation}

\noindent where $C(k, r)$ is the completeness as a function of CMD position and radius and $a(r)$ is the area completeness. We reiterate that $\Phi_{obs}$ is dependent on the synthetic cluster and is calculated \emph{after} the best-fit model is found. It thus serves as a check that the IMF derived in the analysis is indeed a good match to the observations.

\section{Results}
\label{sec:resultsAll}
We find that the Arches cluster is best described by a 1-segment IMF model that is top-heavy ($\alpha$ = 1.80 $\pm$ 0.05 (stat) $\pm$ 0.06 (sys). However, we cannot discount a 2-segment IMF model with a high-mass slope closer to the local IMF value ($\alpha$ = 2.04$^{+0.14}_{-0.19} \pm$ 0.04) but with a break at 5.8$^{+3.2}_{-1.2}$ $\pm$ 0.02 M$_{\odot}$. This section is organized as follows: we describe the best-fit IMF model in $\mathsection$\ref{sec:IMF} and compare the 1-segment and 2-segment IMF model solutions in $\mathsection$\ref{sec:IMFcomp}. In $\mathsection$\ref{sec:IMFimpact} we discuss the impact of our assumptions regarding stellar evolution models and stellar multiplicity.

\subsection{The Arches Cluster IMF: Best-fit Model}
\label{sec:IMF}
The best-fit cluster models for each of the different cases examined in this analysis (1-segment vs. 2-segment IMF, Pisa/Geneva vs. MIST evolution models, with vs. without multiplicity) are given in Table \ref{tab:Results} and a breakdown of the corresponding likelihoods in Table \ref{tab:likelihood}. A detailed comparison of these cases is presented in $\mathsection$\ref{sec:IMFcomp} and $\mathsection$\ref{sec:IMFimpact}, but in summary: 1) the 1-segment IMF model is slightly favored over the 2-segment IMF model, but we cannot rule out the 2-segment IMF model; 2) we cannot distinguish between the Pisa/Geneva and MIST evolution models in the 1-segment IMF case, but the MIST models are favored in the 2-segment IMF case; and 3) the fits without multiplicity are strongly disfavored. As a result, we adopt the 1-segment IMF fit with Pisa/Geneva models and multiplicity as the best-fit IMF for the Arches cluster, and use the MIST model solution to estimate the systematic error. When discussing the 2-segment IMF fit, we adopt the MIST model solution with multiplicity and use the Pisa/Geneva model solution to estimate the systematic error.

\begin{deluxetable*}{l | c c  | c c}
\tablewidth{0pt}
\tabletypesize{\scriptsize}
\tablecaption{Best-fit Cluster Models}
\tablehead{
&  \multicolumn{2}{c}{1-Segment IMF} & \multicolumn{2}{c}{2-Segment IMF} \\
\colhead{Parameter\tablenotemark{a}} &  \colhead{Pisa/Geneva\tablenotemark{b}} & \colhead{MIST v1.0\tablenotemark{c}} & \colhead{Pisa/Geneva\tablenotemark{b}} & \colhead{MIST v1.0\tablenotemark{c}} \\
}
\startdata
$\alpha_1$ & 1.80 $\pm$ 0.05 & 1.68 $\pm$ 0.05 & 2.12 $\pm$ 0.11 & 2.04$^{+0.14}_{-0.19}$  \\
$\alpha_2$ & --- & --- &  0.95 $\pm$ 0.45 & 1.10$^{+0.39}_{-0.31}$ \\
$m_{break}$ & --- & --- &  5.4$^{+2.4}_{-0.8}$ & 5.8$^{+3.2}_{-1.2}$ \\
M$_{cl}$ & 24400$^{+2000}_{-1600}$ & 28600$^{+3000}_{-2800}$  & 19600$^{+2000}_{-1600}$ & 21000$^{+3400}_{-2800}$  \\
log $t$ & 6.57 $\pm$ 0.02 & 6.56 $\pm$ 0.02 & 6.60 $\pm$ 0.05 & 6.55$^{+0.02}_{-0.04}$  \\
$d$ & 7900 $\pm$ 158 & 7900 $\pm$ 160 &  8030 $\pm$ 160 & 8100 $\pm$ 160 \\
A$_{Ks}$ & 2.44 $\pm$ 0.01  &  2.44 $\pm$ 0.01 &  2.46 $\pm$ 0.02  &  2.45 $\pm$ 0.01 \\
$\Delta$A$_{Ks}$ & 0.15 $\pm$ 0.01 & 0.15 $\pm$ 0.01 & 0.13 $\pm$ 0.01 & 0.15 $\pm$ 0.01 \\
\enddata
\tablenotetext{a}{Priors and units are the same as described in Table \ref{tab:Params}. Note that only statistical uncertainties are reported in this table. Systematic uncertainties are estimated to be half the difference between parameter values derived using different stellar evolution models.}
\tablenotetext{b}{Pisa: \citet{Tognelli:2011fr}; Geneva: \citet{Ekstrom:2012qm}}
\tablenotetext{c}{\citet{Choi:2016en, Dotter:2016ve}}
\label{tab:Results}
\end{deluxetable*}

\begin{deluxetable*}{l | c c c | c c c}
\tablewidth{0pt}
\tabletypesize{\scriptsize}
\tablecaption{IMF Model Likelihoods}
\tablehead{
& \multicolumn{3}{c}{1-Segment IMF} & \multicolumn{3}{c}{2-Segment IMF}\\
\colhead{Component} & \colhead{Pisa/Geneva\tablenotemark{a}} & \colhead{MIST v1.0\tablenotemark{b}} & \colhead{No Multiples} & \colhead{Pisa/Geneva\tablenotemark{a}} & \colhead{MIST v1.0\tablenotemark{b}} & \colhead{No Multiples} \\
}
\startdata
CMD & -5058.6 & -5060.2 & -5067.0 & -5055.8 & -5051.9 & -5057.4\\
N$_{stars}$ & -4.48 & -4.15 & -4.23 & -4.23 & -4.22 & -4.18 \\
N$_{WR}$ & -3.24 & -2.46 & -4.23 & -2.21 & -3.45 & -2.26 \\
Spectroscopy & -19.48 & -19.45 & -21.08 & -19.48 & -19.54 & -19.35 \\
\bf{log($\mathcal{L}$)} & \bf{-5103.1} & \bf{-5103.5} & {\bf -5116.9} & \bf{-5101.4} & \bf{-5098.6}  & \bf{-5102.9} \\
\bf{BIC} & \bf{10247.5} & \bf{10248.3} & \bf{10275.1} & \bf{10257.9} & \bf{10252.3}  & \bf{10260.9} \\
\enddata
\tablenotetext{a}{Pisa: \citet{Tognelli:2011fr}; Geneva: \citet{Ekstrom:2012qm}}
\tablenotetext{b}{\citet{Choi:2016en, Dotter:2016ve}}
\label{tab:likelihood}
\end{deluxetable*}

\begin{figure*}
\begin{center}
\includegraphics[scale=0.3]{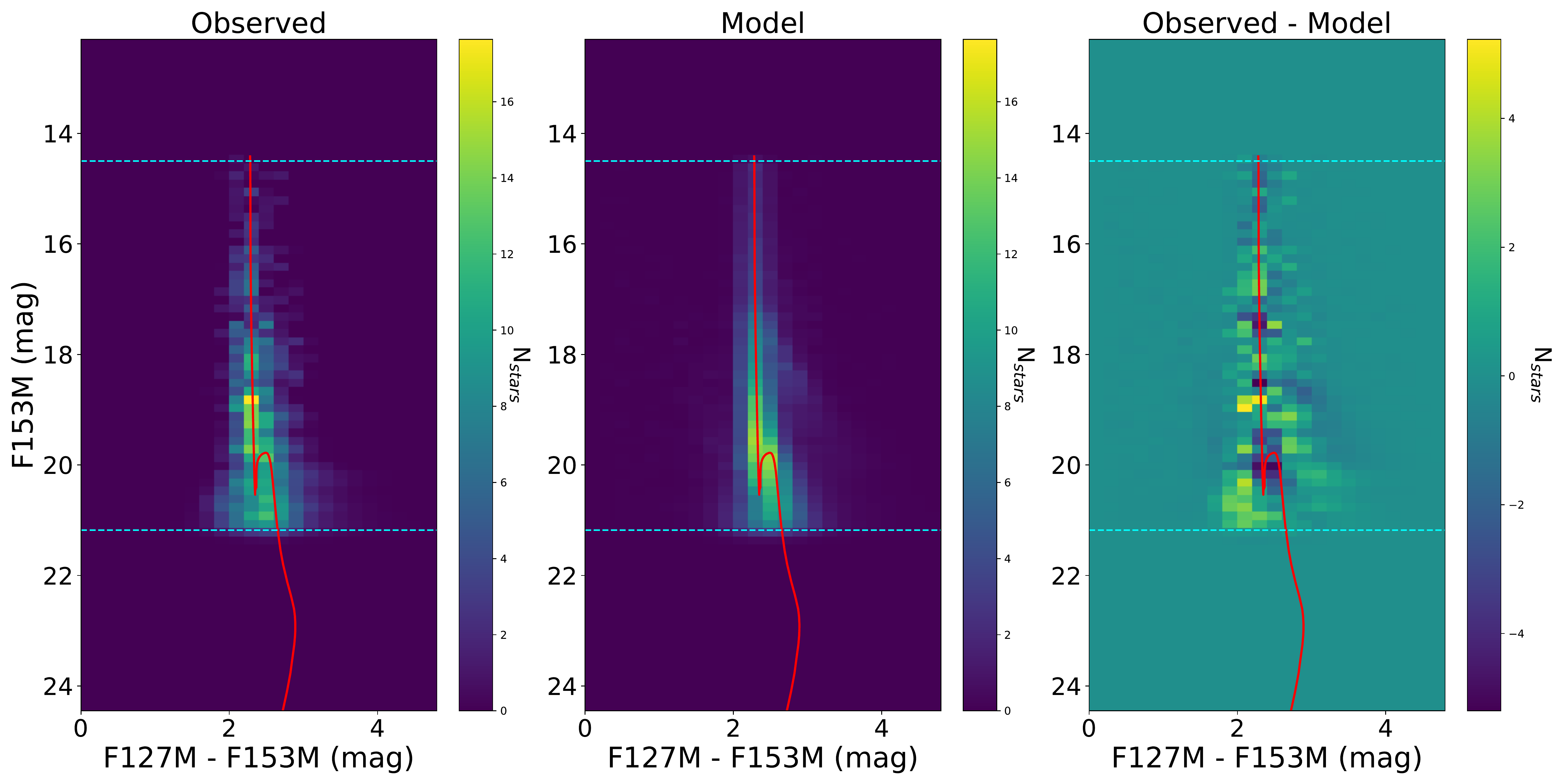}
\caption{A comparison between the observed CMD and the predicted CMD from the best-fit cluster model. The left panel shows the hess diagram for the observed cluster, the middle panel shows the hess diagram of the best-fit cluster model, and the right panel shows the residuals between the two. The cluster model has been convolved with observational uncertainties in this comparison. In all panels the isochrone associated with the best-fit model is plotted as a red line and the F153M magnitude limits are represented by the cyan dashed lines. Note that the cluster model contains both cluster and field components; the impact of the red clump is particularly evident by the slight high-density diagonal feature near F153M $\sim$ 18 mag.}
\label{fig:CMD}
\end{center}
\end{figure*}

\begin{figure}
\begin{center}
\includegraphics[scale=0.3]{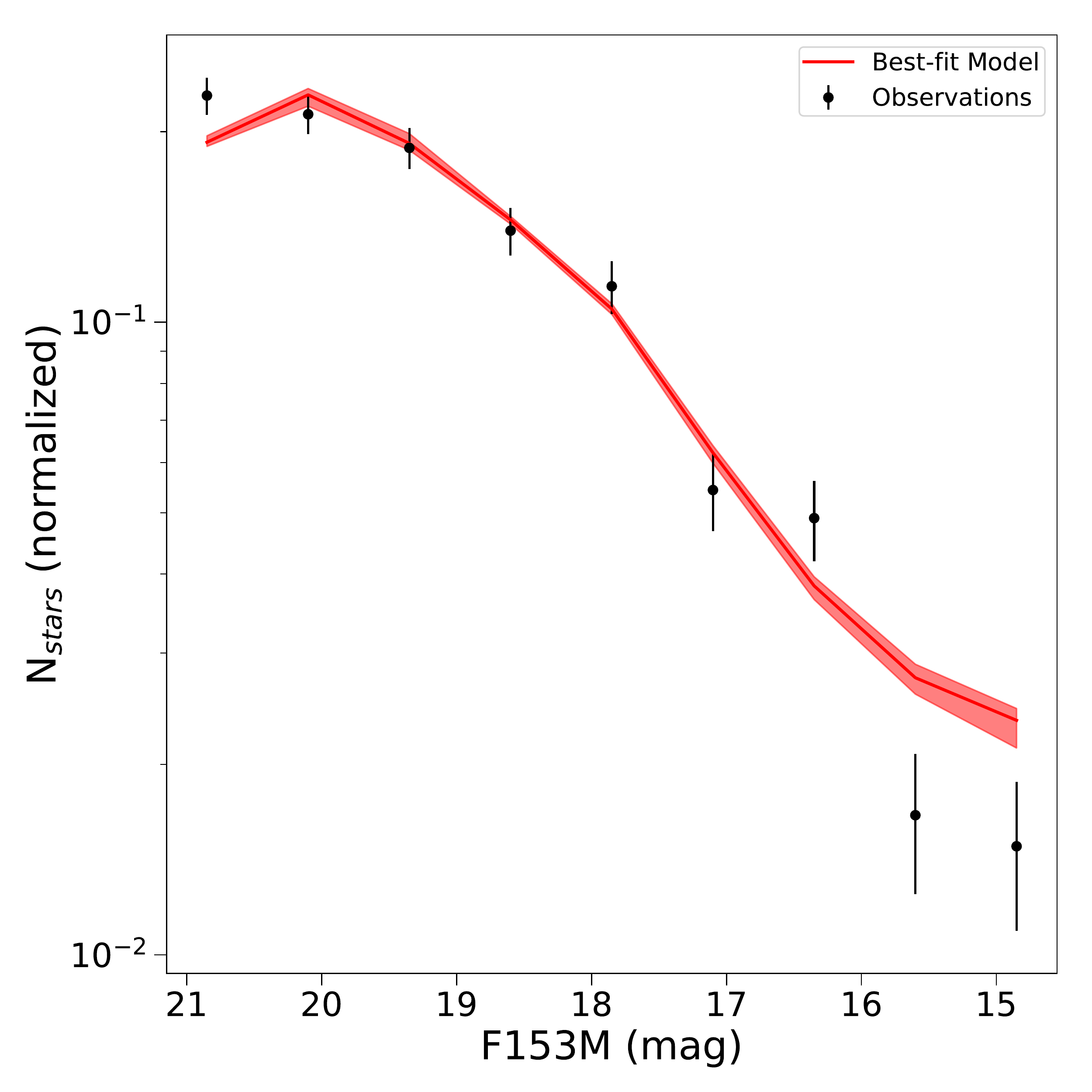}
\caption{A comparison of the observed F153M luminosity function (black points) versus the best-fit model (red line). The 1$\sigma$ envelope of possible models, sampled from the posterior distribution, is shown by the red envelope. Good agreement is found with the exception of a possible excess of model stars in the brightest magnitude bins (F153M $\lesssim$ 16 mag).}
\label{fig:lum_func}
\end{center}
\end{figure}

\begin{figure}
\begin{center}
\includegraphics[scale=0.3]{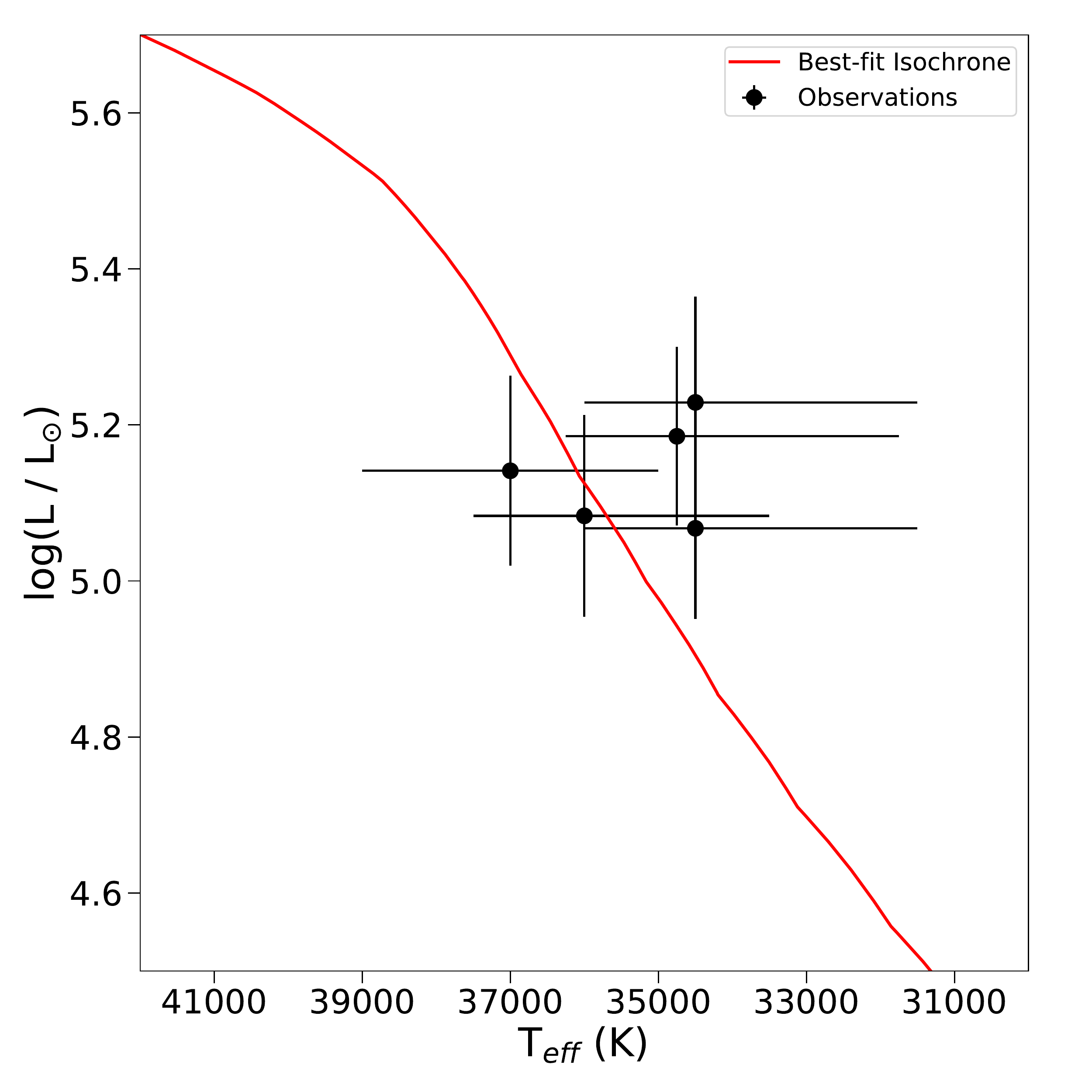}
\caption{The measured T$_{eff}$ and inferred luminosity of the spectroscopic stars (black points) compared to the best-fit model (red line).}
\label{fig:HRD}
\end{center}
\end{figure}

The posteriors for the IMF model parameters are provided in Appendix \ref{app:posteriors}. A comparison between the observed and model CMD is shown in Figure \ref{fig:CMD} and the subsequent F153M luminosity function shown in Figure \ref{fig:lum_func}. Good agreement is generally found between the observations and model, though perhaps with a slight excess of model stars at the bright end of the sample (F153M $\lesssim$ 16 mag). The agreement between the spectroscopic T$_{eff}$ measurements and those predicted by the model is shown in a Hertzsprung-Russell Diagram (HRD), where the (model-dependent) luminosity for each of the observed stars has been derived in the manner described in $\mathsection$\ref{sec:modelDependent} (Figure \ref{fig:HRD}). The luminosities ($\log{L / L_{\odot}}$ $\sim$ 5.0 - 5.2) are noticeably smaller than what has been measured for O-type supergiants of similar spectral type \citep[$\log{L / L_{\odot}}$ $\sim$ 5.6 - 5.95;][]{Najarro:2011nh, Bouret:2012im}, though further work is required to determine if these stars are truly anomalous. The total number of cluster stars predicted by the model (618.9 $\pm$ 33) is in good agreement with the observed value ($\sum$P$_{pm}$ = 636.7), though the expected number of WR stars is $\sim$1.3$\sigma$ higher than observed (18.4 $\pm$ 1.75, compared to N$_{wr}$ = 13).

\begin{figure}
\begin{center}
\includegraphics[scale=0.35]{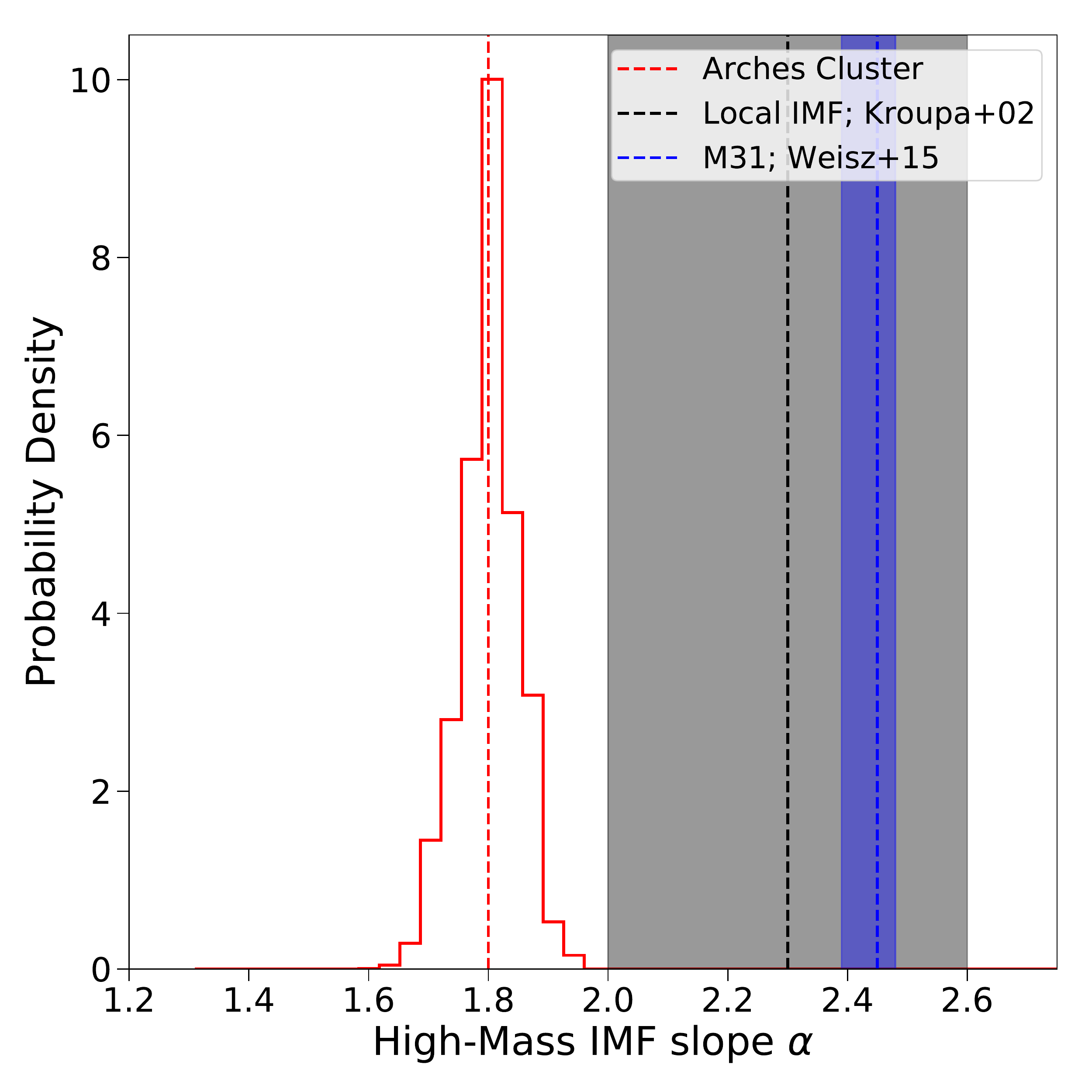}
\caption{The posterior probability distribution for the high-mass IMF slope $\alpha$ in the Arches cluster (red) compared to the local IMF \citep[black dotted line;][]{Kroupa:2002lq} and the IMF of young clusters in M31 \citep[blue dotted line;][]{Weisz:2015sf}, with the 1$\sigma$ statistical uncertainties shown by the respective shaded regions. The Arches IMF slope is significantly lower than the Milky Way or M31, indicating that the cluster has a top-heavy IMF. Note that the uncertainties shown in this figure are statistical in nature. We estimate a systematic uncertainty of $\pm$0.06 in our measurement of $\alpha$.}
\label{fig:IMF_posterior}
\end{center}
\end{figure}

We obtain a high-mass power-law slope of $\alpha$ = 1.80 $\pm$ 0.05, which is either $\sim$1.6$\sigma$ or $\sim$10$\sigma$ lower than the local IMF value, depending on whether the uncertainty on the local IMF is considered. Perhaps a more informative comparison is that our result is $\sim$8.3$\sigma$ lower than the measured IMF of young clusters in M31 \citep[$\alpha = 2.45^{+0.03}_{-0.06}$;][]{Weisz:2015sf}. A comparison of these values is shown in Figure \ref{fig:IMF_posterior}. This suggests that the Arches has a top-heavy IMF, with an overabundance of high-mass stars relative to low-mass stars for M $>$ $\sim$1.8 M$_{\odot}$. The $\alpha$ we derive does somewhat depend on which stellar evolution model we adopt, as the best-fit cluster with the MIST models has $\alpha$ = 1.68 $\pm$ 0.05. We thus add a systematic error term of 0.06 to our $\alpha$ measurement (the difference between the $\alpha$ values of the two fits divided by 2), and so the final constraint becomes $\alpha$ = 1.80 $\pm$ 0.05 (stat) $\pm$ 0.06 (sys). Note that when the statistical and systematic errors are added in quadrature, the Arches result remains 6.6$\sigma$ lower than the M31 result.

The best-fit cluster age is log $t$ = 6.57 $\pm$ 0.02 ($\sim$3.7 $\pm$ 0.17 Myr), with negligible systematic error. This is generally older than previous ages reported in the literature. Past estimates come primarily from spectroscopic studies of the massive stars, with values of 2 - 2.5 Myr based on the observed nitrogen abundances and luminosities of WR stars \citep{Najarro:2004ij}, 2 - 4 Myr based on the locations of WR + O stars on the HR diagram \citep{Martins:2008hl}, 2 - 3.3 Myr based on the spectral types of candidate main-sequence stars \citep{Clark:2018ij}, and 2.6$^{+0.4}_{-0.2}$ Myr based on the properties of an eclipsing binary in the cluster \citep{Lohr:2018gm}. However, a cluster age of 3.7 $\pm$ 0.7 was obtained by \citet{Schneider:2014pj} based on the shape of the PDMF relative to stellar population models with binary star evolution, which is more consistent with our result.

We infer a cluster mass of M$_{cl}$ = (2.44$^{+0.20}_{-0.16} \pm$ 0.21) x 10$^4$ M$_{\odot}$, which represents the intrinsic mass between 0.8 M$_{\odot}$ -- 150 M$_{\odot}$ out to a cluster radius of 3 pc. This assumes that the 1-segment IMF model is valid over the entire mass range and that the radial profile is adequately modeled for r $<$ 0.25 pc, which is outside the boundary of the observed sample ($\mathsection$\ref{sec:caveats}). However, the advantage of this result is that it is jointly constrained with the IMF, while previous photometric mass estimates of the cluster needed to adopt an IMF and extrapolate it to achieve a similar depth \citep[e.g.][]{Serabyn:1998hw, Figer:1999lo, Espinoza:2009bs}.

As a consistency check, we compare the best-fit cluster model to dynamical mass estimates of the cluster made by \citet{Clarkson:2012ty}. Using the velocity dispersion of the cluster core region, they estimate the dynamical mass of the cluster to be $0.9^{+0.40}_{-0.35}$ x $10^{4}$ M$_{\odot}$ for $r_{cl} < 0.4$ pc and $1.5^{+0.74}_{-0.60}$ x $10^{4}$ M$_{\odot}$ for $r_{cl} < 1.0$ pc. Since our model only contains stellar masses down to 0.8 M$_{\odot}$, we would expect the enclosed mass at these radii to be lower than the dynamical estimate. This is indeed the case, with model enclosed masses of (0.74 $\pm$ 0.10) x 10$^{4}$ M$_{\odot}$ and (1.2 $\pm$ 0.14) x 10$^{4}$ M$_{\odot}$ for $r_{cl} < 0.4$ pc and $r_{cl} < 1.0$ pc, respectively.

\begin{figure*}
\begin{center}
\includegraphics[scale=0.35]{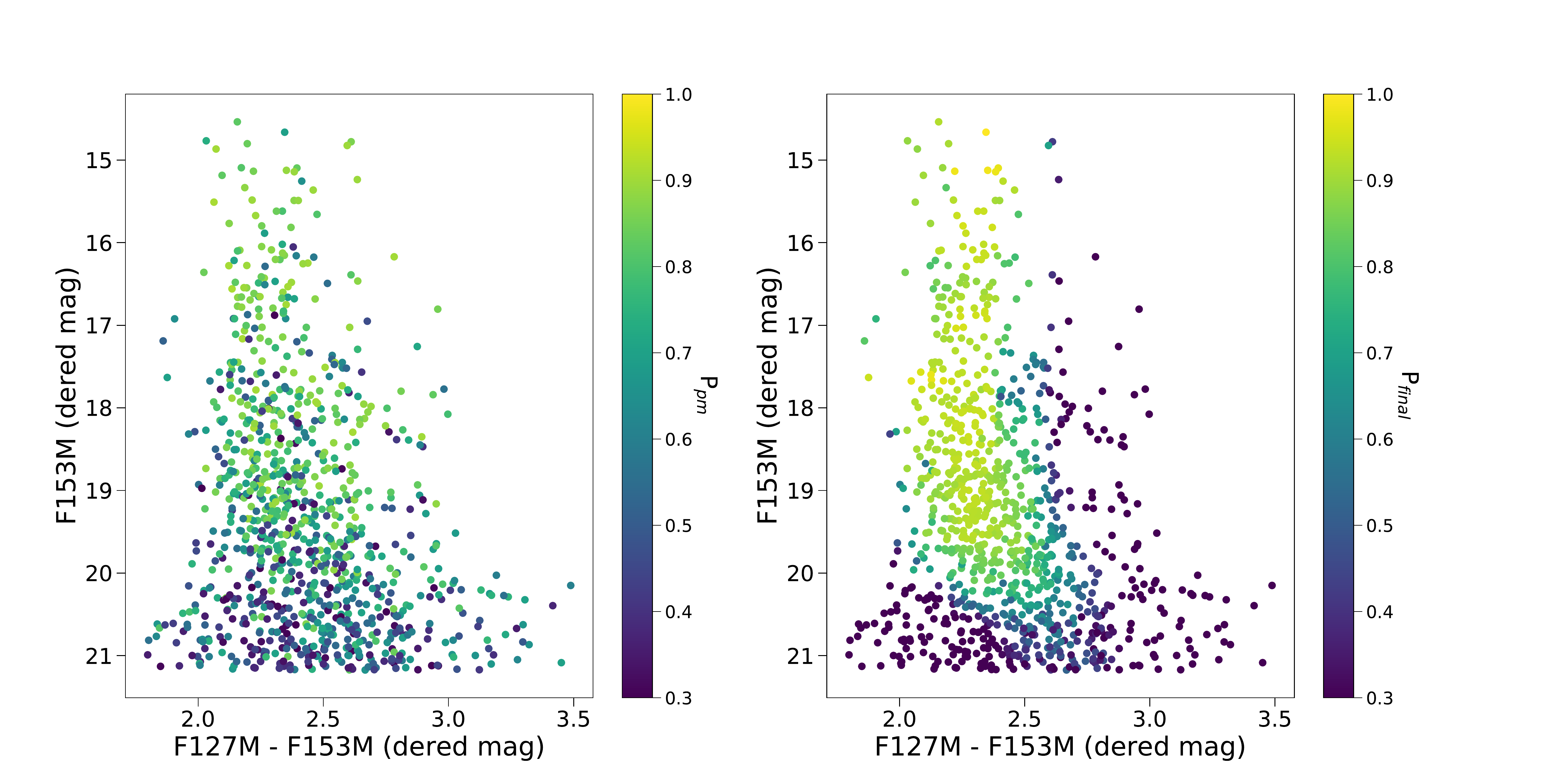}
\caption{P$_{pm}$ (left) and P$_{final}$ (right) for the observed sample, plotted in the CMD. P$_{final}$ is a more accurate determination of the cluster membership probability since it uses both proper motion and photometric information, but is dependent on the best-fit cluster model from the IMF analysis. Regions where P$_{pm} >$ P$_{final}$ reveal field contamination in the proper motion memberships, in particular around the Red Clump (F153M $\sim$ 18 mag, F127M - F153M $>$ $\sim$2.5 mag) and faint field stars (F153M $\geq$ 20 mag, F127M - F153M $<$ $\sim$2.5 mag). All magnitudes have been differentially de-reddened to A$_{Ks}$ = 2.38 mag using the extinction map.}
\label{fig:Pfinal}
\end{center}
\end{figure*}

We use the procedure outlined in $\mathsection$\ref{sec:modelDependent} to calculate revised membership probabilities and  $\Phi_{obs}$. Figure \ref{fig:Pfinal} shows P$_{pm}$ and P$_{clust}$ for the individual stars in the CMD. A comparison of the panels reveals the regions where P$_{pm} >$ P$_{clust}$, suggesting P$_{pm}$ is overestimated due to field contamination, which is especially evident near the Red Clump (the diagonal distribution of stars to the red of the cluster sequence at F153M $\sim$ 18 mag) and faint field star distribution (the stars to the blue of the cluster sequence at F153M $\geq$ 20 mag). The total number of cluster stars based on P$_{clust}$ is 601.3 stars, which is $\sim$6\% smaller than what is calculated from P$_{pm}$. Thus, we estimate that P$_{pm}$ (which was used in the IMF analysis) contains $\sim$6\% field contamination.

\begin{figure}
\begin{center}
\includegraphics[scale=0.35]{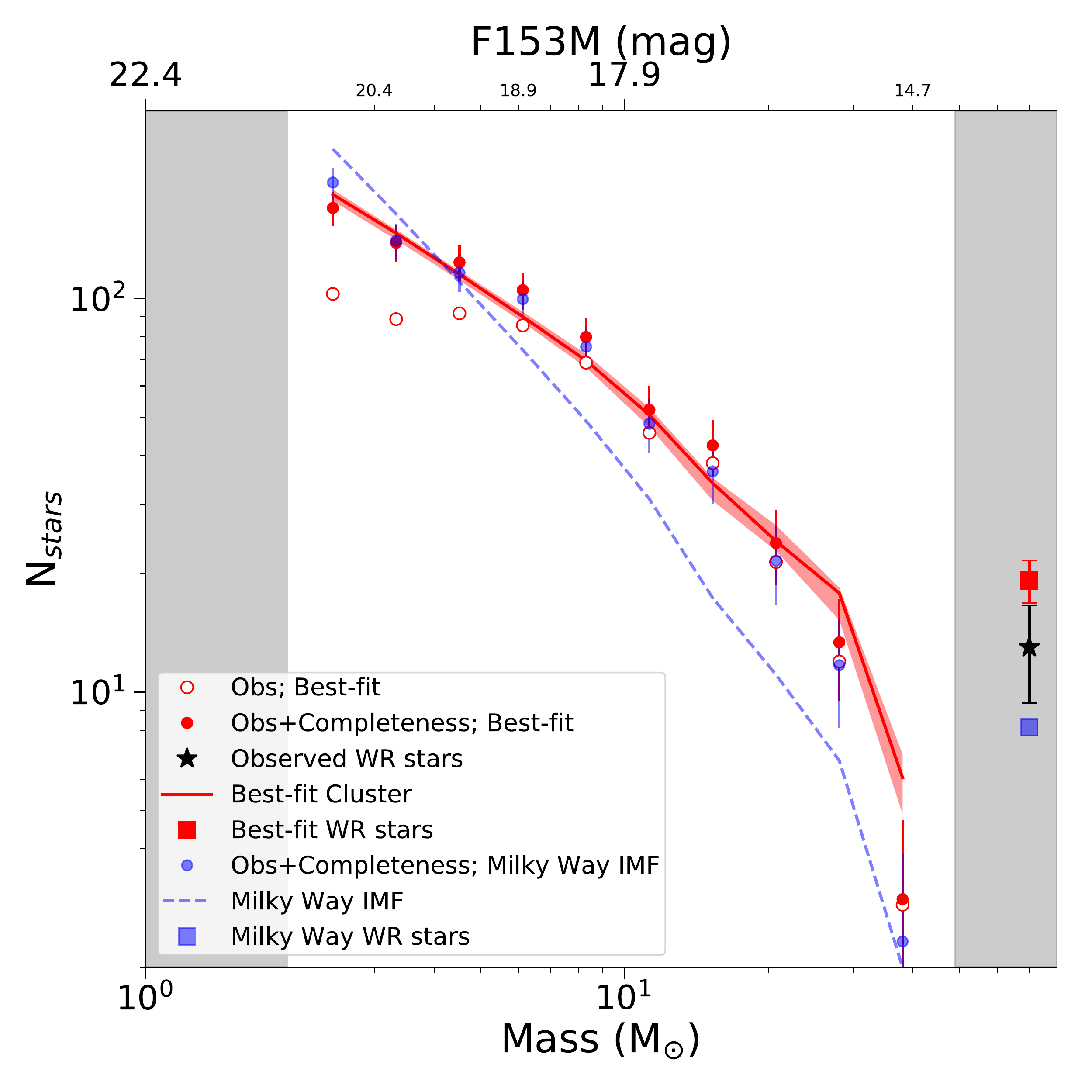}
\caption{The IMF of the Arches cluster constructed using P$_{final}$ and the stellar mass probability distributions derived using the best-fit cluster model. The red points represent the IMF constructed using the observed stellar masses calculated with the model, while the red line is the best-fit IMF itself. The 1$\sigma$ uncertainty in the best-fit cluster model is represented by the red shaded region, which is calculated by drawing different sets of parameter values from the joint posterior distribution. The red box represents the number of WR stars predicted by the best-fit model, compared to the observed number (black star). A good agreement is found between the observed IMF and the cluster model. On the other hand, the blue points represent the IMF constructed using stellar masses derived from a cluster identical to the best-fit but with a Milky Way IMF ($\alpha$ = 2.3), with the intrinsic cluster IMF shown by the blue dashed line. The Milky Way IMF is a poor fit to the data, as it significantly underestimates the number of high-mass stars and overestimates the number of low-mass stars.}
\label{fig:moneyplot}
\end{center}
\end{figure}

The observed initial mass function $\Phi_{obs}$ is shown in Figure \ref{fig:moneyplot}. Also plotted is the $\Phi_{obs}$ we would obtain if we adopted a cluster model identical to the best-fit but with the local IMF. The mass function obtained with the local IMF is significantly inconsistent with the observations, while the mass function obtained from the best-fit model is a good match to the observations.

A catalog of the observed stars with membership probabilities and mass estimates is provided as a machine-readable table with this paper. A sample of the catalog is shown in Table \ref{tab:starTable}.

\begin{deluxetable*}{l c c c c c c c c c c c c c c c}
\tablewidth{0pt}
\tabletypesize{\scriptsize}
\tablecaption{Stellar Parameters}
\tablehead{
\colhead{Name\tablenotemark{a}} & \colhead{F127M\tablenotemark{b}} & \colhead{F153M\tablenotemark{b}} & \colhead{X\tablenotemark{c}} & \colhead{$\sigma_x$} & \colhead{Y\tablenotemark{c}} & \colhead{$\sigma_y$} & \colhead{$\mu_\alpha cos\delta$} &  \colhead{$\sigma_{\mu_{\alpha cos\delta}}$} &\colhead{$\mu_\delta$} &\colhead{$\sigma_{\mu_\delta}$} & \colhead{M} & \colhead{$\sigma_M$} & \colhead{A$_{Ks}$} & \colhead{P$_{pm}$} & \colhead{P$_{clust}$} \\
& mag & mag & '' & '' & '' & '' & mas yr$^{-1}$ & mas yr$^{-1}$ & mas yr$^{-1}$ & mas yr$^{-1}$ & M$_{\odot}$ & M$_{\odot}$ & mag &  &
}
\startdata
Ae035\_001 & 15.99 & 14.12 & -34.5546 & 9.50E-04 & -4.3409 & 1.15E-03 & 0.16 & 0.025 & -0.12 & 0.023 & 26.4 & 3.9 & 2.24 & 0.84 & 0.91 \\ 
Aw061\_002 & 16.38 & 14.47 & 20.3508 & 1.63E-03 & 57.4818 & 1.46E-03 & 0.04 & 0.054 & 0.01 & 0.064 & 22.8 & 3.4 & 2.35 & 0.90 & 0.88 \\ 
Aw048\_004 & 16.39 & 14.49 & 15.0491 & 1.12E-03 & 46.0383 & 1.22E-03 & 0.26 & 0.013 & 0.05 & 0.031 & 22.8 & 5.5 & 2.13 & 0.74 & 0.89 \\ 
As017\_001 & 16.95 & 14.71 & 5.6190 & 1.03E-03 & -16.4889 & 1.15E-03 & -0.16 & 0.019 & 0.05 & 0.064 & 26.4 & 6.4 & 2.05 & 0.82 & 0.92 \\ 
An022\_002 & 17.27 & 14.72 & -6.6602 & 1.03E-03 & 20.8669 & 1.15E-03 & 0.15 & 0.087 & -0.06 & 0.027 & 35.5 & 8.5 & 2.28 & 0.86 & 0.42 \\ 
Aw006\_001 & 17.28 & 14.86 & 3.9843 & 1.03E-03 & 3.9671 & 1.15E-03 & 0.15 & 0.014 & 0.24 & 0.017 & 30.6 & 7.4 & 2.34 & 0.69 & 1.00 \\ 
Ae010\_001 & 17.04 & 14.92 & -10.0664 & 1.22E-03 & -0.3837 & 1.22E-03 & -0.08 & 0.040 & -0.13 & 0.027 & 26.4 & 3.9 & 2.46 & 0.85 & 0.98 \\ 
\enddata
\tablenotetext{a}{Naming convention is as follows: The first letter is always ``A'', followed by ``n/s/e/w'' to designate whether the star is to the north, south, east or west quadrant relative to the central reference star, as determined using 45$^\circ$ and -45$^{\circ}$ line boundaries that intersect at the reference star position. The number immediately following gives the radius of the given star relative to the reference in arcseconds, while the second number (following the ``\_'' separator) is a unique index for the star relative to all others at that same radius.}
\tablenotetext{b}{Observed magnitudes not corrected for differential extinction.}
\tablenotetext{c}{Positions are reported relative to a central reference star, chosen to be star 8 in the \citet{Clarkson:2012ty} catalog. Note that this star isn't in reported in this catalog because it is is outside the cluster radius range used for this study. Reported positional uncertainties are the uncertainties in the star position and the reference star position added in quadrature.}
\tablecomments{Table \ref{tab:starTable} is published in its entirety in the machine-readable format. A portion is shown here for guidance regarding its form and content.}
\label{tab:starTable}
\end{deluxetable*}

\subsection{1-segment vs. 2-segment IMF Model}
\label{sec:IMFcomp}
\begin{figure*}
\begin{center}
\includegraphics[scale=0.45]{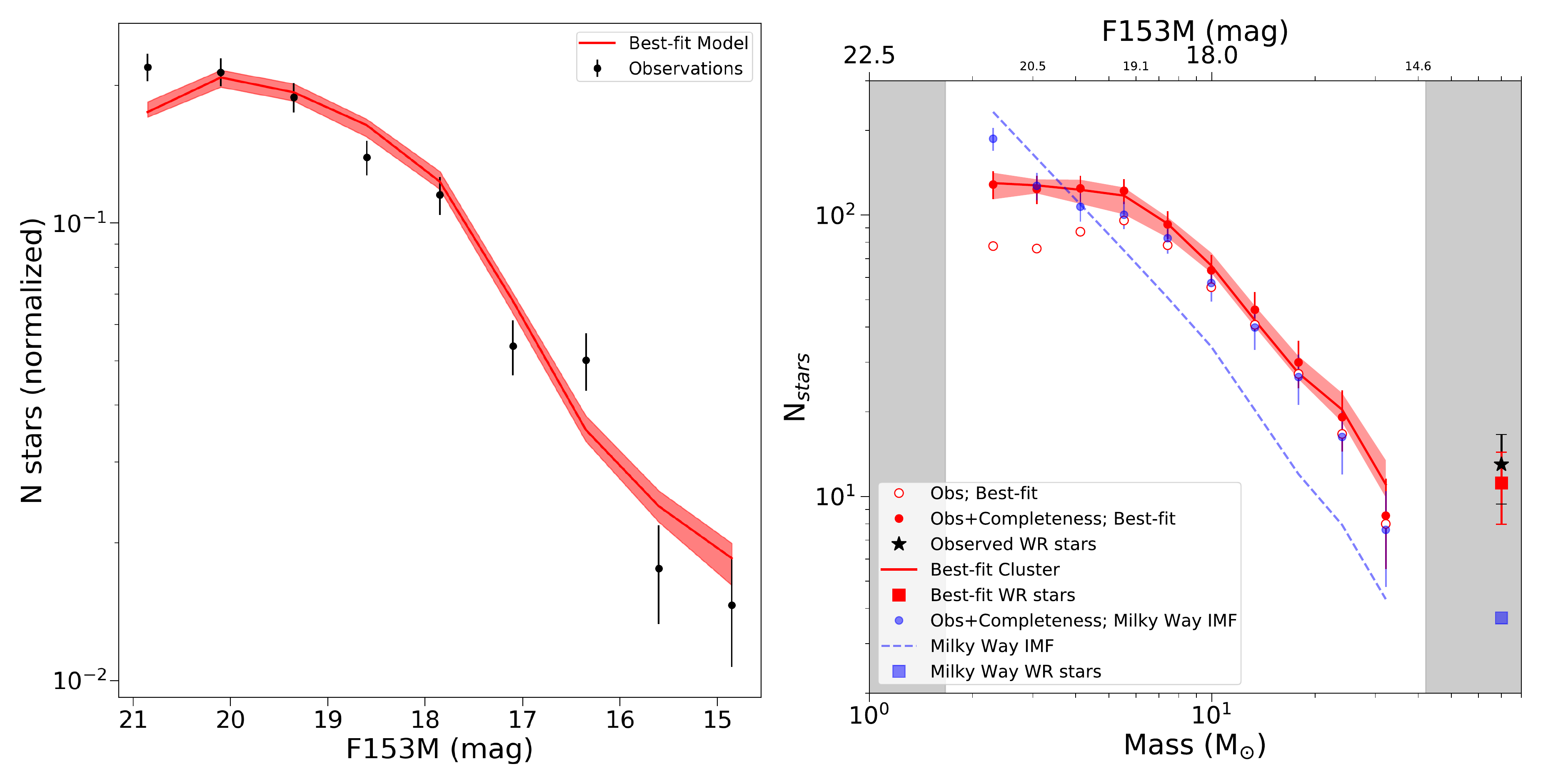}
\caption{A comparison of the best-fit 2-segment IMF model with the observed luminosity function (left) and $\Phi_{obs}$ (right). The features of the plots are the same as described for Figures \ref{fig:lum_func} and \ref{fig:moneyplot}. The 2-segment IMF solution cannot be ruled out based on our data. Additional studies are required to distinguish between the 1-segment and 2-segment IMF models.}
\label{fig:2seg}
\end{center}
\end{figure*}

The best-fit 2-segment cluster model is also significantly different than the local IMF, but in a different manner than the 1-segment IMF model. While the high-mass IMF slope is perhaps slightly shallow ($\alpha_1$ = 2.04$^{+0.14}_{-0.19}$ $\pm$ 0.04), the real discrepancy is in the detection of a significant m$_{break}$ at 5.8$^{+3.2}_{-1.2}$ $\pm$ 0.02 M$_{\odot}$, which is an order of magnitude larger than the local IMF (m$_{break}$ = 0.5 M$_{\odot}$). The power-law slope below m$_{break}$ is $\alpha_2$ = 1.10$^{+0.39}_{-0.31}$ $\pm$ 0.08, which is consistent with the local IMF values of 1.3 $\pm$ 0.3 for 0.08 M$_{\odot} \leq$ M $<$ 0.5 M$_{\odot}$ \citep{Kroupa:2002lq}. As a result of the high m$_{break}$, the 2-segment IMF solution could be characterized as ``bottom-light'', with a deficit of low-mass stars relative to the local IMF. Figure \ref{fig:2seg} shows the 2-segment model compared to the observed luminosity function and the derived $\Phi_{obs}$. 

One of the advantages of the bayesian framework is that we can distinguish between 1-segment and 2-segment IMF models by comparing the likelihoods of the best-fit solutions. We use the Bayesian Information Criterion \citep[BIC;][]{Schwarz+78} for this comparison:

\begin{equation}
\label{eq:BIC}
\text{BIC} = \ln(n)*k - 2*\ln(\mathcal{L})
\end{equation}

\noindent where $n$ is the total number of stars in the sample (980, in this case), $k$ is the number of free parameters in the model (i.e., 6 for 1-segment model and 8 for the 2-segment model), and $\mathcal{L}$ is the best-fit likelihood of the model. When comparing the two models, the model with the lowest BIC is preferred, and the absolute value of the difference between the BIC values ($\Delta$BIC) indicates if the preference is statistically significant. Table \ref{tab:likelihood} contains the likelihoods and BIC values for the 1-segment and 2-segment IMF fits. 

The 1-segment IMF model is slightly preferred over the 2-segment IMF model in both Pisa/Geneva and MIST cases, with $\Delta$BIC = 10.4 and 8.2, respectively. To assess the significance of these values, we generate artificial clusters with 1-segment and 2-segment IMFs as described in Appendix \ref{app:artificialTest} (adopting the best-fit values in the Arches solutions) and fit them in both the 1-segment and 2-segment cases, and then calculate the corresponding $\Delta$BIC values. In our simulations, the $\Delta$BIC values between 1- and 2-segment models are consistently a factor 1.5-7.5 times greater the actual Arches data. Thus, the real data are substantially more agnostic on the distinction between 1- and 2-component IMF models than the synthetic datasets. We conclude that we cannot rule out the 2-segment IMF model,  though we adopt 1-segment IMF model as the overall best-fit. In either case, our results show that the Arches cluster IMF is significantly different from the local IMF.

\subsection{The Impact of Stellar Evolution Models and Stellar Multiplicity}
\label{sec:IMFimpact}
Table \ref{tab:Results} reveals that the best-fit model parameters are only weakly dependent on the choice of stellar evolution model. Similar to $\mathsection$\ref{sec:IMFcomp}, we use the BIC test to determine whether our analysis prefers one set of evolution models over the other. For the 1-segment IMF model, the Pisa/Geneva model solution is slightly favored with $\Delta$BIC = 0.8. However, artificial cluster tests show that this $\Delta$BIC is not significant, as the average difference between evolution model fits is $\Delta$BIC = 2.1 $\pm$ 1.2. Thus we conclude that we cannot distinguish between the two solutions, and adopt the Pisa/Geneva solution as the result and use the MIST solution to estimate the systematic error. In the 2-segment IMF case, the MIST solution is favored with a $\Delta$BIC = 5.6. The simulations show that this discrepancy is indeed significant, with an average difference of $\Delta$BIC = 5.39 $\pm$ 2.56 between 2-segment IMF fits using different evolution models. As a result, we adopt the MIST solution for the 2-segment IMF case, and use the Pisa/Geneva solution to estimate the systematic error.



Whether stellar multiplicity is accounted for in the cluster model is found to significantly impact the quality of the fit. The BIC analysis strongly favors the models that include stellar multiplicity, with $\Delta$BIC values of 27.6 and 8.6 for the 1-segment and 2-segment IMF model cases, respectively. As seen in Table \ref{tab:likelihood}, this difference is primarily driven by the CMD likelihood component. Artificial cluster tests show that the observed $\Delta$BIC values are significant; for artificial clusters that have intrinsic multiplicity, $\Delta$BIC = 12.2 $\pm$ 0.5 in favor of the fit with multiplicity in the 1-segment IMF case and $\Delta$BIC = 8.8 $\pm$ 0.5 in the 2-segment IMF case. Thus, we adopt the model fits with multiplicity included over those without.

\section{Discussion}
\label{sec:discussion}
\subsection{Past IMF Measurements of the Arches Cluster}
\label{sec:IMF_comparison}
\begin{figure}
\begin{center}
\includegraphics[scale=0.5]{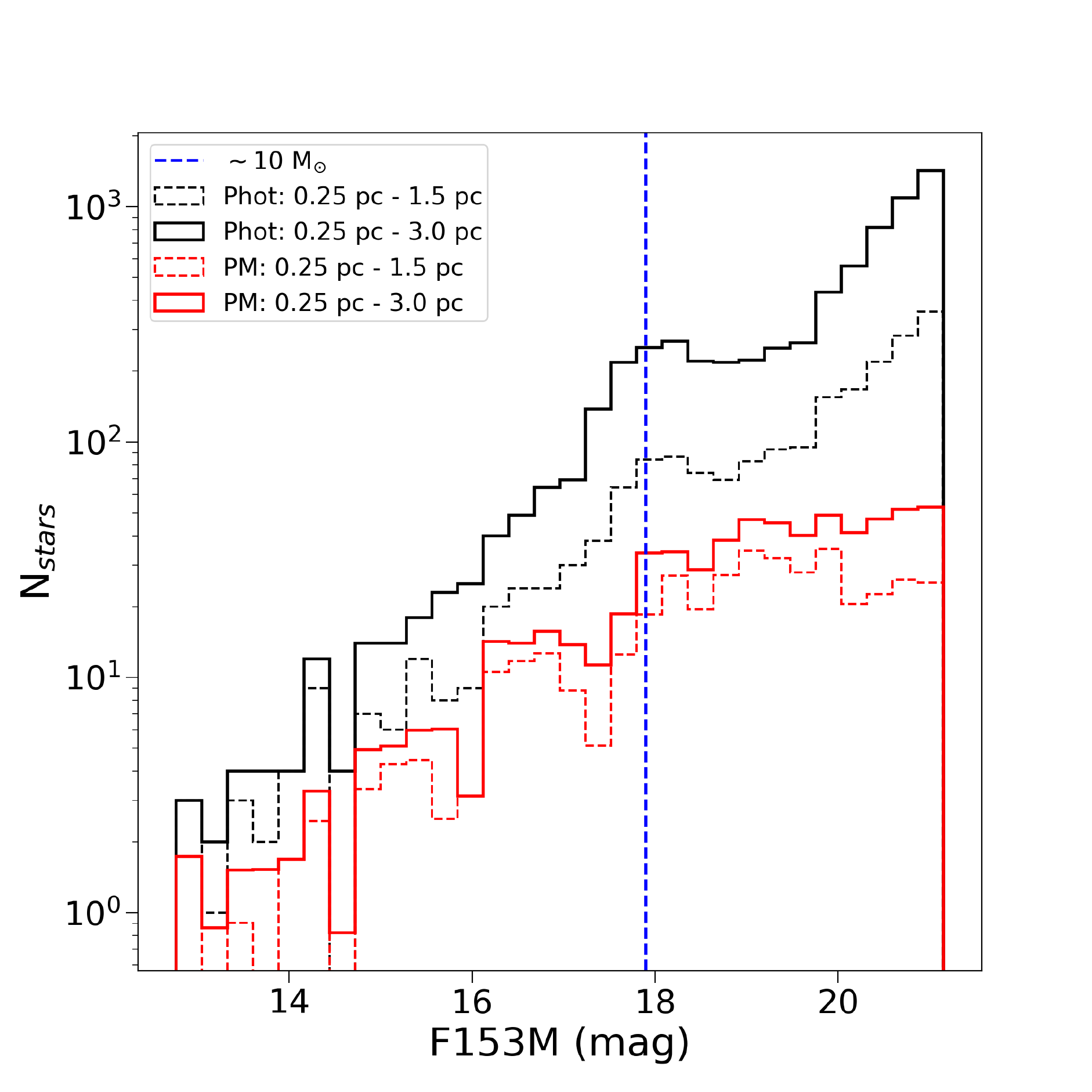}
\caption{A comparison between Arches cluster members selected via proper motion versus a photometric color cut. The proper motion sample, shown as the red solid and dashed lines, contains all stars with P$_{pm}$ $>$ 0.3, where each star is weighted by its membership probability for radius ranges of 0.25 pc $<$ r $<$ 3.0 pc and 0.25 pc $<$ r $<$ 1.5 pc, respectively. The photometric sample is selected as all stars with differentially de-reddened F127M - F153M colors within $\pm$ 0.3 mag of the average color on the main sequence, similar to \citet{Habibi:2013th}. The photometric sample is larger than the proper motion due to field contamination, even at high masses (blue dotted line represents M = 10 M$_{\odot}$).}
\label{fig:photvspm}
\end{center}
\end{figure}

Our result that the high-mass slope of the Arches IMF is significantly top-heavy differs from previous photometric studies of the cluster which have found the IMF to be largely consistent with the local IMF \citep{Kim:2006fy, Espinoza:2009bs, Habibi:2013th, Shin:2015hb}. However, a key advantage of this study is the use of proper motions to calculate cluster membership probabilities, which produces a significantly more accurate sample of cluster members than is possible through photometry alone. For example, Figure \ref{fig:photvspm} shows a comparison between cluster samples obtained using proper motions versus a photometric color-cut similar to \citet{Habibi:2013th}. Even when limited to r $<$ 1.5 pc and M $>$ 10 M$_{\odot}$ \citep[the range PDMF was measured by][]{Habibi:2013th}, the photometric sample is systematically larger than the proper motion selection due to field contamination. On the other hand, adopting stricter color-cuts on photometric samples can be problematic as well, as \citet{Espinoza:2009bs} note that the color-cuts they adopt force them to eliminate stars that could be high-mass (M $>$ 16 M$_{\odot}$) cluster members.

An alternative approach is to statistically subtract the field from the cluster using the field population observed in nearby control fields \citep[e.g.][]{Kim:2006fy, Shin:2015hb}. However, differential extinction can alter both the average extinction and the distribution of extinction values between two fields (e.g., note the detailed extinction structures in Figure \ref{fig:redmap}). As a result, it is challenging to obtain a sufficiently accurate model of the field stars in the cluster observations. In addition, care must be taken that the control fields are beyond the extent of the cluster, which H15 shows extends to a radius of at least 75'' ($\sim$3 pc).

It is interesting to note that several previous studies have reported evidence of an enhancement in the PDMF at $\sim$6 M$_{\odot}$, whether it be evidence of a turnover \citep[][]{Stolte:2005mz} or a localized ``bump'' in the mass function \citep{Kim:2006fy}. The presence of such a feature may be driving the 2-segment IMF model solution. Future studies are needed to extend the proper-motion selected sample to lower masses in order to definitively distinguish between the 1-segment and 2-segment IMF models and determine whether an enhancement at 5-6 M$_{\odot}$ truly exists.

\subsection{A Top-Heavy IMF Near the GC?}
The top-heavy IMF we obtain for the Arches cluster ($\alpha$ = 1.80 $\pm$ 0.05 $\pm$ 0.06) is in good agreement with the YNC \citep[$\alpha$ = 1.7 $\pm$ 0.2 for M $>$ 10 M$_{\odot}$;][]{Lu:2013wo}. This suggests that this unusual IMF extends beyond the central parsec of the Galaxy and into the CMZ, which spans a galactocentric radius of $\sim$200 pc \citep{Morris:1996yq}. Unfortunately, the exact birth location of the Arches is not well constrained due to the range of possible orbits allowed by the three-dimensional motion of the cluster \citep{Stolte:2008qy, Kruijssen:2015fx}. Further, the proper motion of the cluster \emph{in the galactocentric reference frame} is not yet well determined, as current estimates are based on the relative proper motion between the cluster and a single-gaussian kinematic model for the field \citep[e.g.][]{Clarkson:2012ty}. In reality, the field exhibits a more complex kinematic structure (see H15 and Appendix \ref{app:GMM}), and so the cluster motion may need to be revised. This is left to a future paper.

\begin{figure}
\begin{center}
\includegraphics[scale=0.35]{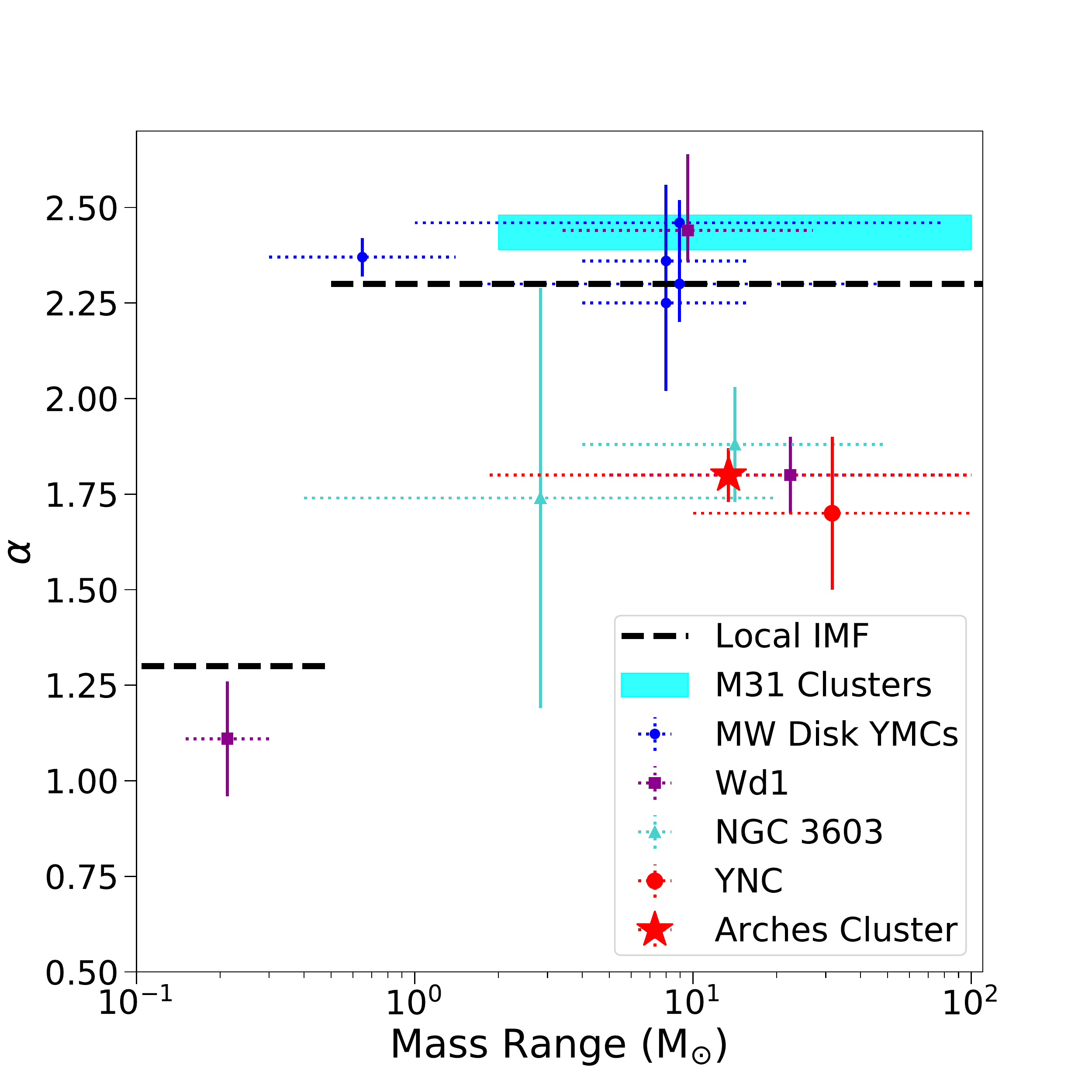}
\caption{A plot of IMF slope $\alpha$ versus mass for YMCs in the Galactic disk (blue points: Wd2, Trumpler 14, Trumpler 16, h and $\chi$ Persei; purple squares: Wd1; turquoise triangles: NGC 3603) and the GC (red circle: YNC; red star: Arches cluster, with statistical and systematic errors added in quadrature). The dotted error bars in the X-direction show the mass range over which the measurement was made, while the solid error bars in the Y-direction show the measurement uncertainty. The references are provided in the text; Wd1 and NGC 3603 have their own symbols in order to represent the multiple values reported in the literature. Also shown is the local IMF (black dashed line) and IMF measured for young cluster in M31 from \citet[][cyan box]{Weisz:2015sf}}
\label{fig:alpha_plot}
\end{center}
\end{figure}

However, this result raises the question of whether the top-heavy IMF is truly due to the GC environment or if it is a general property of young massive clusters \cite[YMCs; see review by][]{Portegies-Zwart:2010xw}. In Figure \ref{fig:alpha_plot} we compare IMF measurements of YMCs in the Milky Way disk to the YNC and Arches cluster at the GC. The YMC sample includes Westerlund 1 \citep[Wd1;][]{Gennaro:2011nx, Lim:2013sf, Andersen:2017aq}, Westerlund 2 \citep[Wd2; ][]{Zeidler:2017lq}, NGC 3603 \citep{Harayama:2008wd, Pang:2013ve}, Trumpler 14 and 16 \citep{Hur:2012gd}, and h and $\chi$ Persei \citep{Slesnick:2002ij}.

Figure \ref{fig:alpha_plot} shows that the YMCs in the Galactic disk are generally consistent with the local IMF, though potential discrepancies exist. In particular, NGC 3603 has been found to be potentially top-heavy \citep[$\alpha$ = 1.74$^{+0.62}_{-0.47}$ and $\alpha$ = 1.88 $\pm$ 0.15;][respectively]{Harayama:2008wd, Pang:2013ve}. However, these results may be biased due to mass segregation, which both studies find to be significant in the cluster. Indeed, the uncertainty in the \citet[][]{Harayama:2008wd} measurement is quite large in order to account for this (as well as other) systematic uncertainties, while \citet{Pang:2013ve} acknowledge that their IMF measurement is restricted to the inner 60" of the cluster. The IMF of Westerlund 1 is potentially discrepant as well, with reported high-mass IMF slopes that are both near-standard \citep[$\alpha$ = 2.44$^{+0.08}_{-0.20}$;][via near-infrared photometry]{Gennaro:2011nx} and top-heavy \citep[$\alpha$ = 1.8 $\pm$ 0.1;][via optical photometry]{Lim:2013sf}. The inconsistency between these studies makes it difficult to draw conclusions about the IMF of Wd1, though the low-mass stellar content of the cluster has been found to be consistent with the local IMF \citep{Andersen:2017aq}. These cases highlight the difficulty of these measurements, as differences in cluster membership selection, stellar models, and methodology may significantly impact results.

Given the uncertainties surrounding the NGC 3603 and Wd1 measurements, the fact that other YMCs in the Galactic disk have been found to be consistent with the local IMF while the Arches and YNC are top-heavy provides tentative evidence that the top-heavy IMF is indeed caused by the extreme GC environment. We discuss the implications of a top-heavy IMF at the GC in $\mathsection$\ref{sec:Theory} and the caveats of our Arches IMF measurement in $\mathsection$\ref{sec:caveats}.

The Quintuplet cluster is a third YMC in the CMZ that provides an additional probe of the IMF at the GC. A proper motion-based analysis of the Quintuplet mass function was carried out by \citet{Husmann:2012lr}, who found a top-heavy PDMF ($\alpha$ = 1.68$^{+0.13}_{-0.09}$) for the inner 0.5 pc of the cluster. However, it is uncertain whether this is due to mass segregation or a top-heavy IMF. A study of the Quintuplet IMF using a similar approach as this work is in progress.

\subsection{Implications for Star Formation}
\label{sec:Theory}
At first, a top-heavy IMF at the GC appears to favor star formation models where the increased thermal Jeans mass leads to the formation of more high-mass stars \citep[e.g.][]{Larson:2005sp, Bonnell:2006os, Klessen:2007nx, Bonnell:2008jl, Papadopoulos:2011qu, Narayanan:2013ec}. However, the main prediction of these models is that the break mass of the IMF should increase, leading to a deficit of low-mass stars, rather than an overabundance of high-mass stars and a shallow high-mass slope. This behavior is similar to the ``bottom-light'' 2-segment IMF solution, but we do not yet have enough evidence to conclude that this is preferred over the top-heavy 1-segment IMF solution.

However, our results are generally inconsistent with models where the IMF is set by the CMF \citep[e.g.][]{Padoan:2002tw, Hopkins:2012eu}. Though the combination of turbulence and gravity naturally produces a CMF with a shape similar to the local IMF, these models predict a steeper mass slope and a bottom-heavy IMF near the GC \citep{Hopkins:2013oz, Chabrier:2014oq}. This suggests that the CMF cannot be directly mapped to the IMF and that additional physical processes are involved. n the other hand, it has been suggested that the gravo-turbulent fragmentation of a molecular cloud may lead to a top-heavy IMF (and possibly a bump around 5-6 M$_{\odot}$) if the coalescence of pre-stellar cores is highly efficient, as might be expected in Arches-like environments \citep[][]{Dib:2007tr}. In addition, recent simulations have shown that a top-heavy IMF can be produced in turbulence-dominated environments if the turbulent probability density distribution can be described as a power law at high densities \citep{Lee:2018lr}.

Previous studies have shown that radiative feedback \citep[e.g.][]{Bate:2009jw, Offner:2009ng, Krumholz:2011ye}, protostellar outflows \citep[e.g.][]{Krumholz:2012qf, Federrath:2014br}, and magnetic fields \citep[e.g.][]{Hennebelle:2011dg, Myers:2013ht} can impact the IMF, as well. Only recently have simulations begun to incorporate all of these processes simultaneously \citep{Myers:2014xc, Krumholz:2016bq, Li:2018yw, Cunningham:2018vs}. However, these simulations have been limited to molecular clouds with initial masses $\leq$1000 M$_{\odot}$, and are only applicable to low mass stars in environmental conditions similar to local star forming regions. Future simulations of higher masses molecular clouds in starburst-like environments are needed in order to determine what physics is behind a shallow high-mass IMF slope in the GC.

\subsection{Caveats}
\label{sec:caveats}
A caveat of our IMF measurement is that we do not take the potential effects of tidal stripping into account. Tidal stripping might play a significant role in the evolution of the Arches cluster given the strong Galactic tidal field near the GC. Since tidal stripping preferentially removes low-mass stars from a cluster \citep[e.g.][]{Kruijssen:2009wq, Lamers:2013lr}, it could bias the mass function to appear top-heavy. However, it is unclear from current dynamical models of the Arches whether tidal stripping would significantly impact the mass range examined in this study (M $\gtrsim$ 1.8 M$_{\odot}$). N-body simulations by \citet{Habibi:2014qf} predict the formation of tidal tails out to 20 pc from the cluster core, which potentially contribute to the population of isolated massive stars observed near the GC. Additional simulations by \citet{Park:2018tg} also predict the formation of tidal tails, but find that $\sim$98\% of the tidally stripped stars have masses less than 2.5 M$_{\odot}$ and that the impact on the mass function above this limit is minor. This is consistent with the observations of H15 that find no evidence of tidal tails down to $\sim$2.5 M$_{\odot}$ and to a cluster radius of 3 pc. Thus, we assume that the effects of tidal stripping can be ignored for the mass range in our sample.

It should be noted that the dynamical models discussed above require assumptions regarding the initial conditions and orbit of the Arches cluster, both of which are quite uncertain. In addition, only stars are considered in the simulations, though the expulsion of excess gas after cluster formation is expected to have a significant impact on the dynamical evolution of the cluster as well \citep[e.g.][]{Goodwin:2006fj, Bastian:2006vf, Farias:2015oz}.

Another caveat is that this analysis does not contain data for r $<$ 0.25 pc, where the observational completeness is low due to stellar crowding. We adopt the radial profile of \citet{Espinoza:2009bs} for this region when modeling the cluster ($\mathsection$\ref{sec:model}), but their profile was derived only for stars with M $>$ 10 M$_{\odot}$. So, while magnitude-dependent radial profiles are used to account for mass segregation between 0.25 pc $<$ r $<$ 3.0 pc, the profile for all stars within the cluster core is assumed to be the same. Combining the \emph{HST} data set from this study with higher resolution ground-based observations of the cluster core would remove the need for this assumption.

Finally, we note that changing the extinction law does not have a significant impact on the IMF results. To demonstrate this, we repeat the analysis using the \citet{Nishiyama:2009fc} and original H18 extinction laws, which are shallower (i.e. lower A$_{\lambda}$ / A$_{Ks}$ values) and steeper (i.e. higher A$_{\lambda}$ / A$_{Ks}$ values) than the law we ultimately adopt, respectively (Appendix \ref{app:EL}). In both cases, the only parameter that is significantly changes is the overall extinction, which decreases H18 law (A$_{Ks}$ = 2.07 $\pm$ 0.01 mag) and increases for the \citet{Nishiyama:2009fc} law (A$_{Ks}$ = 2.47 $\pm$ 0.01 mag). The high-mass IMF slope only changes by $|\Delta\alpha|$ = 0.01 in the 1-segment case and $|\Delta\alpha|$ = 0.04 in the 2-segment case, well within the 1$\sigma$ uncertainties. Additionally, $|\Delta\alpha_2|$ = 0.03 and $|\Delta m_{break}|$ = 0.49 M$_{\odot}$ for the 2-segment case, again well within uncertainties. Therefore the extinction law is not a significant source of uncertainty in this analysis.


\section{Conclusions}
\label{sec:conclusions}
We use multi-epoch \emph{HST} WFC3-IR observations and Keck OSIRIS K-band spectroscopy to measure the IMF of the Arches cluster. Critically, we use proper motions to calculate cluster membership probabilities for stars down to $\sim$1.8 M$_{\odot}$ over a radius range of 0.25 pc $\leq$ r$_{cl}$ $\leq$ 3.0 pc, obtaining a sample with just $\sim$6\% field contamination. This is a significant improvement over purely photometric studies which are compromised by the sever differential extinction across the field. Our proper motion sample contains $\sum$P$_{pm}$ = 638.0 cluster members, which we combine with K-band spectra of 5 O-type giants and supergiants in order to measure the IMF.

We forward model the Arches cluster to simultaneously constrain its IMF with the cluster distance, total mass, average extinction, and residual differential extinction (after a spatially-dependent extinction correction). This approach allows us to account for observational uncertainties, completeness, mass segregation, and stellar multiplicity. We generate synthetic clusters and compare them to the observations using a likelihood equation with four components: the distribution of stars in the color-magnitude diagram, the total number of observed stars, the total total number of Wolf-Rayet stars with r$_{cl} <$ 0.75 pc (taken from spectroscopic surveys in the literature), and the measured T$_{eff}$ of the spectroscopic stars versus those predicted by the cluster model.

We find that the Arches IMF is best described by a 1-segment power law with a slope of $\alpha$ = 1.80 $\pm$ 0.05 (stat) $\pm$ 0.06 (sys), which is significantly more shallow than the local IMF and thus ``top-heavy.'' However, we cannot discount a 2-segment power law model that has a high-mass slope only slightly less than the local IMF slope ($\alpha_1$ = 2.04$^{+0.14}_{-0.19}$ $\pm$ 0.04) but exhibits a break at 5.8$^{+3.2}_{-1.2}$ $\pm$ 0.02 M$_{\odot}$. This would make the Arches IMF deficient in low-mass stars and thus ``bottom-light.'' In either case, the Arches IMF is significantly different than the local IMF common throughout the Milky Way and nearby galaxies.

The unusual nature of the Arches IMF, combined with the top-heavy IMF observed for the Young Nuclear Cluster \citep[$\alpha$ = 1.7 $\pm$ 0.2;][]{Lu:2013wo} suggests that the starburst-like environment at the GC induces variations in the IMF. Other YMCs in the Galactic disk have been found to be generally consistent with the local IMF, suggesting that these variations are truly due to the GC environment rather than an intrinsic property of YMCs. However, several disk YMCs (NGC 3603, Westerlund 1) have been found to be potentially discrepant with the local IMF, and so future studies must clarify the nature of their IMFs in order to strengthen this conclusion.

We note that the potential impact of tidal stripping is not included in our analysis. Measurements of the stellar radial density profile \citep{Hosek:2015cs} and the N-body simulations of the Arches \citep{Park:2018tg} suggest that tidal stripping has not significantly impacted the mass function over the mass range examined in this study. However, better constraints on the cluster orbit \citep[e.g.][]{Stolte:2008qy} and full dynamical modeling of the stars and primordial gas is needed to fully explore the effects of tidal stripping. This is beyond the scope of the current study.

New observational capabilities will offer exciting opportunities for future study of the Arches cluster IMF. In particular, the \emph{James Webb Space Telescope} (JWST) will provide the increased sensitivity and spatial resolution required to extend the IMF significantly beyond the current completeness limits \citep[e.g.][]{Kalirai:2018fk}, allowing us to distinguish between the 1-segment and 2-segment IMF solutions. In addition, the larger field-of-view offered by JWST will facilitate the detection of tidal tails, which will yield critical new insight into the cluster's dynamical evolution.

\acknowledgements
The authors thank Kelly Lockhart and Tuan Do for help with OSIRIS data reduction, as well as the anonymous referee for their feedback. M.W.H. and J.R.L. acknowledge support from NSF AAG (AST-1518273) and HST GO-13809. F.N. acknowledges financial support through Spanish grants ESP2015-65597-C4-1-R and ESP2017-86582-C4-1-R (MINECO/FEDER). This work is based on observations made with the NASA/ESA Hubble Space Telescope, obtained at the Space Telescope Science Institute, which is operated by the Association of Universities for Research in Astronomy, Inc., under NASA contract NAS 5-26555. The observations are associated with programs 11671, 12318, 12667, and 14613. Additional observations were obtained at the W.M. Keck Observatory, which is operated as a scientific partnership among the California Institute of Technology, the University of California and the National Aeronautics and Space Administration. The Observatory was made possible by the generous financial support of the W.M. Keck Foundation. The authors wish to recognize and acknowledge the very significant cultural role and reverence that the summit of Mauna Kea has always had within the indigenous Hawaiian community. We are most fortunate to have the opportunity to conduct observations from this mountain. This research has made extensive use of the NASA Astrophysical Data System. 

\facilities{HST (WFC3-IR), Keck:II (OSIRIS)}

\software{AstroPy \citep{Astropy-Collaboration:2013kx}, extlaw\_H18.py (https://doi.org/10.5281/zenodo.1063708), \texttt{KS2} \citep{Anderson:2008qy}, Matplotlib \citep{Hunter:2007}, \texttt{Molecfit} \citep{Smette:2015ge, Kausch:2015fe}, SciPy \citep{SciPy}, \texttt{Skycorr} \citep{Noll:2014gr}}

\bibliography{Thesis}

\clearpage

\appendix

\section{Gaussian Mixture Model}
\label{app:GMM}
The Gaussian Mixture Model used to describe the cluster and field kinematics is described in Table \ref{PopModel}. Cluster membership probabilities are calculated using this model as discussed in $\mathsection$\ref{sec:Membership}.

\begin{deluxetable}{c |c c |c c | c c | c c | c c }
\tabletypesize{\scriptsize}
\tablecaption{Cluster and Field Population Model: Free Parameters, Priors, and Results}
\tablehead{
& \multicolumn{2}{c}{Cluster Gaussian} & \multicolumn{2}{c}{Field Gaussian 1} & \multicolumn{2}{c}{Field Gaussian 2} & \multicolumn{2}{c}{Field Gaussian 3} & \multicolumn{2}{c}{Field Gaussian 4} \\
\colhead{Parameter\tablenotemark{a}} & \colhead{Prior\tablenotemark{b}} & \colhead{Result} & \colhead{Prior} & \colhead{Result} & \colhead{Prior} & \colhead{Result} & \colhead{Prior} & \colhead{Result} & \colhead{Prior} & \colhead{Result}
}
\startdata
$\pi_{k}$ & U(0, 1) & 0.047 $\pm$ 0.003 & U(0, 1) & 0.182 $\pm$ 0.019 & U(0, 1) & 0.467 $\pm$ 0.026 & U(0, 1) & 0.296 $\pm$ 0.023 & U(0, 1) & 0.008 $\pm$ 0.001 \\
$\mu_{\alpha, k}$ (mas yr$^{-1}$) & G(0, 0.2) & -0.01 $\pm$ 0.014 & U(-7, 4) & -0.69 $\pm$ 0.05 & U(-7, 4) & -1.75 $\pm$ 0.07 & U(-7, 4) & -1.90 $\pm$ 0.08 & U(-7, 4) & -0.76 $\pm$ 1.36 \\
$\mu_{\delta, k}$ (mas yr$^{-1}$) & G(0, 0.2) & -0.34 $\pm$ 0.014 & U(-7, 4) & -1.01 $\pm$ 0.06 & U(-7, 4) & -2.65 $\pm$ 0.10 & U(-7, 4) & -2.89 $\pm$ 0.10 & U(-7, 4) & -0.20 $\pm$ 1.44\\
$\sigma_{a, k}$ (mas yr$^{-1}$) & U(0, 4) & 0.15 $\pm$ 0.01 & U(0, 20) & 1.27 $\pm$ 0.08 & U(0, 20) & 2.68 $\pm$ 0.05 & U(0, 20) & 3.46 $\pm$ 0.09 & U(0, 20) & 14.41 $\pm$ 1.24 \\
$\sigma_{b, k}$ (mas yr$^{-1}$) & $\sigma_b$ = $\sigma_a$ & 0.15 $\pm$ 0.01 & U(0, $\sigma_{a, k}$) & 0.66 $\pm$ 0.05 & U(0, $\sigma_{a, k}$)  & 1.39 $\pm$ 0.06 & U(0, $\sigma_{a, k}$) & 3.21 $\pm$ 0.09 & U(0, $\sigma_{a, k}$) & 11.24 $\pm$ 1.01\\
$\theta_{k}$ (rad) & --- & 0 & U(0, $\pi$) & 0.93 $\pm$ 0.04 & U(0, $\pi$) & 0.99 $\pm$ 0.02 & U(0, $\pi$) & 1.01 $\pm$ 0.14 & U(0, $\pi$) & 0.79 $\pm$ 0.21\\
\enddata
\label{PopModel}
\tablenotetext{a}{Description of parameters: $\pi_k$ = fraction of stars in Gaussian; $\mu_{\alpha, k}$ = RA-velocity centroid of Gaussian; $\mu_{\delta,k}$ = DEC-velocity centroid of Gaussian; $\sigma_{a,k}$ = semi-major axis of Gaussian; $\sigma_{b,k}$ = semi-minor axis of Gaussian; $\theta_{k}$ = angle between $\sigma_{a,k}$ and the RA-axis}
\tablenotetext{b}{Uniform distributions: U(min, max), where min and max are bounds of the distribution; Gaussian distributions: G($\mu$, $\sigma$), where $\mu$ is the mean and $\sigma$ is the standard deviation}
\end{deluxetable}

\section{Revised Extinction Law}
\label{app:EL}
The extinction law used in this analysis is a revised version of the one presented in H18, which is derived from \emph{HST} observations of Red Clump stars in the Arches field and proper motion-selected main sequence stars in Westerlund 1. These samples represent highly reddened stellar populations located in the Galactic Plane. The revision is necessary because an error was discovered in the application of the photometric zeropoints to the F160W and F814W photometry in the H18 analysis, resulting in magnitudes that were systematically too faint by 0.072 mag and 0.134 mag, respectively. No other filters were effected, and since the error was restricted to how the zeropoints were applied, the zeropoints presented in Figure 3 of H18 are still correct. The revised extinction law is derived using the same methodology and model (free parameters, priors, etc) described in H18, but with the corrected F160W and F814W photometry. 

\begin{figure}[h]
\begin{center}
\includegraphics[scale=0.4]{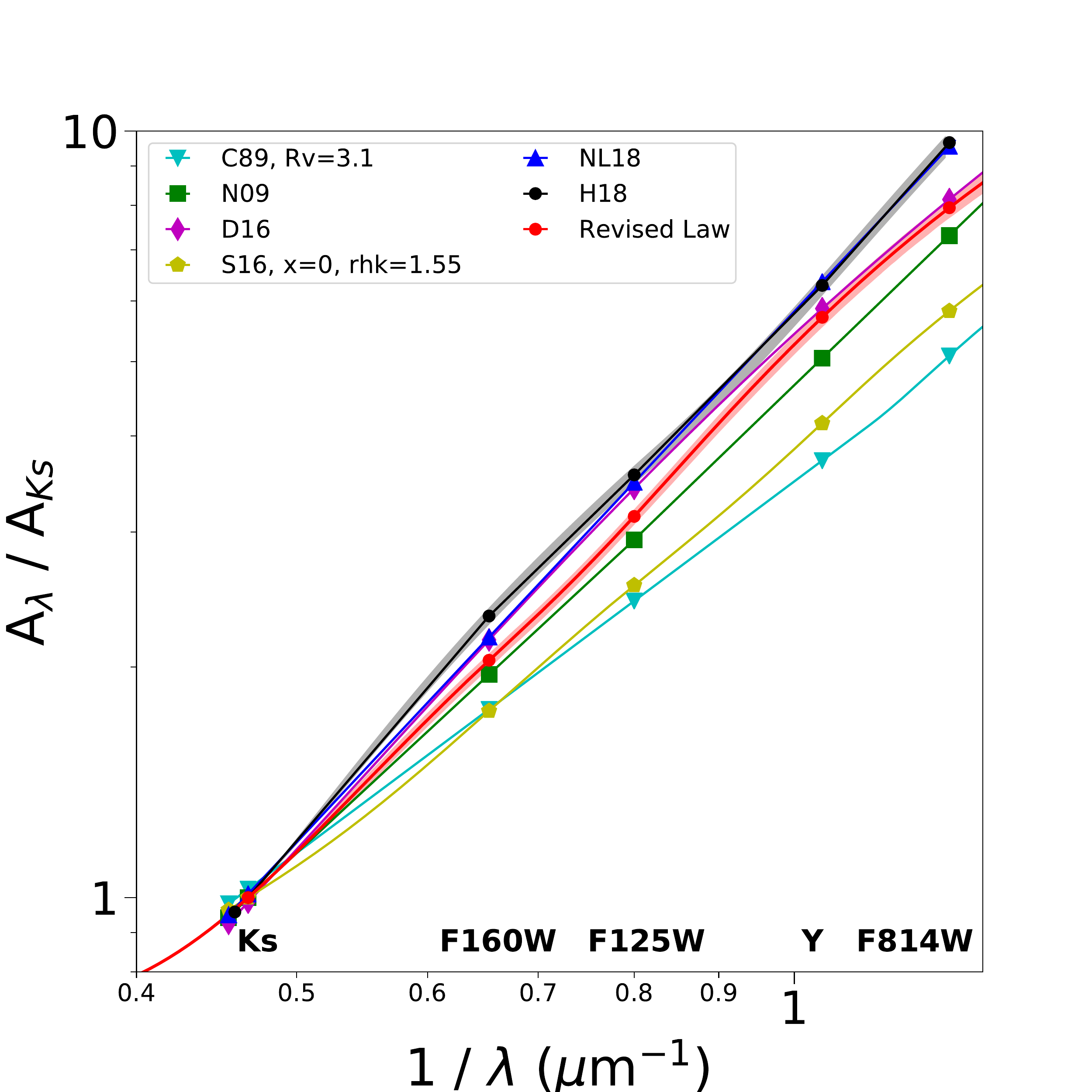}
\caption{The revised extinction law used in this analysis (red points) compared to other laws in the literature. The 1$\sigma$ uncertainty in the revised law is shown by the red shaded region. The other laws shown are from \citet[][cyan triangles]{Cardelli:1989qf}, \citet[][green squares]{Nishiyama:2009fc}, \citet[][magenta diamonds]{Damineli:2016no}, \citet[][yellow pentagons]{Schlafly:2016cr},  \citet[][blue triangles]{Nogueras-Lara:2018uq}, and \citet[][black points]{Hosek:2018lr}.
}
\label{fig:EL}
\end{center}
\end{figure}

\begin{figure}[h]
\begin{center}
\includegraphics[scale=0.4]{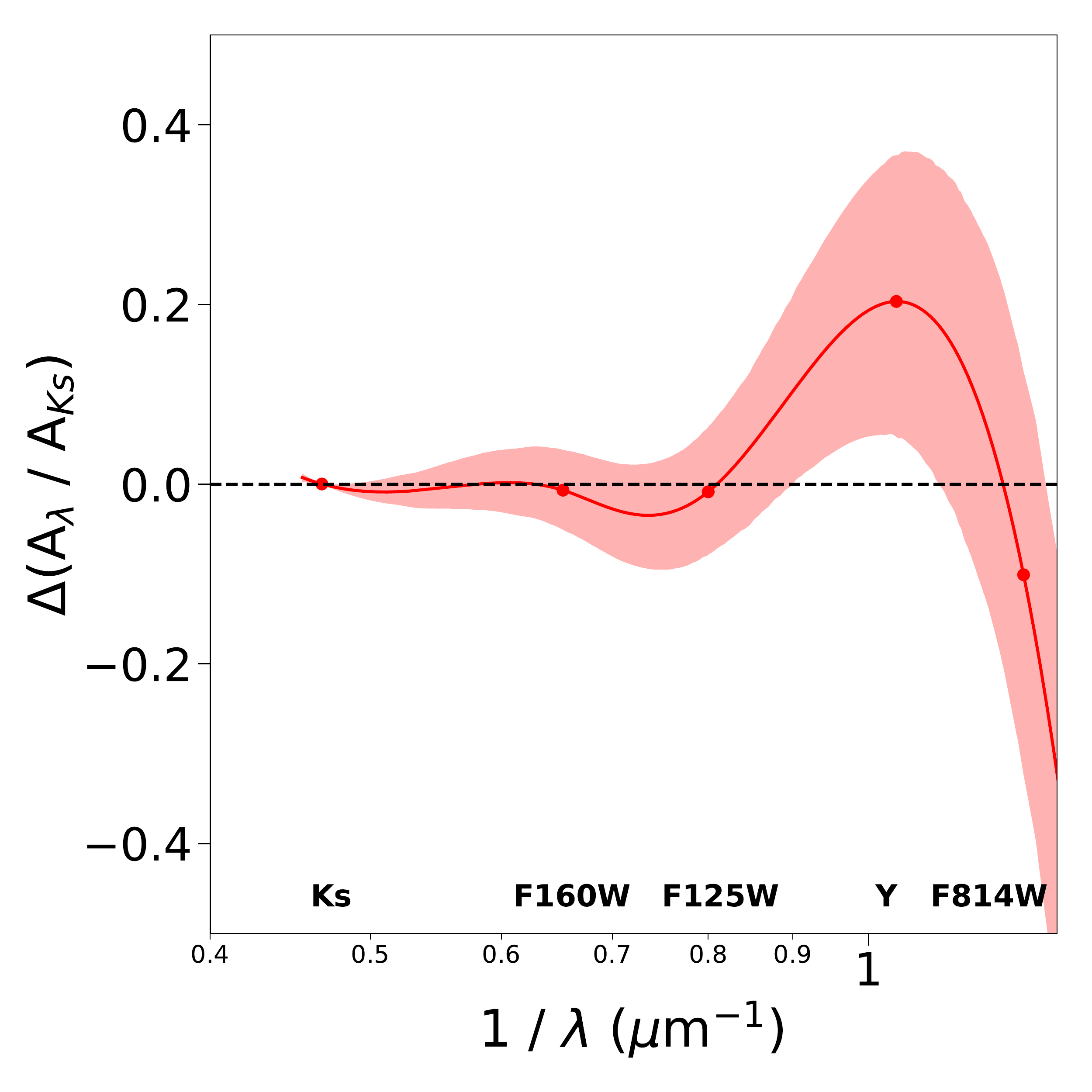}
\caption{The residuals between the best-fit power law (exponent: $\beta$ = 2.14) and the revised extinction law, as a function of wavelength. The 1$\sigma$ uncertainty in the law is shown by the red shaded region. The law is consistent with a power law for $\lambda >$ 1.25 $\mu$m but possibly deviates from a power law for $\lambda <$ 1.25 $\mu$m. 
}
\label{fig:PL}
\end{center}
\end{figure}

A comparison between the revised extinction law and other laws in the literature is shown in Figure \ref{fig:EL}. The revised law is shallower (i.e., lower A$_{\lambda}$ / A$_{Ks}$ values) than the original H18 law and the \citet{Nogueras-Lara:2018uq} law derived at the Galactic Center, but remains steeper than the  \citet{Nishiyama:2009fc} law often used for stars in the inner bulge. Longward of 1.25 $\mu$m, the revised law is consistent with a power law (A$_{\lambda}$ / A$_{Ks}$ $\propto$ $\lambda^{-\beta}$) with $\beta$ = 2.14, though there is some evidence that the law deviates from this function shortward of 1.25 $\mu$m (Figure \ref{fig:PL}). However, additional studies of the extinction law in this wavelength range are required to investigate further. The revised A$_{\lambda}$ / A$_{Ks}$ values and their corresponding errors are shown in Table \ref{tab:NewLaw}. 

As discussed in $\mathsection$\ref{sec:caveats}, the IMF results are found to be insensitive to variations in the extinction law, and so the choice of extinction law does not significantly impact the results in this paper.

\begin{deluxetable}{l l c c }
\tabletypesize{\footnotesize}
\tablecaption{Revised Extinction Law}
\tablehead{
\colhead{Parameter} & \colhead{$\lambda$\tablenotemark{a} ($\mu$m)} & \colhead{Prior\tablenotemark{b}} & \colhead{Result}
}
\startdata
 A$_{F814W}$ / A$_{K_s}$ &  0.806 & U(4, 14) & 7.94 $\pm$ 0.21 \\ 
 A$_{y}$ / A$_{K_s}$ & 0.962 & U(4, 14) & 5.72 $\pm$ 0.16 \\
A$_{F125W}$ / A$_{K_s}$ &  1.25 & U(1, 6) & 3.14 $\pm$ 0.07 \\
A$_{F160W}$ / A$_{K_s}$ &  1.53 & U(1, 6) & 2.04 $\pm$ 0.04 \\  
A$_{[3.6]}$ / A$_{K_s}$ & 3.545 & G(0.5, 0.05) & 0.50 $\pm$ 0.04 \\  
\enddata
\label{tab:NewLaw}
\tablenotetext{a}{HST + PanSTARRS filters: Pivot wavelengths of filter; IRAC [3.6] filter: isophotal wavelength from \citet{Nishiyama:2009fc}}
\tablenotetext{b}{Uniform distributions: U(min, max), where min and max are bounds of the distribution; Gaussian distributions: G($\mu$, $\sigma$), where $\mu$ is the mean and $\sigma$ is the standard deviation}  
\end{deluxetable}

\section{Arches Cluster: Model Posteriors}
\label{app:posteriors}
\begin{figure}[h]
\begin{center}
\includegraphics[scale=0.35]{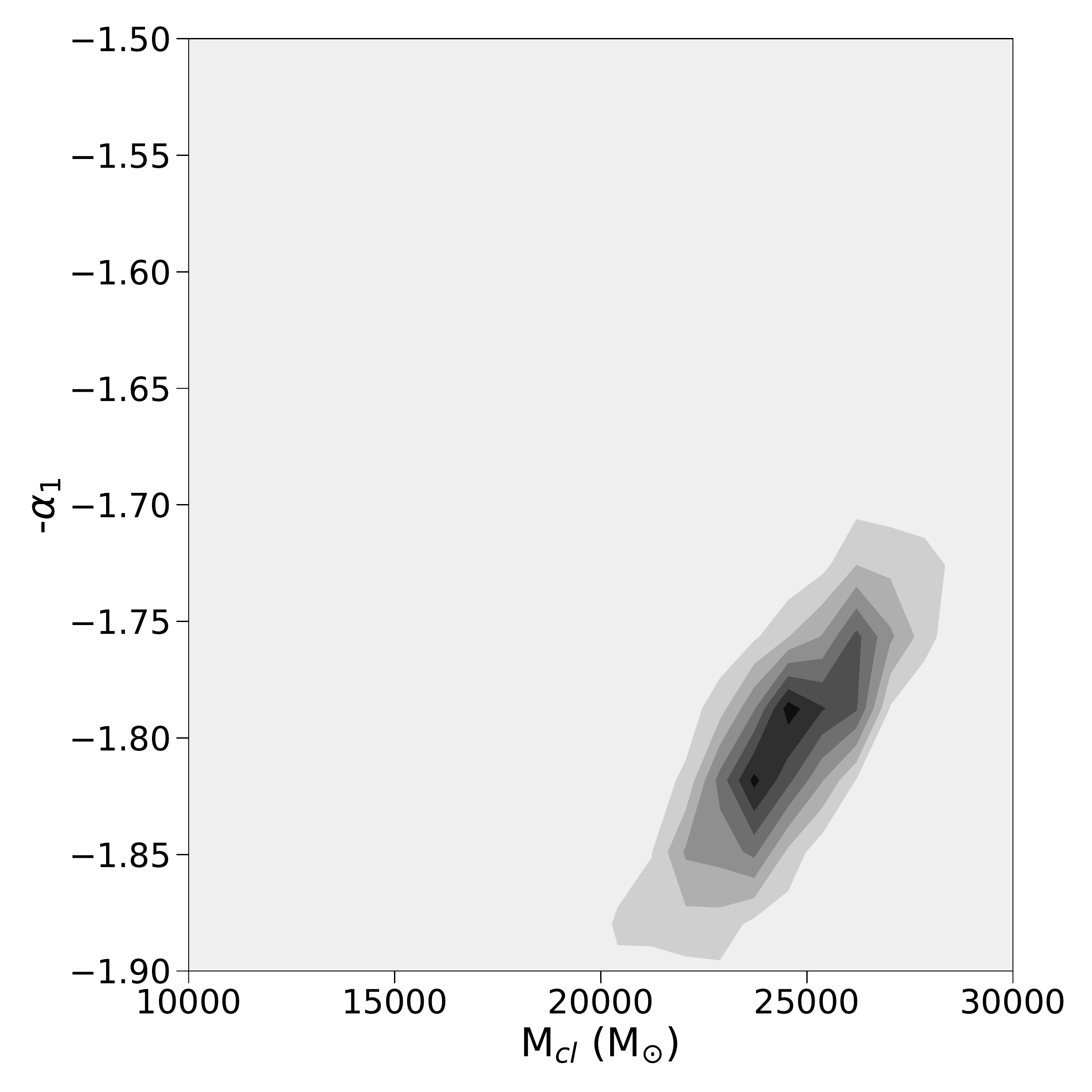}
\caption{The 2D posterior probability distribution for -$\alpha_1$ and M$_{cl}$ for the 1-segment IMF analysis for the Arches cluster.}
\label{fig:1seg2D}
\end{center}
\end{figure}

\begin{figure}[h]
\begin{center}
\includegraphics[scale=0.25]{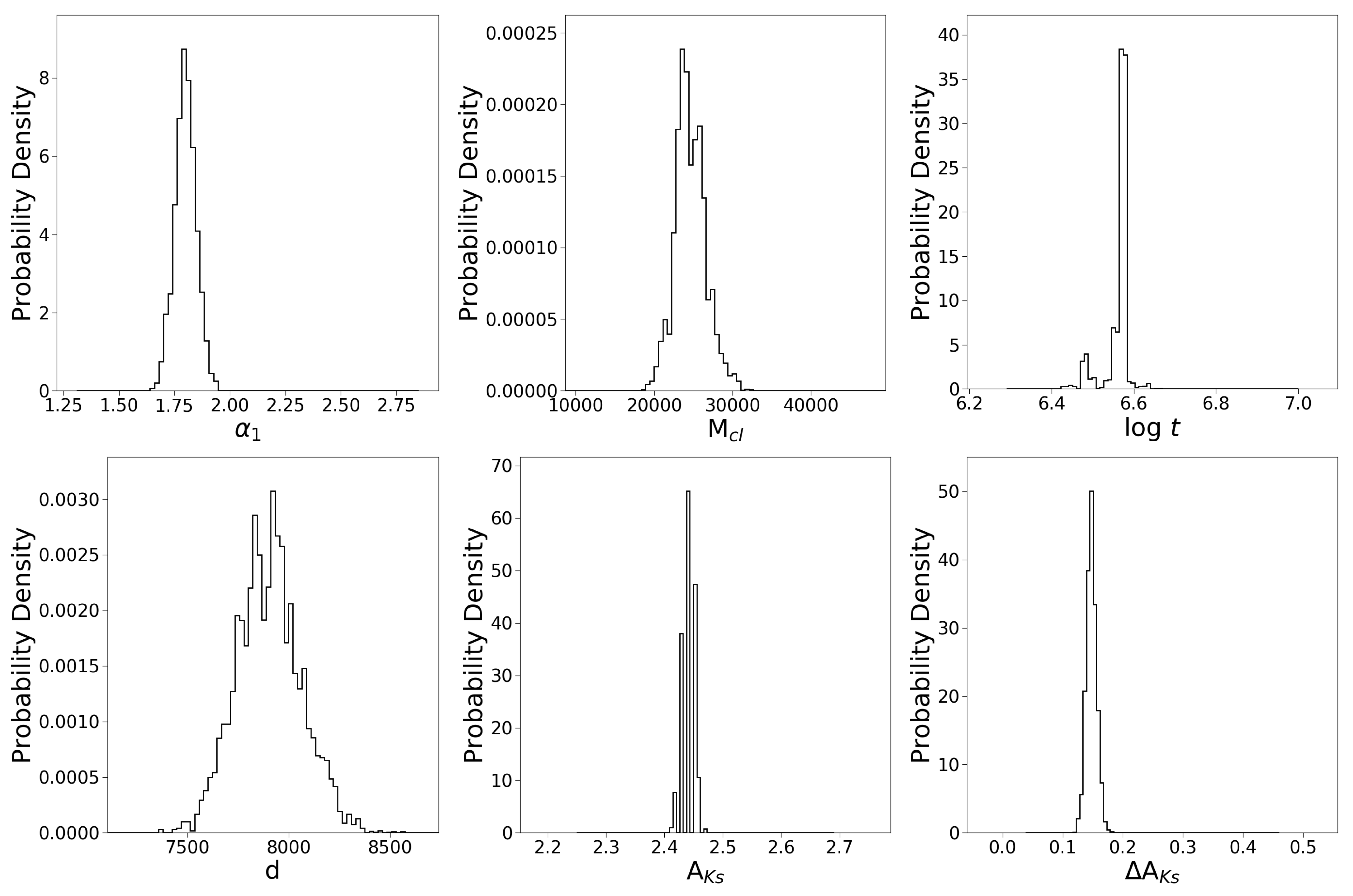}
\caption{The 1D posterior probability distributions for the 1-segment IMF model for the Arches cluster.}
\label{fig:1seg1D}
\end{center}
\end{figure}

\begin{figure}
\begin{center}
\includegraphics[scale=0.35]{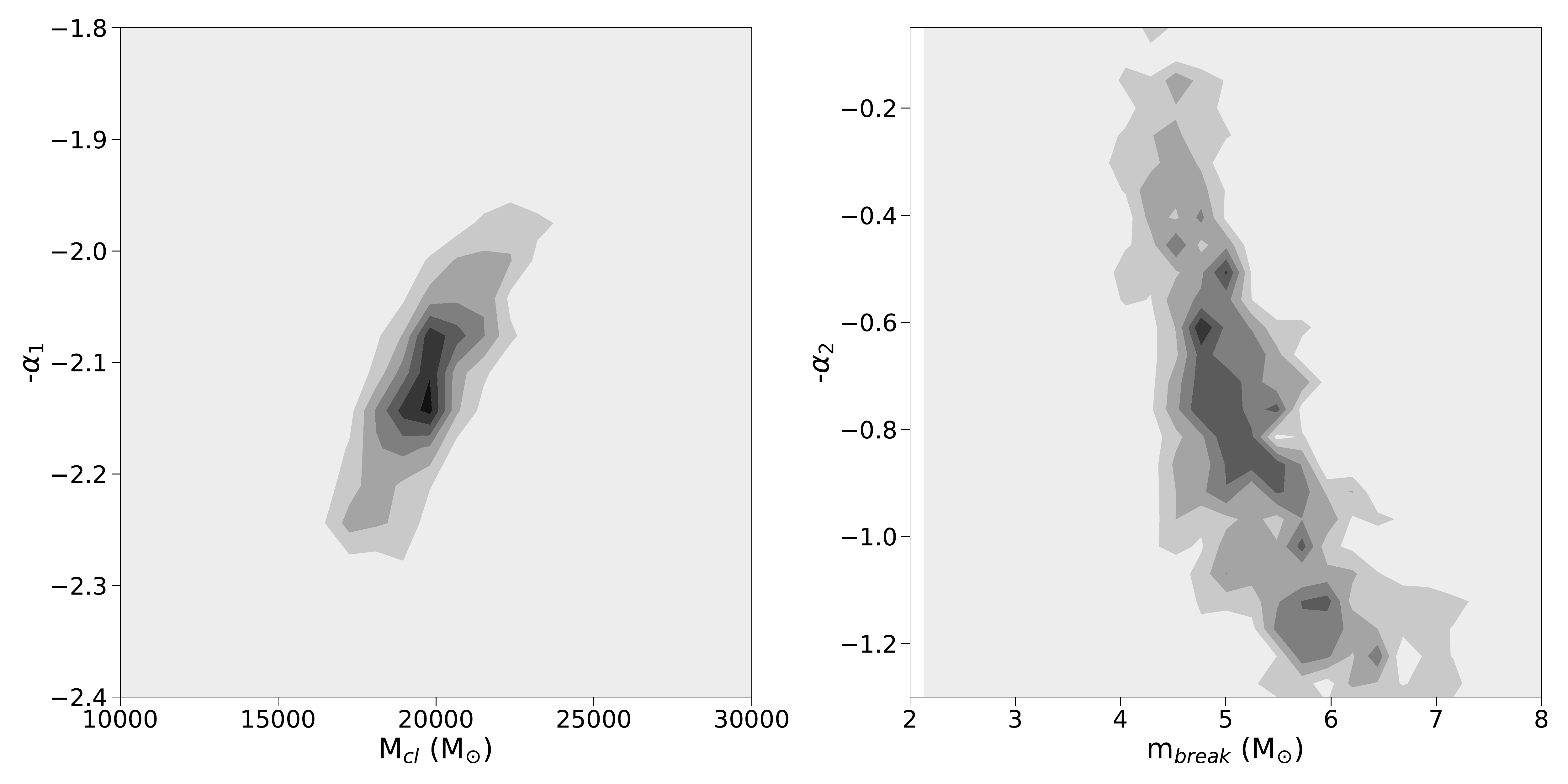}
\caption{The 2D posterior probability distribution for -$\alpha_1$ and M$_{cl}$ and -$\alpha_2$ and m$_{break}$ for the 2-segment IMF analysis for the Arches cluster.}
\label{fig:2seg2D}
\end{center}
\end{figure}

\begin{figure}
\begin{center}
\includegraphics[scale=0.2]{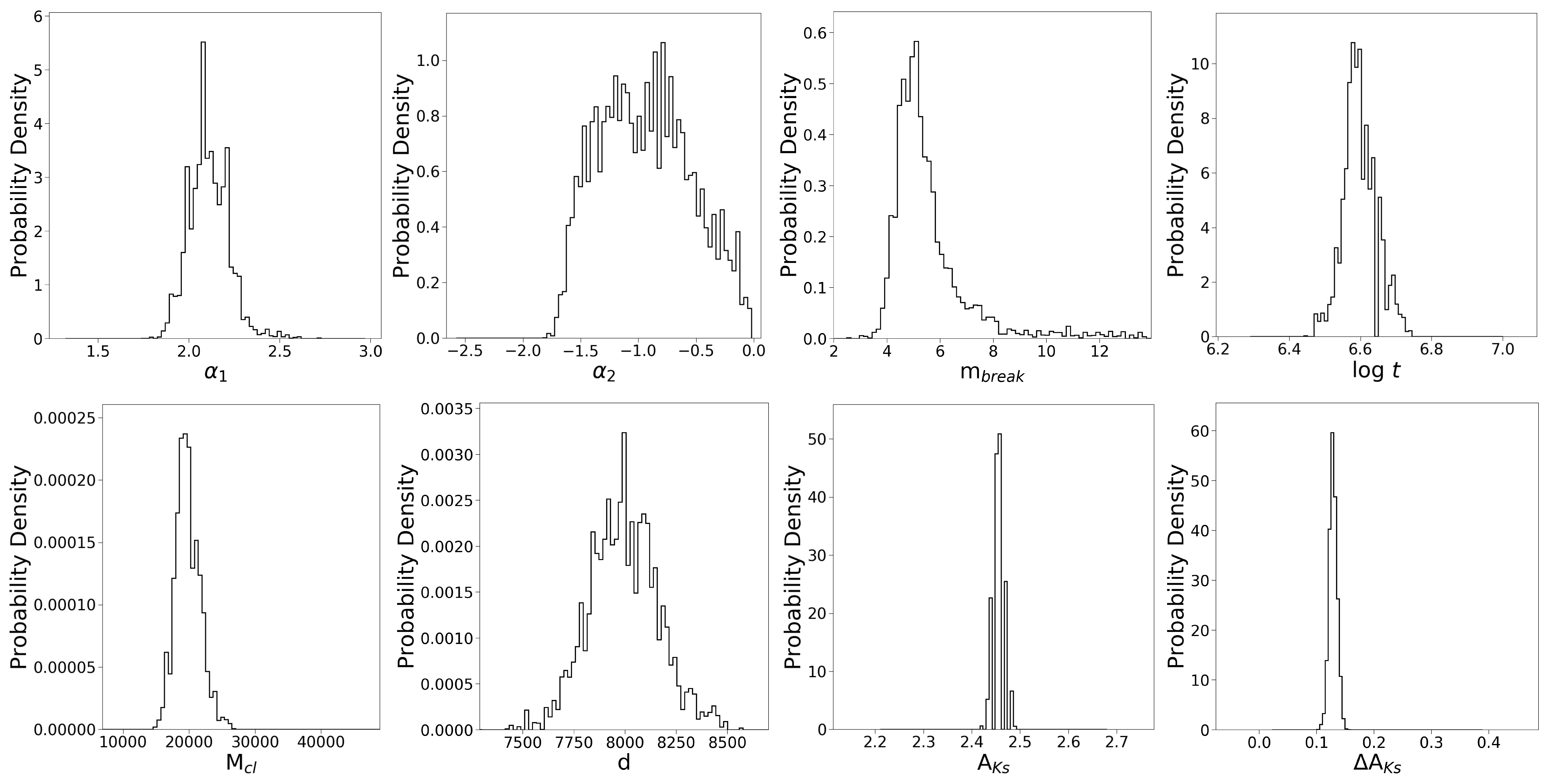}
\caption{The 1D posterior probability distributions for the 2-segment IMF model for the Arches cluster.}
\label{fig:2seg1D}
\end{center}
\end{figure}

In this appendix we show the posterior probability distributions for the 1-segment IMF and 2-segment IMF analyses. For the 1-segment IMF fit, we show the joint posterior distribution for $\alpha_1$ and M$_{cl}$ in Figure \ref{fig:1seg2D} and the 1D posteriors for each model parameter in Figure \ref{fig:1seg1D}. The corresponding posteriors for the 2-segment IMF fit posteriors are shown in Figures \ref{fig:2seg2D} and \ref{fig:2seg1D}.

\section{Testing the IMF Analysis with Synthetic Clusters}
\label{app:artificialTest}
To verify the accuracy of the IMF analysis, we apply it to simulated observations of a synthetic cluster and compare the output best-fit parameters with the input ones. The synthetic cluster is created as described in $\mathsection$\ref{sec:model} and observational completeness applied as a function of position in the CMD and cluster radius. To simulate observational errors, the synthetic photometry for each star is perturbed by a random amount drawn from a normal distribution with a width equal to the median photometric error of the observed stars at the synthetic star's magnitude. These stars are assigned $P_{pm}$ = 1. To simulate field stars, a number of stars are drawn from the observed field star population used to calculate $p(k | \Theta)_{f, obs}$ in Equation \ref{eq:field} and are assigned $P_{pm}$ = 0. The number of field stars drawn is chosen such that the combined sample contains 80\% cluster stars and 20\% field stars. The spectroscopic sample is simulated by selecting 6 random stars with 14.5 mag $\leq$ F153M $\leq$ 15.0 mag and assigning them T$_{eff}$ uncertainties similar to those found in $\mathsection$\ref{sec:specResults}.

The combined synthetic catalog is run through the Bayesian analysis in $\mathsection$\ref{sec:bayes} in the same way as the real observed catalog, with two exceptions: no differential de-reddening correction is applied, since the cluster is already generated with a realistic value of $\Delta$A$_{Ks}$, and no minimum $P_{pm}$ value is enforced. The number of WR stars within $r_{cl} <$ 0.75 pc is calculated and input to the fitter, mimicking the information gained from the real spectroscopic surveys of the Arches. The priors are the same as the real analysis, as described in Table \ref{tab:Params}.

\begin{figure}
\begin{center}
\includegraphics[scale=0.35]{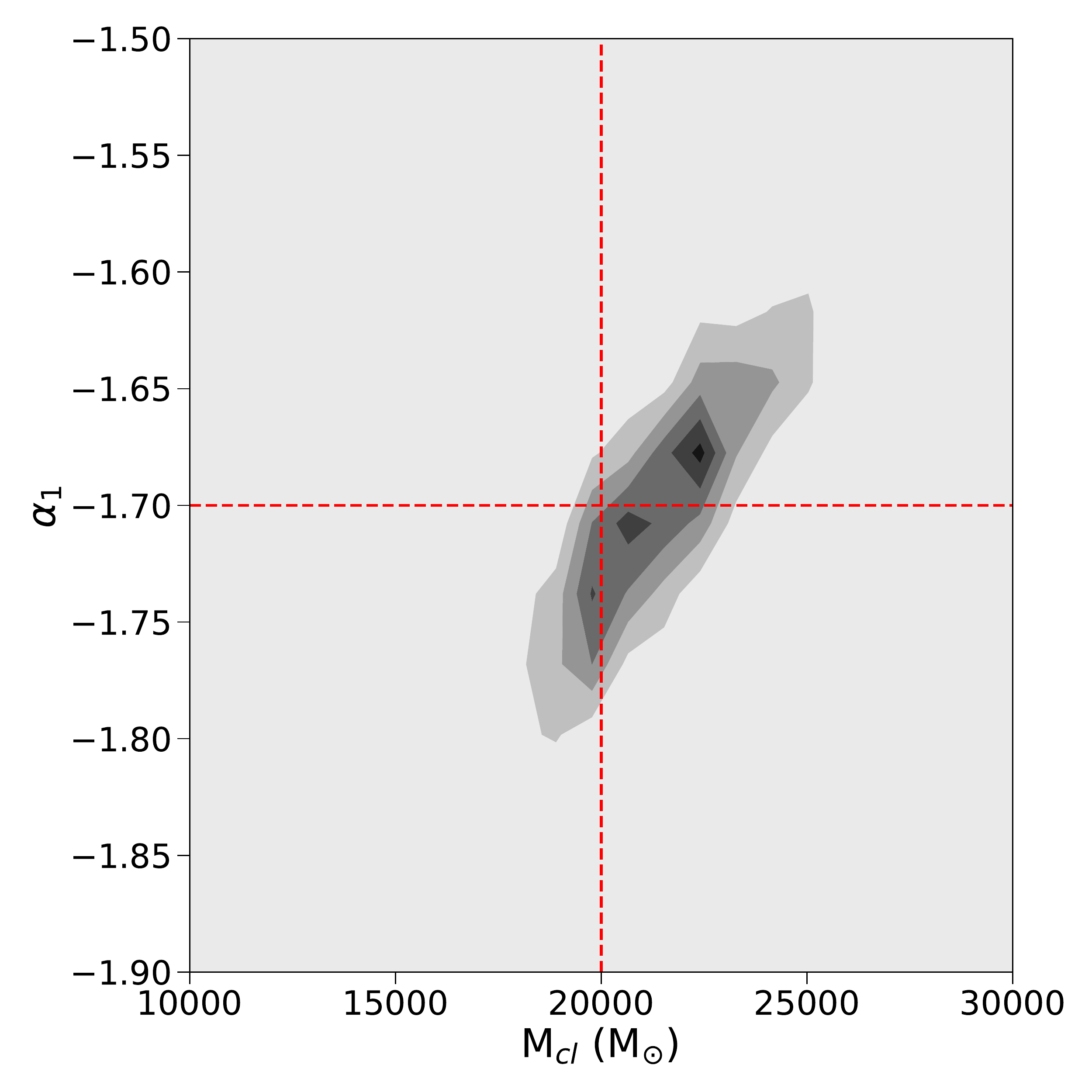}
\caption{The 2D posterior probability distribution for -$\alpha_1$ and M$_{cl}$ for the 1-segment IMF simulated cluster analysis. The input values are represented by the red dotted lines.}
\label{fig:1segTest}
\end{center}

\end{figure}

\begin{figure}
\begin{center}
\includegraphics[scale=0.35]{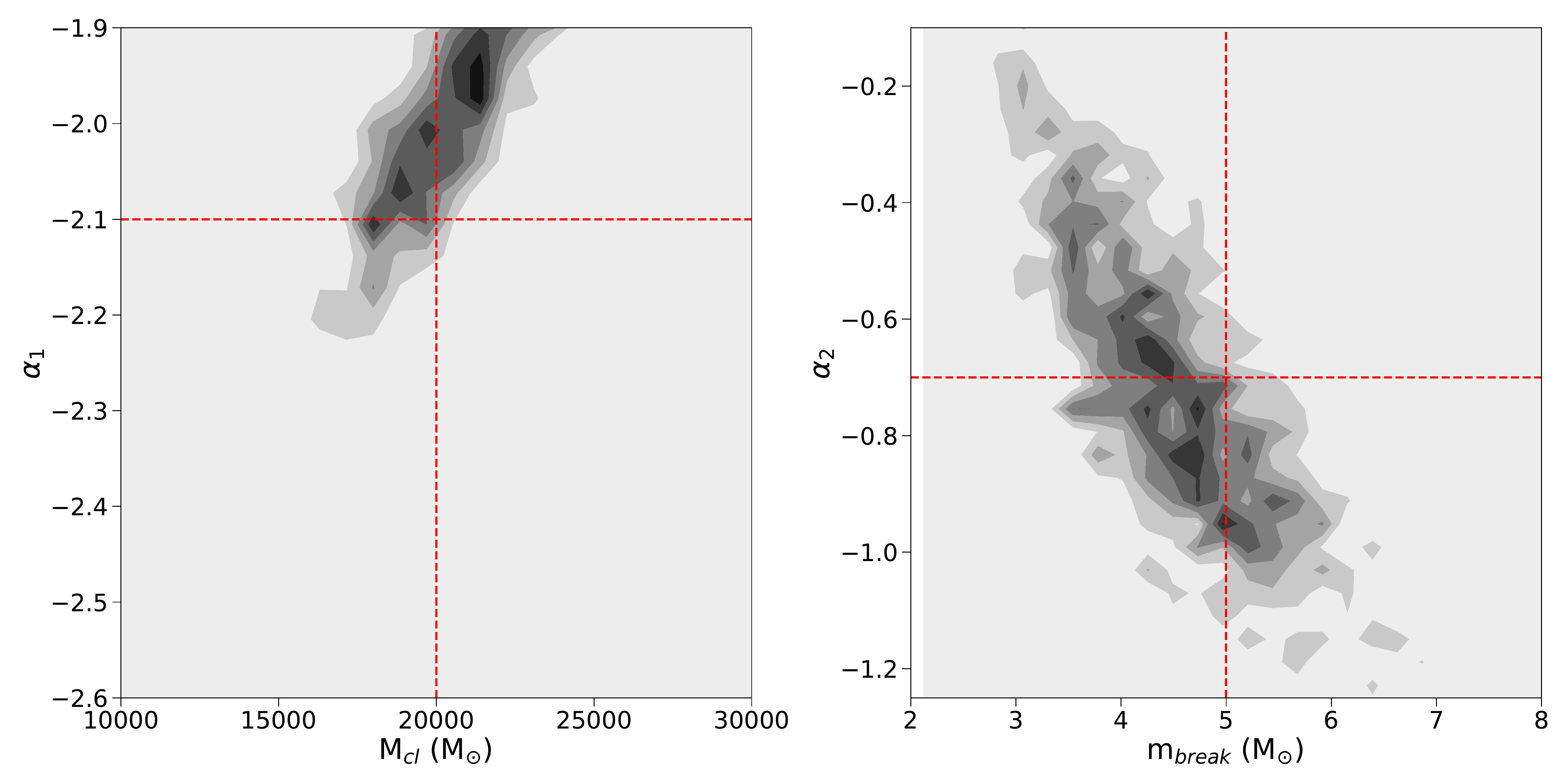}
\caption{The joint posterior probability distribution for -$\alpha_1$ and M$_{cl}$ (left) and -$\alpha_2$ and m$_{break}$ (right) for the 2-segment IMF simulated cluster analysis. The input values are represented by the red dotted lines.}
\label{fig:2segTest}
\end{center}
\end{figure}

The results of the tests are shown in Table \ref{tab:artificial}, which found the output values to match the input values to within 1$\sigma$. The joint posterior probability distributions for $\alpha_1$ and M$_{cl}$ in the 1-segment IMF fit is shown in Figure \ref{fig:1segTest}, while the joint posterior probability distributions for $\alpha_1$ and M$_{cl}$ and $\alpha_2$ and m$_{break}$ in the 2-segment IMF fit is shown in Figure \ref{fig:2segTest}.

\begin{deluxetable}{l | c c | c c}
\tablewidth{0pt}
\tabletypesize{\footnotesize}
\tablecaption{Simulated Cluster Analyses\tablenotemark{a}}
\tablehead{
& \multicolumn{2}{c}{1-segment IMF} &  \multicolumn{2}{c}{2-segment IMF} \\
\colhead{Parameter} & \colhead{Input Value} & \colhead{Recovered Value} & \colhead{Input Value} & \colhead{Recovered Value}
}
\startdata
$\alpha_1$ & 1.7 & 1.7 $\pm$ 0.06 & 2.1 & 1.99 $\pm$ 0.13\\
$\alpha_2$ & --- & ---  & 0.7 & 0.74 $\pm$ 0.27 \\
m$_{break}$ & --- & --- & 5.0 & 4.43 $\pm$ 0.91 \\
M$_{cl}$ & 20000 & 21400 $\pm$ 1900 & 20000 & 20400 $\pm$ 2300 \\
log $t$ & 6.40 & 6.41 $\pm$ 0.03 & 6.40 & 6.39 $\pm$ 0.01 \\
$d$ & 8000 & 7865 $\pm$ 146 & 8000 & 8101 $\pm$ 139 \\
A$_{Ks}$ & 2.07 & 2.07 $\pm$ 0.01 & 2.07 & 2.06 $\pm$ 0.01 \\
$\Delta$A$_{Ks}$ & 0.15 & 0.13 $\pm$ 0.01 & 0.15  & 0.14 $\pm$ 0.01 \\
\enddata
\tablenotetext{a}{Parameter priors and units are the same as Table \ref{tab:Params}}
\label{tab:artificial}
\end{deluxetable}


\end{document}